\input BoxedEPS
   \SetepsfEPSFSpecial %% ******* will work for many:
\HideDisplacementBoxes

\documentstyle[aps,prd]{revtex}

\newcommand{\bi}{\begin{itemize}}
\newcommand{\ei}{\end{itemize}}
\newcommand{\be}{\begin{equation}}
\newcommand{\ee}{\end{equation}}
\newcommand{\ba}{\begin{eqnarray}}
\newcommand{\ea}{\end{eqnarray}}

\newcommand{\tb}{t_{_{bb}}}
\newcommand{\Rhoav}{\langle \rho \rangle}
\newcommand{\rhoi}{\rho_i}
\newcommand{\rhoav}{\langle \rho_i \rangle}
\newcommand{\Ri}{^{(3)}{\cal{R}}_i}
\newcommand{\Riav}{\langle ^{(3)}{\cal{R}}_i\rangle}
\newcommand{\R}{{^{(3)}\cal{R}}}
\newcommand{\Rav}{\langle ^{(3)}{\cal{R}}\rangle}

\newcommand{\Dmi}{\Delta_i^{(m)}}
\newcommand{\Dki}{\Delta_i^{(k)}}
\newcommand{\M}{{\cal{M}}}
\newcommand{\K}{{\cal{K}}}

\newcommand{\bRi}{{}^{(3)}{\bar{\cal{ R}}}_i}
\newcommand{\Ti}{{\cal{T}}_i}
\newcommand{\rmax}{r_{\hbox{\tiny{max}}}}

\begin{document}

\title{New variables for the Lema\^{\i}tre-Tolman-Bondi dust
solutions.}
\author{Roberto A. Sussman$^\dagger$ and Luis Garc\'{\i}a Trujillo$^\ddagger$}
\address
{$^\dagger$
Instituto de Ciencias Nucleares,  Apartado Postal 70543, UNAM, M\'exico DF,
04510, M\'exico.\\ $^\ddagger$ Instituto de F\'{\i}sica, Universidad de Guanajuato, Leon,
Guanajuato, M\'exico. }

\maketitle

\begin{abstract}
We re-examine the Lem\^aitre-Tolman-Bondi (LTB) solutions with a dust source admitting 
symmetry centers. We consider as free parameters of the solutions the initial
value functions: $Y_i,\,\rho_i,\,\Ri$,  obtained by restricting the curvature radius, $Y\equiv
\sqrt{g_{\theta\theta}}$, the rest mass density, $\rho$, and the 3-dimensional Ricci scalar of
the rest frames, $\R$, to an arbitrary regular Cauchy hypersurface, $\Ti$, marked by constant
cosmic time ($t=t_i$). Using $Y_i$ to fix the radial coordinate and the topology
(homeomorphic class) of $\Ti$, and scaling the time evolution in
terms of an adimensional scale factor $y=Y/Y_i$, we show that the dynamics, regularity
conditions and geometric features of the models are determined by $\rhoi,\,\Ri$
and by suitably constructed volume averages and contrast functions expressible in
terms of invariant scalars defined in $\Ti$. These quantities lead to a
straightforward  characterization of initial conditions in terms of the nature of the
inhomogeneity of  $\Ti$, as density and/or curvature overdensities (``lumps'') and
underdensities (''voids'') around a symmetry center. In general, only models with initial
density and curvature lumps evolve without shell crossing singularities, though special classes
of initial conditions, associated with a simmultaneous big bang, allow for a regular evolution
for initial density and curvature voids. Specific restrictions are found so that a regular
evolution for $t>t_i$ is possible for initial voids. A step-by-step guideline is provided for
using the new variables in the construction of LTB models and for plotting all relevant
quantities.   
\end{abstract}

\section {Introduction}\label{intro}

Lema\^{\i}tre-Tolman-Bondi metrics with a dust source and a comoving and geodesic
4-velocity constitute a well known  class of exact solutions of
Einstein's field equations\cite{KSMH}\cite{krasinski}. Under the prevailing
conditions of a present day, standard, matter domminated universe, a dust
momentum-energy tensor is a good approximation to non-relativistic matter
sources. This fact, together with their mathematical simplicity, makes LTB
solutions adequate models of inhomogeneous configurations of non-relativistic
and collisionless matter (either barions or CDM). The relevance of such models
is evident as they complement the usual perturbative approach by allowing one to
study a case of exact non-linear evolution of inhomogeneities 
\cite{goodwain} \cite{silk} \cite{krasinski} \cite{HMM2}. Also, recent
literature\cite{cel} \cite{mustapha} illustrates how a nearly isotropic CMB might not
rule out an inhomogeneous universe compatible with current CBR observations.

The process of integration of the field equations for an LTB metric with a dust
source in a comoving frame is very straightforward. This process, reviewed in
section II, leads to a Friedmann-like dynamical equation for the curvatue
radius $Y=\sqrt{g_{\theta\theta}}$, containing two arbitrary functions of the
radial coordinate: $M(r)$ and $E(r)$. The integration of the dynamical equation
yields a third arbitrary function, $\tb(r)$. The first two of these functions
($M$ and $E$) are usualy identified as the relativistic analogues of the
newtonian mass and the local energy associated with the comoving dust shells, the
third function ($\tb$) can be interpreted as the ``bang time'', marking the
proper time for the initial (or final) singularity for each comoving oberver. The
functions  $M,\,E,\,\tb$ are then the free parameters of the solutions, though
any one of them can always be used to fix the radial coordinate, thus there are
realy only two free functions.  Regularity conditions, related to well behaved
symmetry centers, abscence of shell crossing singularities and surface layers,
can be given as restrictions on these functions and their radial gradients.  
These functions are usualy selected by means of convenient ansatzes based on the
interpretations described above and/or the compatibility with regularity
conditions or observational criteria. Once the free parameters have been selected we
have a completely determined LTB model. 

Since the standard free functions lead to a consistent description of LTB models,
these free functions are used in most literature dealing with these solutions (see
 \cite{krasinski} for a comprehensive and autoritative review of this literature).
There are papers considering specific modifications of the standard free functions in
order to formulate initial conditions for studying the censorship of singularities
\cite{joshi}, or aiming at an ``intial value'' formulation of LTB
models \cite{gromov}.  We believe that new variables, defined along an arbitrary and
regular ``initial'' hypersurface $t=t_i$, lead to a more intuitive understanding of
LTB solutions. Therefore, we propose in this paper using a new set of basic free
functions along the following steps 

\bi
\item{} Consider the rest mass density, $\rho$, the 3-dimensional Ricci scalar, $\R
$, and the curvature radius, $Y$, all of them evaluated along an arbitrary Cauchy
hypersurface, $\Ti$, marked by constant cosmic time $t=t_i$. This leads to:
$\rho_i(r)$, $\Ri(r)$ and $Y_i(r)$, where the subindex $_i$ indicates evaluation along
at $t=t_i$ (we shall use this convention henceforth). 

\item{} Rescale $Y$ with $Y_i$, leading to an
adimensional scale factor $y=Y/Y_i$, so that we can distinguish between the
regular vanishing of $Y$ (at a symmetry center, now given as $Y_i=0$) and a
central singularity (now characterized as $y=0$). Likewise, we can distinguish
between the two different situations when $Y'=0$: regular vanishing or surface layers
if $Y_i'=0$ for a single value of $r$ (thus restricting gradients of $\rhoi$ and
$\Ri$) and shell crossings when $\Gamma=(Y'/Y)/(Y_i'/Y_i)=0$. 

\item{} Among the three free functions, $\rho_i(r)$, $\Ri(r)$ and $Y_i(r)$, we fix the
radial coordinate (as well as the topology of $\Ti$) by a convenient selection of
$Y_i$, using the remaining two functions as initial value functions. 

\item{} From the latter functions we can construct suitable volume averages and
contrast functions, all of which expressible in terms of invariant scalars defined in
$\Ti$. These auxiliary functions provide a clear cut and precise characterization of
the nature of the initial inhomogeneity around a symmetry center, as overdensities
(``lumps'') or underdensities (``voids'') of $\rhoi$ and $\Ri$.     
\ei
The old functions $M,\,E,\,t_{_{bb}}$ and their gradients can always be recovered and
expressed in terms of the above mentioned volume averages and contrast functions.
Hence, all known  regularity conditions given in the old variables can
easily be recast in terms of the new ones. In particular, the known conditions for
avoiding shell crossing singularities \cite{he_la2} (see also \cite{humph_phd} and
\cite{HM1}) have a more appealing and elegant form with the new variables, since with
the old variables the signs of the gradients $M',\,E',\,\tb'$ do not distinguish
manifestly between lumps or voids. In terms of the new variables, it is evident that,
in general, shell crossings are avoided only for lumps of $\rhoi$ and $\Ri$. However,
some particular cases of initial conditions, such as a ``simmultaneous  bang
time'' ($\tb'=0$), do allow for some initial conditions given as voids to evolve
without shell crossings. If we consider an evolution in the time range $t>t_i$, then
(under specific restrictions) initial voids can also evolve without shell crossings.
All these features could also emerge with the old free functions, but the new
variables provide a more straightforward and intuitive characterization of the
effect of initial conditions in the regular evolution of LTB models.

The new approach to LTB dust solutions presented in this paper accounts, not only for
an initial value treatment on a regular Cauchy hypersurface $\Ti$, but to a new
formalism based on rescaling the evolution of the models to variables defined in this
hypersurface.  Some aspects of this formalism (average and contrast functions) have
been developed elsewhere (see \cite{pavsuss} and \cite{suss} for applications of LTB metrics
with viscous fluid sources).  Variables similar to the ones introduced here have been defined
in \cite{bon1}, \cite{joshi}, \cite{gromov}, \cite{liu} and \cite{goodwain} (the latter in
connection with Szekeres dust solutions, see equations 2.4.8 to 2.4.14 of \cite{krasinski}).
Mena and Tavakol \cite{ellisvan} have claimed that some of these variables are not
covariant, thus suggesting covariant expressions for density contrasts (see
also \cite{mentav}). Although we show that the new variables can be
expressed in terms of invariant scalars, and so are coordinate independent and can be
computed for any coordinate system, we do not claim that the density and curvature
averages and contrast functions introduced here have a clear ``covariant
interpretation'', or that they are unique. However, what we can certainly claim is
that these new variables emerge naturaly from the field equations, provide an
adequate initial value treatment and yield a clear and concise description of
regularity conditions.   

The paper is organized as follows: section \ref{oldvars} reviews the basic
expressions (metric, evolution equation and its solutions) characteristic of LTB
models in the usual variables.  The new variables  are introduced
in section \ref{newvars}, leading to new forms for the metric, the rest mass density
and the solutions of the evolution equation. Volume averages and contrast functions
along $\Ti$ are defined in section \ref{ave_contrast}. Section \ref{int_newvars}
examines in detail generic properties of the new variables: regularity, symmetry centers and
geometric features of the hypersurface $\Ti$ (subsection \ref{init_Ti}),
characterization of singularities (subsection \ref{comp}), choice of radial coordinate and
homeomorphic class (topology) of $\Ti$ (subsection \ref{Yi_etc}), expressing the new 
variables in terms of invariant scalars (subsection \ref{inv_expr}), FLRW and Schwarzschild
(subsection limits \ref{limits}). In section \ref{lumps_voids} we look at the characterization
of the inhomogneity of $\Ti$ in terms density and 3-curvature ``lumps'' and ``voids'' and its
relation with the contrast and average functions. In section \ref{Gamma_etc} we use the new
variables in order to look at the conditions to avoid shell crossings, big bang times,
regularity of the  selected hypersurface $\Ti$ and simmultaneous big bang times. A summary
of the no-shell-crossing conditions is given in Table 1. Section
\ref{ex_LTB} provides simple functional ansatzes for the new variables and their auxiliary
quantities (volume averages, contrast functions, as well as the old variables and the big bang
times). A simple step-by-step guideline is provided for the construction of LTB models and for
plotting all relevant quantities derived from these ansatzes. The characterization of
intial conditions as ``lumps'' and ``voids'', as well as the fulfilment of the no-shell-crossing
conditions is illustrated and tested graphicaly in 12 figures that complement the information
provided in the text. Finaly, the Appendix provides a useful summary (in terms of the usual
variables) of the generic geometric features of the solutions (regularity conditions,
singularities, symmetry centers, surface layers).  

Some important topics have been left out in this paper: time evolution of lumps and voids by
generalizing volume averages and contrast functions to arbitrary times ($t\ne t_i$), models
without centers, models with a mixed dynamics (elliptic and hyperbolic, etc), matchings of
various LTB models, etc. The examination of these topics with the formalism presented here is
currently in progress \cite{sussgar}.

\section{LTB solutions, the usual approach.}\label{oldvars}

 Lemaitre-Tolman-Bondi (LTB) solutions are characterized by the line element 
\be
ds^2 \ = \ -c^2dt^2 \ + \ \frac{Y'^{2}}{1\,+\,E} \,dr^2
\ + \ Y^2\left[d\theta^2+\sin^2 (\theta) d\phi^2
\right]\,\label{ltbmetric}
\ee
where $Y = Y(t,r)$,\, $E=E(r)$ and a prime denotes derivative with respect to
$r$. The usual dust source associated with (\ref{ltbmetric}) is
\be
T^{ab}= \rho \ u^au^b, \label{Tdust}
\ee
where the 4-velocity is comoving  $u^a=c\delta^a_t $.  
and  $\dot{Y}=u^aY_{,a}=Y_{,ct}$.  Einstein's field equations for (\ref{ltbmetric})
and (\ref{Tdust}) reduce to  
\be\dot Y^2 \ = \ \frac{2\,M}{Y} \ + \ E,\label{freq1}
\ee
\be \frac{4\pi G}{c^4}\,\rho \ = \ \frac{M'}{Y^2\,Y'},\label{rho1} \ee
where 
\be M \ \equiv \ \frac{4\pi G}{c^4}\,m(r),\label{defm}\ee
and $m(r)$ is an arbitrary function (with units of energy)
that emerges as an ``integration constant''. The form of the integrals (ie
antiderivatives) of (\ref{freq1}) depends on the sign of the function $E(r)$.
Since (\ref{freq1})  is the dynamical equation for LTB solutions, the usual
convention classifies their possible dynamical evolution according to the three
possible signs of $E$, as ``parabolic'' ($E=0$), ``elliptic'' ($E<0$) and
``hyperbolic'' ($E>0$). Two possible situations arise: either $E$ has a specific
(positive, negative or zero) sign for the full domain of regularity of
$r$, or only for an open subset, $r_1 < r < r_2$, possibly changing sign for other
subsets of this domain. Whenever is necessary we will distinguish these two
situations by using the terms ``parabolic, elliptic or hyperbolic solutions''
(full domain) and ``parabolic, elliptic or hyperbolic regions'' (subset). A
convenient terminology that includes genericaly both situations is that of
``parabolic, elliptic or hyperbolic dynamics''. 

Only for parabolic dynamics we can obtain a closed integral of (\ref{freq1})
expressible in the form $Y=Y(t,r)$. For elliptic and hyperbolic dynamics we have
either an implicit canonical solution of the form $c\,t-c\,\tb(r)=F(Y,r)$, were $F$ is
the integral quadrature of (\ref{freq1}) and $\tb$ is an arbitrary function (ie
an ``integration constant''), or a parametric solution of the form
$[Y(\eta,r),\, c\,t(\eta,r)-c\,\tb(r)]$. The function $\tb$ marks the proper
time value corresponding to $Y=0$ for each comoving observer, hence it is
customarily refered to as the ``bang time''. The solutions of (\ref{freq1}) are

\bi
\item Parabolic solution, $E=0$,\, $Y>0$: 
\\
\be Y^{3/2} \ = \ \pm\,\frac{3}{2}\sqrt{2 \, M} \
c\,\left[ \ t-\tb(r) \ \right],\qquad\qquad t>\tb (+),\quad t<\tb
(-)\label{parab_can1}\ee

\item Elliptic solution, $E<0$, $0<Y<2M/|E|  $:
\\
\ba c\,t(Y,r) \ = \ c\,\tb - \frac{\sqrt {Y(2M-|E|Y)}}{|E|} + 
\frac{M}{|E|^{3/2}}\arccos
\left( {1-{{|E|Y} \over M}} \right),\cr\cr \dot
Y>0,\qquad c\,\tb<c\,t<\frac{\pi\,M}{|E|^{3/2}},\nonumber\ea 
\ba
c\,t(Y,r)\ = \ c\,\tb+\frac{\sqrt {Y(2M-|E|Y)}}{|E|} +
\frac{M}{|E|^{3/2}}\,\left[2\pi-\arccos 
\left( {1-{{|E|Y} \over M}} \right)\right],\cr\cr 
\dot Y<0,\quad
\frac{\pi\,M}{|E|^{3/2}}<c\,t<c\,\tb+\frac{2\pi\,M}{|E|^{3/2}},
\label{ell_can1}
 \ea
\be Y(\eta,r) \ = \ \frac{M}{|E|} \ \left[ \ 1 \ - \ \cos\,\eta \
\right],\label{ell_par_Y1}\ee
\be c\,t(\eta,r) \ = \ c\,\tb \ + \ \frac{M}{|E|^{3/2}} \ \left[ \ \eta \ - \
\sin\,\eta
\
\right],\label{ell_par_t1}\ee

\item Hyperbolic solution, $E>0$:
 
\be \pm\,c\,(t-\tb) \ = \ \frac{\sqrt{Y\,(\,2M+E\, Y\,)}}{E} \ - \
\frac{M}{E^{3/2}} \
\hbox{arccosh}\left(1+\frac{E\, Y}{M}\right),\label{hyp_can1}\ee
\be Y(\eta,r) \ = \ \frac{M}{E} \ \left[ \ \cosh\,\eta \ - \ 1 \
\right],\label{hyp_par_Y1}\ee
\be c\,t(\eta,r) \ = \ c\,\tb \ \pm\,\frac{M}{E^{3/2}} \ \left[ \ \sinh\,\eta \ -
\ \eta \ \right],\label{hyp_par_t1}\ee

\ei
where the signs $\pm$ in the parabolic and hyperbolic cases distinguish between
``expanding'' or ``collapsing'' dust layers, that is $Y$ increasing ($+$) or
decreasing ($-$) as $t$ increases, so that for any given layer $r=$ const., we
have $t>\tb $ (expansion) and $t<\tb$ (collapse). Notice that the sign
$\pm$ is not necessary in the elliptic case, since dust layers bounce (ie
$\dot Y=0$) as $Y=2M/|E|$ and so both expansion and collapse are described by the
same equations (\ref{ell_par_Y1}) and (\ref{ell_par_t1}) and by each branch in
the canonical solution (\ref{ell_can1}).

As we mentioned before, $\tb(r)$ marks the proper time associated with $Y=0$
for each comoving layer ({\it ie} the ``bang time'').  The conventional 
interpretation for the other two free functions $M$ and $E$ is based on the analogue
between the dynamical equation (\ref{freq1}) and an energy equation in newtonian
hydrodynamics (see \cite{bondi}), in which $M$ and $E$ are usualy understood,
respectively,  as the ``efective gravitational mass'' and the ``local energy
per unit mass'' associated with a given comoving layer. An alternative
interpretation for $E$ has been suggested as the local ``embedding''
angle of hypersurfaces of constant $t$ (\cite{hellaby1}, \cite{hellaby2},
\cite{ellisvan}). The specification or prescription of $M$ and $E$ is often
made in the literature by selecting ``convenient'' mathematical ansatzes that are
somehow intuitively based on these interpretations, loosely based on newtonian
anlogues. However, configurations without centers \cite{hellaby2} have no newtonian analogue and
so the interpretation of $M$ and $E$ in this case is still an open question. Also, there
is no  newtonian equivalent for $\tb$ and no simple and intuitive way in which
this function should be prescribed, though the gradient of this function has been
shown to relate to growing/decreasing modes of dust perturbations in a FLRW
background \cite{goodwain} \cite{silk}\cite{he_la1}\cite{krasinski}, and thus the
functional form of $\tb$ can be suggested from this feature. 

\section{New variables}\label{newvars}

A useful formulation of LTB solutions follows by redefining the free parameters
$M,\,E,\,\tb$ in terms of an alternative set of new variables defined at a given
arbitrary and fuly regular Cauchy hypersurface, $\Ti$, marked by $t=t_i$ (all such
hypersurfaces are Cauchy hypersurfaces in LTB solutions \cite{burnet}). In order to
relate the free function $E$ to the scalar 3-curvature of $\Ti$, we remark
from (\ref{def_R}) that the Ricci scalar of these hypersurfaces is given by 
\be \R \ = \ -\frac{2\,(E\,Y)'}{Y^2Y'},\label{def3R} \ee
a relation expressible in terms of invariant scalars (see (\ref{def_R}) in the 
Appendix). We introduce
\be  K \ \equiv \ -E,\label{E_K}\ee
so that the sign of $\R$ in (\ref{def3R}) coincides with the sign of $K$. From here
onwards we shall systematicaly replace $E$ by $-K$, so that the type of dynamical
evolution will now be denoted by the sign of $K$ as: parabolic ($K=0$), hyperbolic
($K<0$) and elliptic ($K>0$).  Bearing in mind (\ref{rho1}), (\ref{def3R}) and
(\ref{E_K}), we consider the following functions: 
\be Y_i(r) \ \equiv \ Y(t_i, r),\label{def_Yi}\ee
\be \frac{4\pi G}{c^4}\,\rho_i(r) \ \equiv \ \frac{4\pi G}{c^4}\,\rho(t_i, r) \ = \
\frac{M'}{Y_i^2\,Y_i'}, \label{def_rhoi}
\ee
\be \Ri(r) \ \equiv \ {\R}(t_i, r) \ = \
\frac{2\,(K\,Y_i)'}{Y_i^2\,Y_i'},\label{def_Ri}
\ee
as initial value functions defined in $\Ti$. The subindex $_i$ will
indicate henceforth evaluation along $t=t_i$. For the moment we will assume this
hypersurface to be fuly regular, we discuss the corresponding regularity conditions
in sections \ref{init_Ti} and \ref{Gamma_etc}. 

In order to recast (\ref{freq1}) and its solutions in terms of the new variables
it is convenient to scale $Y$ with respect to $Y_i$. By defining the adimensional
scale variable 
\be y \ \equiv \ \frac{Y}{Y_i},\label{defy}\ee 
and using (\ref{E_K}), (\ref{def_Yi}), (\ref{def_rhoi}) and (\ref{def_Ri}),
equations (\ref{ltbmetric}),  (\ref{freq1}) and (\ref{rho1}) become
\be ds^2 \ = \ -c^2\,dt^2 \ + \ y^2\left[ \ \frac{\Gamma^2\,
(Y_i')^2}{1-\kappa\,Y_i^2}\, dr^2 \ + \ Y_i^2\left(d\,\theta^2
\ + \ \sin^2\theta\,d\phi^2\right)\right],\label{ltbmetric2}
\ee
\be \dot y^2 \ = \ \frac{2\, \mu}{y} \ - \ \kappa. \label{freq2}\ee
\be \rho \ = \  \rho_i\,\frac{Y_i^2\,Y_i'}{Y^2\,Y'} \ = \
\frac{\rho_i}{y^3\,\Gamma},\label{rho2}
\ee
where
\be\Gamma \ \equiv \ \frac{Y'/Y}{Y_i'/Y_i} \ = \ 1 \ + \
\frac{y\,'/y}{Y_i'/Y_i},\label{Gamma}
\ee 
\be \mu \ \equiv \ \frac{M}{Y_i^3},\label{defmu}\ee
\be \kappa \ \equiv \ \frac{K}{Y_i^2},\label{defkappa}\ee
The solutions for the evolution equation (\ref{freq2}) are those of (\ref{freq1}),
that is, equations   (\ref{parab_can1}) to (\ref{hyp_par_t1}) expressed in terms
of the new variables
$y,\,\mu,\, \kappa$ by means of (\ref{defy}), (\ref{defmu}),
(\ref{defkappa}) and by eliminating $\tb$ by setting $t=t_i$ and $Y=Y_i$ into
equations (\ref{parab_can1}) to (\ref{hyp_par_t1}). The solutions are  
\bi
\item Parabolic solution 
\be y(t,r) \ = \ \left[\ 1 \ \pm \ \frac{3}{2}\,\sqrt{2\,\mu} \ c\,(\ t \ - \ t_i
\ ) \ \right]^{2/3},\label{parab_can2}\ee
\item Elliptic solution

\ba  c\,t(y,r) \ = \ c\,t_i
+ \ \frac{\left[\ 2\mu-\kappa \ \right]^{1/2}-\left[\ y\,(2\mu-\kappa
\,y) \ \right]^{1/2}}{\kappa} \ + \
\frac{\mu}{\kappa^{3/2}}\left[\ \arccos\left(1-\frac{\kappa\,y}
{\mu}\right)\ - \ \arccos\left(1-\frac{\kappa}
{\mu}\right) \ \right], \cr\cr\cr \dot y >0, \quad c\,\tb < c\,t_i <
\frac{\pi\,\mu}{\kappa^{3/2}}\qquad 
\label{ell_can21}
\ea
\ba
  t(y,r) \ = \ c\,t_i +\,\frac{\left[\
2\mu-\kappa \ \right]^{1/2}+\left[\ y\,(2\mu-\kappa
\,y)\ \right]^{1/2}}{\kappa} \ - \
\frac{\mu}{\kappa^{3/2}}\left[\ \arccos\left(1-\frac{\kappa\,y}
{\mu}\right)\ + \ \arccos\left(1-\frac{\kappa}
{\mu}\right)-2\pi \ \right], \cr\cr\cr \dot y <0, \quad
\frac{\pi\,\mu}{\kappa^{3/2}} < c\,t <
c\,\tb+\frac{2\pi\,\mu}{\kappa^{3/2}},\qquad 
\label{ell_can22}\ea
\be y(\eta,r) \ = \ \frac{\mu}{\kappa}\,\left[ \ 1 \ - \cos\,\eta \
\right],\label{ell_par_Y2}\ee
\be c\,t(\eta,r) \ = \ c\,t_i\  + \ \frac{\mu}{\kappa^{3/2}}\,\left[ \ (\eta \ -
\sin\,\eta)
\ -
\ (\eta_i \ - \ \sin\,\eta_i) \ \right],\label{ell_par_t2}\ee
\be \eta\  = \ \arccos\left(1-\frac{\kappa\,y}
{\mu}\right),\qquad  \eta_i\  = \ \arccos\left(1-\frac{\kappa}
{\mu}\right),\label{ell_def_eta}\ee
\item Hyperbolic solution
\be  c\,t(y,r) \ = \ c\,t_i \ \pm\,\frac{\left[\ y\,(2\mu+|\kappa|
\,y)\ \right]^{1/2}-\left[\ 2\mu+|\kappa|\, \ \right]^{1/2}}{|\kappa|} \ \pm \
\frac{\mu}{|\kappa|^{3/2}}\left[\ \hbox{arccosh}\left(1+\frac{|\kappa|\,y}
{\mu}\right)\ - \ \hbox{arccosh}\left(1+\frac{|\kappa|}
{\mu}\right) \ \right],\label{hyp_can2}\ee
\be y(\eta,r) \ = \ \frac{\mu}{|\kappa|}\,\left[ \ \cosh\,\eta  \ -  1\
\right],\label{hyp_par_Y2}\ee
\be c\,t(\eta,r) \ = \ c\,t_i \ \pm\,\frac{\mu}{|\kappa|^{3/2}}\,\left[ \ 
\ (\sinh\,\eta \ - \ \eta) \ - \ (\sinh\,\eta_i \ - \eta_i) \
\right],\label{hyp_par_t2}\ee
\be \eta\  = \ \hbox{arccosh}\left(1+\frac{|\kappa|\,y}
{\mu}\right),\qquad  \eta_i\  = \ \hbox{arccosh}\left(1+\frac{|\kappa|}
{\mu}\right),\label{hyp_def_eta}\ee
\ei 
where, as before, the signs $\pm$ distinguish between expanding layers ($-$
sign, $\tb$ in the past of $t_i$) and collapsing ones ($+$ sign, $\tb$ in the
future of $t_i$). Explicit forms for $\tb$ in terms of $\mu,\,\kappa,\,\eta_i$
and $t_i$ are given in section \ref{Gamma_etc}.   

\begin{figure}
\begin{center}
\BoxedEPSF{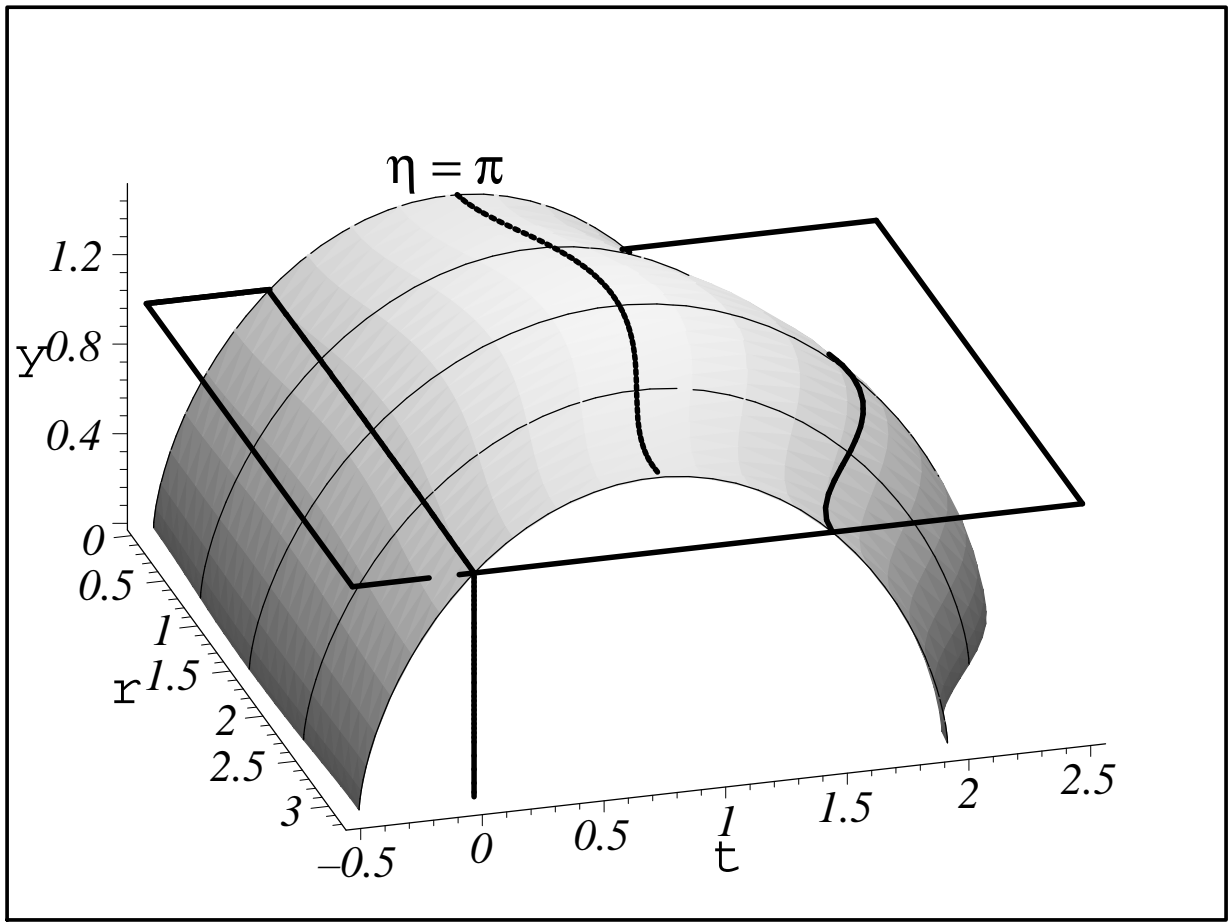 scaled 1600}
\end{center}
\caption{$\underline{\hbox{Comoving layers in elliptic solutions}}$. 
This figure displays $y(t,r)$ as a parametric 3-d plot $[r,t(r,\eta),y(r,\eta)]$ (see
equation (130)) for a regular elliptic model with two SC's. We have considered as
initial hypersurface $t_i=0$ and all initial value functions were obtained from  section
VIII with $f=\sin\,r$ and $a_1=1,\,a_2=2,\,b_1=1,\,b_2=3$.  Notice how the plane $y=1$
coincides with $t=t_i=0$ for all layers in the expanding phase, but in the collapsing phase
the locuus $y=1$ does not correspond to a single hypersurface of constant $t$. The figure also
illustrates how the layers in the expanding and collapsing phases in elliptic dynamics are (in
general) not time-symmetric with respect to any hypersurface of constant $t$. They are symmetric
with respect to the hypersurface $\eta=\pi$ where $\dot y=0$ (hypersurfaces of constant
$\eta$ do not coincide, in general, with those of constant $t$). These features also occur for
other models with elliptic dynamics, such as models with one SC (see figure 2e) and
models with a simmultaneous big bang (figure 12c).} 
\label{layers_e}
\end{figure}

It is important to remark that $y(t_i,r)=y_i=1$, by definition, and so
$y=1$ corresponds to $t=t_i$. This is clear for the parabolic (\ref{parab_can2})
and hyperbolic solutions (\ref{hyp_can2}) and (\ref{hyp_par_t2}).  However, it
is important to notice that $c\,t_i<\pi\mu/\kappa^{3/2}$ in (\ref{ell_can21}) and so
the initial hypersurface must be defined in the expanding phase in the elliptic
solution. In this solution the correspondence between $t=t_i$ and
$y=1$ only holds in the expanding phase of (\ref{ell_can21}) but not in its
collapsing phase. The explanation is simple: in general the expanding and
collapsing phases of the elliptic solution are not time-symmetric, as in the
homogeneous FLRW case, therefore $y=1$ will also be reached in the collapsing
phase but it will not (in general)  correspond to a single hypersurface $t=$
constant.  Figure \ref{layers_e} illustrates this feature. 

\section{Volume averages and contrast functions along the initial
hypersurface.}\label{ave_contrast}

The definitions (\ref{def_rhoi}) and (\ref{def_Ri}) of the initial value functions
$\rho_i$ and $\Ri$ lead to the following integrals 
\be M(r) \ = \ \frac{4\pi G}{c^4}\,\int_{r_c}^r{\rho_i\,Y_i^2\,Y'_i\,dr}\label{defM}\ee
\be K(r) \ = \
\frac{1}{2\,Y_i}\int_{r_c}^r{\Ri\,Y_i^2\,Y'_i\,dr}\label{defK}\ee
where the lower integration limits in these integrals have been fixed by the conditions
$M(r_c)=K(r_c)=0$, where $r_1=r_c$ marks a {\em $\underline{\hbox{Symmetry
Center}}$}\,\footnote{See section 
\ref{init_Ti} for more details. See the Appendix for a formal defintition of a Symmetry Center}
(to be denoted henceforth as ``SC'' or as ``center'', see section \ref{init_Ti} and the
Appendix). For the remaining of this paper we shall assume, unless explicitly specified
otherwise, that the spacetime manifold admits at least one SC and so the integrals
(\ref{defM}) and (\ref{defK}) become functions that depend only on the upper integration limit
$r$.

Since $4\pi\int{Y_i^2Y_i'dr}$ is the volume associated with the orbits of SO(3) in
the hypersurface $\Ti$ (though it is not a proper volume related to (\ref{def_V})), 
equations (\ref{defM}) and (\ref{defK}) suggest then
the introduction of the following volume averages 
\be \rhoav(r) \ = \
\frac{\int^r{\rho_i\,Y_i^2\,Y_i'\,dr}}{\int^r{Y_i^2\,Y_i'\,dr}}
\ =
\ \frac{3\, m(r)}{Y_i^3},\label{def_rhoi_av2}
\ee 
\be \Riav(r) \ = \
\frac{\int^r{\Ri\,Y_i^2\,Y_i'\,dr}}{\int^r{Y_i^2\,Y_i'\,dr}}
\ =
\ \frac{6\, K(r)}{Y_i^2},\label{def_Ri_av2}
\ee
In terms of these volume averages the variables $\mu$ and $\kappa$ in
(\ref{defmu}) and (\ref{defkappa}) become
\be \mu \ = \ \frac{4\pi G}{3c^4}\,\rhoav,\label{defmu2} \ee
\be \kappa \ = \ \frac{1}{6}\, \Riav,\label{defkappa2} \ee
Setting $\rhoi$ and $\Ri$ to constants in equations (\ref{def_rhoi_av2}) and
(\ref{def_Ri_av2}) yields $\rhoi=\rhoav$ and $\Ri=\Riav$, therefore the
comparison between $\rhoi,\, \Ri$ and $\rhoav,\, \Riav$ must provide a ``gauge''
that measures the inhomogeneity of $\rhoi$ and $\Ri$ in the closed interval     
$r\geq r_c$. This motivates defining the following ``inhomogeneity
gauges'' or contrast functions
\be \Dmi \ = \ \frac{\rho_i}{\rhoav} \ - \ 1, \qquad \Rightarrow \qquad \rho_i \
= \ \rhoav\, \left[\ 1 \ + \ \Dmi\right],\label{defDmi}
\ee 
\be \Dki \ = \ \frac{\Ri}{\Riav} \ - \ 1, \qquad \Rightarrow \qquad \Ri \
= \ \Riav\, \left[\ 1 \ + \ \Dki\right],\label{defDki}
\ee   
These functions  can be expressed in terms of $M$, $K$, $\mu$ and $\kappa$ and their
radial gradients. For example:

\bi
\item{} $\Dmi$ and $\Dki$ in terms of $\mu,\,M,\,\kappa,\,K$
\ba 1 \ + \ \Dmi \ = \ \frac{4\pi G}{3c^4}\,\frac{\rhoi}{\mu} \ = \ \frac{4\pi
G}{3c^4}\,\frac{\rhoi\,Y_i^3}{M},\cr\cr\cr  1 \ + \ \Dki \ = \
\frac{\Ri}{6\,\kappa} \ = \ \frac{\Ri\,Y_i^2}{6\,K},\label{D_vs_vbls}\ea
\item{} $\Dmi$ and $\Dki$ in terms of $\mu',\,M',\,\kappa',\,K'$ and $Y_i'$

\be 1 \ + \
\Dmi \ = \ \frac{M'/M}{3Y_i'/Y_i}, \qquad \frac{\mu'/\mu}{3Y_i'/Y_i} \ = \
\Dmi,\label{grad_m}
\ee
\be 1 \ + \ \frac{3}{2}\Dki \ = \ \frac{K'/K}{2Y_i'/Y_i}, \qquad
\frac{\kappa'/\kappa}{3Y_i'/Y_i} \ = \ \Dki,\label{grad_k}\ee
\ei
Another useful relation is
\be \frac{M'}{M} \ - \frac{3}{2}\,\frac{K'}{K}\ = \ \frac{\mu'}{\mu} \ -
\frac{3}{2}\,\frac{\kappa'}{\kappa} \ = \ \frac{Y_i'}{Y_i}\,\left(\Dmi
\ - \ \frac{3}{2}\,\Dki\right),\label{grad_m32k}\ee
The interpretation of $\Dmi$ and $\Dki$ as contrast functions will be discussed in section
\ref{lumps_voids}. Using (\ref{grad_m}), (\ref{grad_k}) and (\ref{grad_m32k}) we can
eliminate all radial gradients like $\mu',\,\kappa'$ (or $M',\,K'$) in terms of $\Dmi$ and
$\Dki$. Since these gradients appear in the evaluation of $\Gamma$ from (\ref{Gamma}) and
the solutions of (\ref{freq2}) (see section \ref{Gamma_etc}), it is certainly useful to be
able to relate an important quantity like $\Gamma$ to initial conditions expressed  in terms
of quantities like $\Dmi$ and $\Dki$, quantities that have a more intuitive and appealing
interpretation than $M'$ and $K'$.

\section{Generic properies of the new variables}\label{int_newvars}

We discuss in this section the regularity properties and generic features of the new
variables, as well as their relation with $M$, $K$ defined by (\ref{def_rhoi_av2}),
(\ref{def_Ri_av2}). The averages and contrast functions, (\ref{defDmi}) and (\ref{defDki}),
are examined in section \ref{lumps_voids}. Frequent reference will be made
to expressions presented and summarized in the Appendix. 

\subsection{Initial hypersurfaces.}\label{init_Ti}

A hypersurface $\Ti$ is the 3-dimensional submanifold obtained by 
restricting an LTB spacetime to $t=t_i$. This submanifold can be invariantly
characterized as the set of rest frames of comoving observers associated with proper time
$t=t_i$ (see the Appendix). Its induced 3-dimensional metric is
\be ds_i^2 \ = \ \frac{Y_i'^2}{1-\kappa\,Y_i^2}\, dr^2 \ + \
Y_i^2\,\left[\,d\theta^2 \ + \ \sin^2\theta\, d\phi^2\right],\label{3-metric}\ee 
where  $0\leq\theta\leq\pi$ and $0\leq \phi\leq 2\pi$ and $r$ extends along its domain
of regularity (see subsection \ref{Yi_etc} and the parragraph below). For a regular $\Ti$,
all invariant scalars like (\ref{def_R}), (\ref{def_scalars}) and (\ref{def_E}) resticted to
$t=t_i$, or invariant scalars computed directly from (\ref{3-metric}), must be bounded. The
metric (\ref{3-metric}) is also subjected to the regularity condition (\ref{cond_K_2}) which
now reads  
\be 1 \ - \ K \ = \ 1 \ - \ \kappa\,Y_i^2 \ \geq \ 0\qquad
\Rightarrow\qquad
\frac{1}{6}\Ri\,Y_i^2 \ = \ \frac{\Ri}{\Riav}\,K \ \leq \ 1 \ + \
\Dki,\label{cond_K2}
\ee
where the equality must coincide with a zero of $Y_i'$ (see below) and we have used
(\ref{E_K}), (\ref{defkappa}), (\ref{def_Ri_av2}), (\ref{defkappa2}) and
(\ref{defDki}). The hypersurface $\Ti$ can be foliated by the orbits of SO(3) marked
bt $r=$ const., thus the regular domain of $r$ is then limited, either by a SC, or by
those maximal values $r=\rmax$ defined by (\ref{ell1})
restricted to $t=t_i$   
\be \ell_i(r_1,r) \ = \ \int_{r_1}^r{\frac{Y_i'\,dr}{\sqrt{1-K}}} \ = \ 
\int_{r_1}^r{\frac{Y_i'\,dr}{\sqrt{1-\kappa\,Y_i^2}}} \ = \
\int_{r_1}^r{\frac{\left[1+\Dki\right]^{1/2}\,Y_i'\,dr}
{\left[1+\Dki+\frac{1}{6}\Ri
\,Y_i^2\right]^{1/2}}} \ \to \ \infty\quad\hbox{as}\quad r\to
\rmax\label{ell2}\ee
where $\K$ and $\kappa$ have been eliminated in terms of $\Ri$ and $\Dki$ by means
of (\ref{defkappa2}) and (\ref{defDki}). These maximal values could correspond to
$r\to\infty$ or $r\to \pm\infty$, but this is not necessary, as it is always
possible to define the radial coordinate so that maximal values $\rmax$ correspond to finite
values of $r$ (see section \ref{Yi_etc}).

\bi

\item{{$\underline{\hbox{Regularity and differentiability.}}$}}

From their definition in (\ref{def_rhoi}), (\ref{def_Ri}) and (\ref{def_Yi}), the
functions $\rhoi,\,\Ri,\,Y_i$ satisfy the same regularity and differentiability
conditions as $\rho,\,\R,\,Y$ in the radial direction (see the Appendix): 
$Y_i\geq 0$ must be continuous and $\rho_i\geq 0,\,\Ri$ can be piecewise continuous but
with a countable number of finite jump discontinuities. The average functions,
$\rhoav,\,\Riav $, constructed from intergration of $\rho_i,\,\Ri$,  will
be, in general continuous (like $M$ and $K$). The contrast functions, $\Dmi$ and
$\Dki$, explicitly contain $\rhoi$ and $\Ri$ (from (\ref{D_vs_vbls})), hence they are,
at least, piecewise continuous.

\item{{$\underline{\hbox{Zeroes of $Y_i$ and $Y_i'$.}}$}}

The zeroes of $Y_i$ and $Y_i'$ are also zeroes of $Y$ and $Y'$, though the converse
is not (in general) true. Because of (\ref{def_Yi}) all regular zeroes of $Y$ and
$Y'$, associated with a comoving $r$ (symmetry centers and turning values of $Y'$)
will be shared by $Y_i$, but a regular $\Ti$ must avoid the zeroes of $Y$ and $Y'$
that do not correspond to a comoving $r$ (singular zeroes).  Therefore, we have along
$\Ti$

\bi
\item{{$\underline{\hbox{Symmetry Centers, SC.}}$}}.\\
Let $r=r_c$ mark a SC (see the Appendix). The functions
$Y_i,\,\rho_i,\,\Ri$, from their definitions in (\ref{def_Yi}), (\ref{def_rhoi}) and
(\ref{def_Ri}),  must satisfy 
\be\matrix{{Y_i(r_c)=0,}&{Y'_i(r_c)=S_i \ne 0,}&{Y_i\approx S_i(r-r_c)}\cr\cr
{\rho _i(r_c)>0}&{\rho\, '_i(r_c)=0}&{\rho _i\approx \rho _i(r_c)+{\textstyle{1
\over 2}}\rho\, ''_i(r_c)(r-r_c)^2}\cr\cr {\Ri(r_c)\ne
0}&{\Ri'(r_c)=0}&{\Ri\approx\, \Ri(r_c)+{\textstyle{1 \over
2}}\Ri''(r_c)(r-r_c)^2}\cr\cr {\Ri(r_c)=
0}&{\Ri'(r_c)=0}&{\Ri(r)=0,\quad \hbox{parabolic dynamics}}\cr
}\label{central1}\ee
where $S_i=S(t_i)=Y'(t_i,r_c)>0$ and the fourth subcase above denotes the
case of parabolic dynamics in which the condition $K=\kappa=0$ holds in an interval $r_c
\leq r$, thus implying that  $\Ri(r)=0$ for all the interval. The central
behavior of  $M,\,K$ in (\ref{defM}) and (\ref{defK}) is given by the leading
terms   
\ba M \ \approx \ \frac{4\pi G}{3c^4}\,\rho_i(r_c)\,S_i^3\,(r-r_c)^3,
\cr\cr K \ \approx \
\frac{1}{6}\,\,\Ri(r_c)\,S_i^2\,(r-r_c)^2,\label{central2}   
\ea
From (\ref{defmu}), (\ref{defkappa}), (\ref{defmu2}), (\ref{defkappa2}),
(\ref{defDmi}) and (\ref{defDki}), the value of 
$\mu,\,\kappa$ and the remaining initial value functions at $r=r_c$ are given by
\ba \mu(r_c) \ = \ \frac{4\,\pi\,G}{3c^4}\,\rho_i(r_c),
\qquad \kappa(r_c) \ = \ \frac{1}{6}\,\,\Ri(r_c),\qquad\qquad\qquad\qquad
\cr\cr \rhoav(r_c) \ = \ \rhoi(r_c),   
\qquad 
\Riav(r_c) \ = \ \Ri(r_c),\qquad\Rightarrow\qquad  \Dmi(r_c) \ = \ \Dki(r_c) \ = \
0, \label{central3}   
\ea
\item{{$\underline{\hbox{Signs and turning values of $Y_i'$.}}$}}\\
From  (\ref{def_Yi}) and comparing (\ref{rho1}) with (\ref{def_rhoi}) and (\ref{def3R})
with (\ref{def_Ri}), it is evident that regularity of $\rho$ and $\rhoi$ requires the
sign of $Y_i'$ to be the same as the sign of $Y'$. Hence condition (\ref{wec}) can be
restated as
\be \rhoi \  \geq \ 0 \qquad \Leftrightarrow \qquad \hbox{sign}(\,M') \ = \ \hbox{sign}
(\,Y_i'),\label{signsYiM}
\ee 
so that regularity of $\rhoi$, $\Ri$ and of the metric (\ref{3-metric}) leads to the
condition that all the following equations 
\be Y_i'(r^*) \ = \ 0, \qquad M'(r^*) \ = \ 0,\qquad \ 1 \ -
\ K(r^*) \ = \ 0,\qquad (K\,Y_i)' \ = \ 0,\label{cond_K_3}\ee
must have common zeroes (turning values), $r=r^*$, of the same order in $r-r^*$.
Failure to comply with (\ref{cond_K_3}), not only reflects ``bad'' choices of
$\rhoi,\,\Ri$, but leads to surface layers in $r=r^*$ associated with a discontinuous
extrinsic curvature of the submanifold $r=r^*$ \cite{ZG}, \cite{bon3},
\cite{humph_phd}, \cite{HM1}, \cite{HMM1}.   
\ei

\ei

\subsection{Comparison of $y$ vs $Y$ and $\Gamma$ vs $Y'$.}\label{comp}

Since the zeroes of $\{Y=0,\,Y'=0\}$ do not (in general) imply $\{Y_i=0,\,Y_i'=0\}$,
but all comoving zeroes of $\{Y=0,\,Y'=0\}$ are also zeroes of $\{Y_i=0,\,Y_i'=0\}$,
those zeroes of $\{Y=0,\,Y'=0\}$ that are not common to $\{Y_i=0,\,Y_i'=0\}$ must
correspond to curvature singularities (central or shell crossing). This removes the
ambigueity of identifying these singularities by the constraints $Y=0$ and $Y'=0$,
placing this identification on $y$ and $\Gamma$.  From its definition in (\ref{defy}),
it is evident that
$y(t_i,r)=1$, while its central behavior is given by  
\ba y(t,r_c) \ = \ \frac{S(t)}{S_i}, \qquad y(t,r) \ \approx \
\frac{S(t)}{S_i}\,\left[\ 1 \ + \ O(r-r_c)
\ +
\ O(t-t_i)\right],\label{central_y}\ea
where $S(t)=Y'(t,r_c)$, so that (in general) $y$ does not vanish at a SC. This
means that $y=0$ implies $Y=0$, but the converse is false, since $Y=0$ does not
imply $y=0$ (for example at $r=r_c$). Therefore, the coordinate locus for a
central  singularity (either big bang or big crunch) becomes 
\be y(t,r) \ = \ 0,\qquad \hbox{Central singularity.}\label{bb_sing}\ee
an unambiguous characterization, as opposed to $Y=0$ which can be associated
either to such a singularity or to a SC. In fact, as mentioned in the previous
section and from the form of the metric (\ref{ltbmetric2}), the function $y$ plays the role
of a local scale factor, in contrast to $Y$ defined non-localy in terms of the
proper surface of the orbits of SO(3) around a SC that might not even be
contained in the spacetime manifold. The difference between $y$ and $Y$ is
illustrated in figures \ref{y_vs_Y} displaying parametric 3d plots $Y(t,r)$ and $y(t,r)$ 
for various LTB models. These figures clearly show that $y(t_i)=1$ for
all layers.

From (\ref{central_1}), the gradient $Y'/Y$ is singular at a SC, however, from
(\ref{rho2}), (\ref{Gamma}) and (\ref{central_y}), we have  
\be \Gamma(t,r_c) \ = \ 1,\qquad \rho(t,r_c) \ = \
\rho_i(r_c)\,\frac{S_i^3}{S^3(t)},\qquad \sigma(t,r_c) \ = \ 0,\qquad
\Theta(t,r_c) \ = \ \frac{3\,\dot S(t)}{S(t)}\ee 
while the coordinate locus for shell crossing singularities can be given as 
\be \Gamma \ = \ 0,\quad \rho_i \ > \ 0,\ \qquad \hbox{Shell crossing
singularity.}\label{shx}\ee 
a characterization that avoids the ambigueity of $Y'=0$, since, as mentioned 
in the Appendix, $Y'$ might vanish and change sign regularly for a single $r=r^*\ne
r_c$ if (\ref{cond_K_1}) holds, preventing surface layers. This situation is now taken
care by conditions (\ref{cond_K_3}) acting on $Y_i'$.

\subsection{The function $Y_i$: choice of topology and radial coordinate.}\label{Yi_etc}

From the common geometric meaning of $Y$ and $Y_i$ (see Appendix), it is evident that
the homeomorphic class (``open'' or ``closed'' topology) of the regular hypersurfaces
of constant $t$ follows from the homeomorphic class of $\Ti$. The latter can be
determined by applying (\ref{def_V}) to (\ref{3-metric}) for a given
choice of $Y_i(r)$, along the domain of regularity of $r$ limited between SC's (real
roots of $Y_i(r)=0$) and/or a maximal value $\rmax$ defined by
(\ref{ell2}). Because we can always rescale $r$, it is always possible to select this
coordinate so that $Y_i$ takes the simplest form, hence the freedom to select $Y_i$ can
also be the freedom to find a convenient radial coordinate for a desired LTB
configuration.  As mentioned in the Appendix, the compatibility of any one of the
possible choices of homeomorphic class with any one of the three choices of dynamical
evolution (parabolic, hyperbolic or elliptic) is only restricted by the regularity
conditions (\ref{cond_K2}), (\ref{cond_K_3}) and (\ref{ell2}). We ellaborate these
issues below looking separately the case with one and two SC's. 

\bi
\item{{$\underline{\hbox{One symmetry center.}}$}}

Since we can always mark the SC by $r=0$, then $Y_i(0)=0$ and if
$Y_i$ is a monotonously increasing (one-to-one) function ($Y_i'>0$) for all $r$,
we can then choose the radial coordinate so that $Y_i=S_i\,r$, where $S_i$ is a
characteristic constant length. This is equivalent to using $Y_i/S_i$ as a new
radial adimensional coordinate, but we can also set $Y_i=S_i\,f(r)$, for any monotonously
increasing function with $f(r_c)=0$. The range of $r$ is $0\leq r < \rmax$, so that proper
distances $\ell_i$ increase as $r$ grows and $\ell_i\to\infty$ as $r\to \rmax$. A useful
coordinate choice is $f=\tan\,r$, so that maximal values associated with $Y_i\to\infty$ can be
mapped to $r=\pi/2$. Fulfilment of  conditions (\ref{cond_K2}), (\ref{cond_K_3}) and (\ref{ell2})
is trivial for parabolic and hyperbolic dynamics, but for elliptic dynamics we must have $K\to 1$
as $r\to \rmax$. The homeomorphic class is ${\bf R^3}$.  

\item{{$\underline{\hbox{Two symmetry centers.}}$}}

Let the centers be marked by $r=r_{c_1}$ and $r=r_{c_2}$. Then
$Y_i(r_{c_1})=Y_i(r_{c_2})=0$, so that $Y_i'$ vanishes for some $r^*$ in the open
interval $r_{c_1} <r^* < r_{c_2} $. In this case, it is not possible to rescale
$Y_i$ with $r$ because $Y_i$ is not a one-to-one function of $r$. However,
depending on the model under consideration, it is always possible to select the
radial coordinate so that $Y_i=S_i\,f(r)$, where $f(r)$ is a $C^0$
function satisfying $f(r_{c_1})=f(r_{c_2})=0$ and $f'(r^*)=0$, so 
that $\rhoi\,',\Ri',\,M',\,K',\,1-K=1-\kappa Y_i^2$ all vanish at $r^*$ and have zeroes of
the same order on $r-r^* $.  Since the hypersurface $t=t_i$ has a finite proper volume,
$\ell_i$ evaluated from one SC to the other (the maximal range of $r$) must be finite.
Although it is possible to set $K\leq 0$ (parabolic and hyperbolic evolution), only for
elliptic evolution it is possible to fulfil condition (\ref{cond_K_3}) in order 
to avoid surface layers in $r=r^*$ (see \cite{bon3}, \cite{humph_phd} and
\cite{HMM1}). The hypersurface $\Ti$ is homeomorphic to a 3-sphere, ${\bf
S^3}$.   
\ei

\subsection{Invariant expressions.}\label{inv_expr}

The volume averages and contrast functions introduced in section \ref{ave_contrast}
were constructed by averaging $\rhoi$ and $\Ri$ with the volume
$(4/3)\,\pi\,Y_i^3=4\,\pi\int{Y_i^2\,Y_i'\,dr}$, a quantity that can be related to the
proper area, $4\pi\,Y_i^2$, of the group orbits of SO(3). However, it can be argued
that these volume averages and contrast functions are coordinate dependent expressions.
We can show that these quantities are equivalent to coordinate independent
expressions given in terms of invariant scalars evaluated along
$\Ti$. From (\ref{def_scalars}) and (\ref{def_E}) applied to $t=t_i$, together with
(\ref{defDmi}), we obtain 
\ba \frac{8\pi G}{c^4}\,\rhoav \ = \ \frac{8\pi G}{c^4}\,\rhoi \ + \ 
6\,[\psi_2]_i\ = \ {\cal{R}}_i \ + \ 6\,[\psi_2]_i\cr\cr 1 \ + \ \Dmi \ = \
\frac{4\pi G\,\rhoi}{4\pi G\,\rhoi\,+\,3\,c^4\,[\psi_2]_i} \ = \
\frac{{\cal{R}}_i}{{\cal{R}}_i\,+\,6\,[\psi_2]_i},\label{m_scalars}\ea
where ${\cal{R}}_i$ and $[\psi_2]_i$ are the 4-dimensional Ricci scalar and the
conformal invariant $\psi_2$ evaluated at $t=t_i$. The evolution equation
(\ref{freq2}) applied to $t=t_i$, together with (\ref{sigma1}), (\ref{def_R}),
(\ref{defmu2}), (\ref{defkappa2}), (\ref{defDki}) and (\ref{m_scalars}) yields
\ba \frac{c^2}{2}\,\Riav \ = \ \frac{c^2}{2}\,\Ri \ - \
6\,\sigma_i\,\left(\sigma_i+\frac{\Theta_i}{3}\right) \ = \ c^2\,{\cal{R}}_i \ +
\ 6\,c^2\,[\psi_2]_i \ - \
3\,\left(\sigma_i+\frac{\Theta_i}{3}\right)^2,\cr\cr\cr
1 \ + \ \Dki \ = \
\frac{c^2\,{\cal{R}}_i\,+\,3\,[\,\sigma_i^2-(\frac{1}{3}\Theta_i)^2\,]}{{
c^2\,\cal{R}}_i\,+\,6\,c^2\,[\psi_2]_i\,-\,(\,\sigma_i\,+\,\frac{1}{3}\Theta_i\,)^2} \
=
\
\frac{c^2\,\Ri}{c^2\,\Ri\,-\,12\,\sigma_i\,(\sigma_i+\frac{1}{3}\Theta_i)},
\label{k_scalars}
\ea
where $\sigma_i$ and $\Theta_i$ are the kinematic invariants appearing in
(\ref{Theta1}) and (\ref{sigma1}) (the expansion scalar and the degenerate eigenvalue
of the shear tensor) evaluated along $\Ti$. The expressions (\ref{m_scalars}) and
(\ref{k_scalars}) are not the definitive ``covariant interpretations'' for the average
and contrast functions. They have been presented in order to emphasize that these
quantities can be expressed in terms of invariant scalars and so can be computed in
any coordinate system. An alternative volume averaging could have been attempted using
the proper volume (\ref{def_V}), leading to quantities like
\be
 \frac{\int{\rhoi\,(1-K)^{-1/2}\,Y_i^2\,Y_i'\,dr}}
{\int{(1-K)^{-1/2}\,Y_i^2\,Y_i'\,dr}},\qquad 
\frac{\int{\Ri\,(1-K)^{-1/2}\,Y_i^2\,Y_i'\,dr}}
{\int{(1-K)^{-1/2}\,Y_i^2\,Y_i'\,dr}},\label{propvols}\ee  
instead of the volume averages in (\ref{def_rhoi_av2}) and (\ref{def_Ri_av2}). Notice
that only for parabolic dynamics all these averages coincide. We have chosen to use
(\ref{def_rhoi_av2}) and (\ref{def_Ri_av2}) instead of (\ref{propvols}) mostly because
the former are mathematicaly simpler than the latter and because they follow
naturaly from the field and evolution equations. For example, the functions
$M,E=-K$ or $\mu,\kappa$ that appear in the solutions of (\ref{freq1}) or (\ref{freq2})
immediately relate to (\ref{def_rhoi_av2}) and (\ref{def_Ri_av2}), but their relation
with (\ref{propvols}) is (except for parabolic dynamics) mathematicaly difficult,
making it very hard to accomodate quantities like (\ref{propvols}) to the dynamical
study of the models. Since all the invariant scalars defined in
$\Ti$ by (\ref{m_scalars}) and (\ref{k_scalars}) can be generalized for arbitrary times
$t\ne t_i$, the volume averages and contrast functions can also be generalized in this
manner. For example, the fact that $M=M(r)$ and
$K=K(r)$, and so these functions are constants for every comoving oberver,
follows (from (\ref{def_rhoi_av2}) and (\ref{def_Ri_av2})) from the
conservation (along the 4-velocity flow) of quantities like $\Rhoav Y^3$ and $\Rav
Y^2$, were $\Rhoav$ and $\Rav$ are the generalization to arbitrary times of
$\rhoav$ and $\Riav$. This and other conservation laws are being examined elsewhere
\cite{sussgar}.

\subsection{FLRW and Schwarzschild limits, matchings and hybrid models.}\label{limits}

It is a well known fact that LTB solutions contain as particular
cases FLRW cosmologies with dust source (homogeneous and isotropic
limit) and the Schwarzschild-Kruskal spacetime (vacuum limit). We
examine these limits below

\bi
\item{{$\underline{\hbox{FLRW limit}}$}}. If for all the
domain of regularity of $r$ we have $\rhoi=\rhoi^{(0)}$ and
$\Ri=\Ri^{(0)}$, where $\rhoi^{(0)}\geq 0$ and $\Ri^{(0)}$ are constants, then
equations (\ref{def_rhoi_av2}), (\ref{def_Ri_av2}), (\ref{defDmi}) and
(\ref{defDki}) imply
\ba \rhoi \ = \ \rhoav \ = \ \rhoi^{(0)},\qquad \Rightarrow\qquad \Dmi \ = \ 0,
\qquad M \ = \
\frac{4\pi G}{3 c^4}\,\rhoi^{(0)}\,Y_i^3\cr\cr
\Ri \ = \ \Riav \ = \ \Ri^{(0)},\qquad \Rightarrow\qquad \Dki \ = \ 0, \qquad
K \ = \
\frac{1}{6}\,\Ri^{(0)}\,Y_i^2,\label{flrw_limit}\ea
Since $\mu$ and $\kappa$ in (\ref{defmu}) and (\ref{defkappa}) are now constants,
$y$ (determined by (\ref{freq2})) must be a function of $t$ only, and so, from 
(\ref{defy}) and (\ref{Gamma}), we have $\Gamma=1$ and $Y=(S(t)/S_i)\,Y_i$,
identifying $y=S(t)/S_i$ with an adimensional FLRW scale factor. Rest mass density
and scalar 3-curvature take the FLRW forms: $\rho= \rhoi^{(0)}\,(S_i/S)^{3}$ and
$\R=\Ri^{(0)}\,(S_i/S)^{2}$, while the metric (\ref{ltbmetric2}) becomes a FLRW
metric if we identify $Y_i=S_i\,r$ so that $(1/6)\Ri^{(0)}S_i^2 $ becomes
an adimensional curvature index.  Equations (\ref{flrw_limit})
reafirm the  role of inhomogeneity gauges for $\Dmi$ and $\Dki$, making also evident
that initial conditions with homogeneous $\rhoi$ and $\Ri$ leads only to FLRW
evolution. 

\item{{$\underline{\hbox{Schwarzschild limit}}$}}. For an intial density
given by a normalized Dirac delta (a point mass)
\be \rhoi = \rhoi^{(0)}\,\delta(Y_i^3)=\rhoi^{(0)}\,S_i^3\,\delta(r^3)
\ee
we obtain:  $m=\rhoi^{(0)}\,S_i^3/3$ from (\ref{defM}), so that we can
identify the ``Schwarzschild mass'' with $2M$ given by (\ref{defm}). The
metric (\ref{ltbmetric}), together with (\ref{freq1}) with $M=$ const.
describes the Schwarzchild-Kruskal spacetime in terms of the worldlines of
geodesic test observers with 4-velocity $u^a$. The function $\R$ is the
scalar 3-curvature of the hypersurfaces of simmultaneity of these
congruences.       
\ei
If only one of $\rhoi$ or $\Ri$ is constant (but nonzero) in a domain of $r$ containing
a SC, then only $\Dmi$ or $\Dki$ vanishes in this domain. For hyperbolic or
elliptic dynamics, these are special cases of initial conditions but do not correspond
to the FLRW limit (for parabolic dynamics $\Ri=0$ and so the FLRW limit follows from
$\rhoi=\rhoi^{(0)}>0$). The cases in which $\rhoi$ and $\Ri$ are, in general, not
constant but become constant (even zero) only in an interval of the domain of $r$ can
be also understood as initial value functions of ``hybrid'' spacetimes obtained by
matching a given LTB region (with $\rhoi,\,\Ri$ not constants) with a section of FLRW or
Schwarzschild spacetimes in another part. These matchings can also be defined among various
types of LTB (non-FLRW or non-Schwarzschild) regions (see \cite{humph_phd}, \cite{HM1}). We
will not examine these hybrid spacetimes in this paper (see \cite{sussgar}).

\section{Lumps and voids in the initial hypersurface.}\label{lumps_voids}
\subsection{Initial density lumps and voids.}\label{dens_lv}

The relation of $\rho_i$ with $\rhoav $ and $\Dmi$ is straightforward.
Consider the following auxiliary equations obtained by 
applying integration by parts to (\ref{def_rhoi_av2})
\ba \rho_i \ - \ \rhoav \ = \ \frac{1}{Y_i^3}\,\int^r{\rho\,'_i\,Y_i^3\,
dr} \cr\cr
\Dmi \ = \frac{\int^r{\rho\,'_i\,Y_i^3\,dr}}
{\rho_i\,Y_i^3-\int^r{\rho\,'_i\,Y_i^3\,dr}}
\label{aux_rhoi}\ea
In all intervals of $r$ containing a SC, $\rho\,_i'\ne 0$ implies $\Dmi\ne 0$, except 
at the SC itself where both $\rho\,'_i$ and $\Dmi$ must vanish (from (\ref{central1})
and (\ref{central1})). We still have $\Dmi\ne 0$, even if $\rho_i$ and/or
$\rho\,'_i$ vanish in a subset of this domain that contains no SC.  With the
help of  (\ref{def_rhoi_av2}) and (\ref{aux_rhoi}), we discuss below the cases with
one and two symmetry centers. Unless specificaly mentioned otherwise, we assume that either
$\rhoi'\geq 0$ or $\rhoi'\leq 0$ throughout the interval under consideration. The discussion
below is illustrated in figures \ref{initfunc_dens}. 

\bi
\item{{$\underline{\hbox{One symmetry center.}}$} See figures \ref{initfunc_dens}a and
\ref{initfunc_dens}b.}

In this case  a density ``lump'' or ``void'' is characterized
by $\rho_i$ having a maximum/minimum at the center worldline $r=r_c$.
Hence, for a lump or void, we have  (from (\ref{aux_rhoi})) the following 
possible ranges of $\Dmi$   
\ba \hbox{initial lump:}\qquad  \rho\,'_i \ \leq \ 0, \qquad \rho_i \ \leq \
\rhoav,  \qquad\Rightarrow \qquad -1 \ \leq \ \Dmi \ \leq 0,
\cr\cr\hbox{initial void:}\qquad  \rho\,'_i \ \geq \ 0, \qquad \rho_i \ \geq \
\rhoav,\qquad\Rightarrow \qquad \Dmi\ \geq\  0,\label{Dmi_1sc}
\ea  
where the lower bound on $\Dmi$ for a lump is $\Dmi=-1$ if
$\rho_i$ vanishes for $r>r_c$ (from(\ref{defDmi})).

\item{{$\underline{\hbox{Two symmetry centers.}}$} See figures \ref{initfunc_dens}c and
\ref{initfunc_dens}d.}

From (\ref{central1}), (\ref{central2}) and (\ref{central3}), all of $Y_i,\,M,\,
M',\,\rho_i\,',\,Y_i$, $\rho_i-\rhoav$ and $\Dmi$ must vanish at both SC,
$r=r_{c_1}$ and $r=r_{c_2}$, with zeroes of the same order in $r-r_{c_1}$ and $r-r_{c_2}$.
This means that there must exist a turning value $r=r^*$ in the open domain $r_{c_1}< r <
r_{c_2}$ for which $Y_i',\,M',\,\rho_i\,'$ all have a common zero of the same order at
$r=r^*$. Besides these conditions, $\rho_i$
and
$Y_i$ must satisfy 
\be M(r_{c_2}) \ = \ \frac{4\pi G}{3c^4}\,\int_{r_{c_1}}^{r_{c_2}}{\rho_i Y_i^2Y_i'
dr}=0\ee
Therefore, for every $r=a$ in the open domain $r_{c_1}< r < r_{c_2}$  we have
\be \int_{r_{c_1}}^a{\rho_i Y_i^2Y_i' dr} \ = \ -\int_a^{r_{c_2}}{\rho_i
Y_i^2Y_i' dr} \ = \ \int_{r_{c_2}}^a{\rho_i Y_i^2Y_i'
dr}
\ee
Since $Y_i$ vanishes at both SC's, this means that, for whatever $\rho_i$ we
choose, the average $\rhoav$ evaluated from one center (say $r=r_{c_1}$) to $r=a$
must be the same as that average evaluated from the other center ($r=r_{c_2}$).
Also, following the usual interpretation of $M$ as the ``effective mass inside''
a comoving layer $r$, this ``effective mass'' for the whole hypersurface (from
$r=r_{c_1}$ ro
$r=r_{c_2}$) is zero.   

We have two possible regular configurations (see figures \ref{initfunc_dens}c and
\ref{initfunc_dens}d), depending whether $\rho_i$ is: (i) a local maximum at each center (each
center is a density lump) and (ii) a local minimum at each center (each center is a density
void). For each case,  $\Dmi$ vanishes at the centers, while (from
(\ref{aux_rhoi})) and using the same arguments as in the case of one center, we
have for the cases (a) and (b).
\ba \hbox{an initial lump in each center:}\qquad   \rho_i \ \leq \ \rhoav, 
\qquad\Rightarrow \qquad -1 \ \leq \ \Dmi \ \leq 0,\cr\cr
\hbox{an initial void in each center:}\qquad  \qquad\rho_i \ \geq \
\rhoav,\qquad\Rightarrow \qquad \Dmi\ \geq\  0,\label{Dmi_2sc}
\ea  

\ei

If allow $\rhoi\,'$ to change its sign in a given domain containing a SC, we
obtain an initial density that can be associated with ``density ripples'' (see
figure \ref{initfunc_other}a). In this case we have $-1\leq \Dmi\leq 0$ for all $r$ for which 
$\rhoi\,'$ is negative and $\Dmi\geq 0$ for positive $\rhoi\,'$.  

\subsection{Initial curvature lumps and voids.}\label{curv_lv}

Unlike $\rho_i$, the intial 3-curvature $\Ri$ can be zero,  positive, negative or
change signs in any given domain of $r$. This makes the relation between $\Ri$,
$\Riav$, $K$ and $\Dki$ more complicated than in the case of $\rho_i$. Also,
the regularity conditions (\ref{cond_K2}) and (\ref{cond_K_3}), as well as the distinction
between the type of dynamics (parabolic, hyperbolic or elliptic), all involve specific
choices of $\Ri$ and $K$ but not of $\rhoi$ and $M$. 

\begin{figure}

\ba\matrix{{\BoxedEPSF{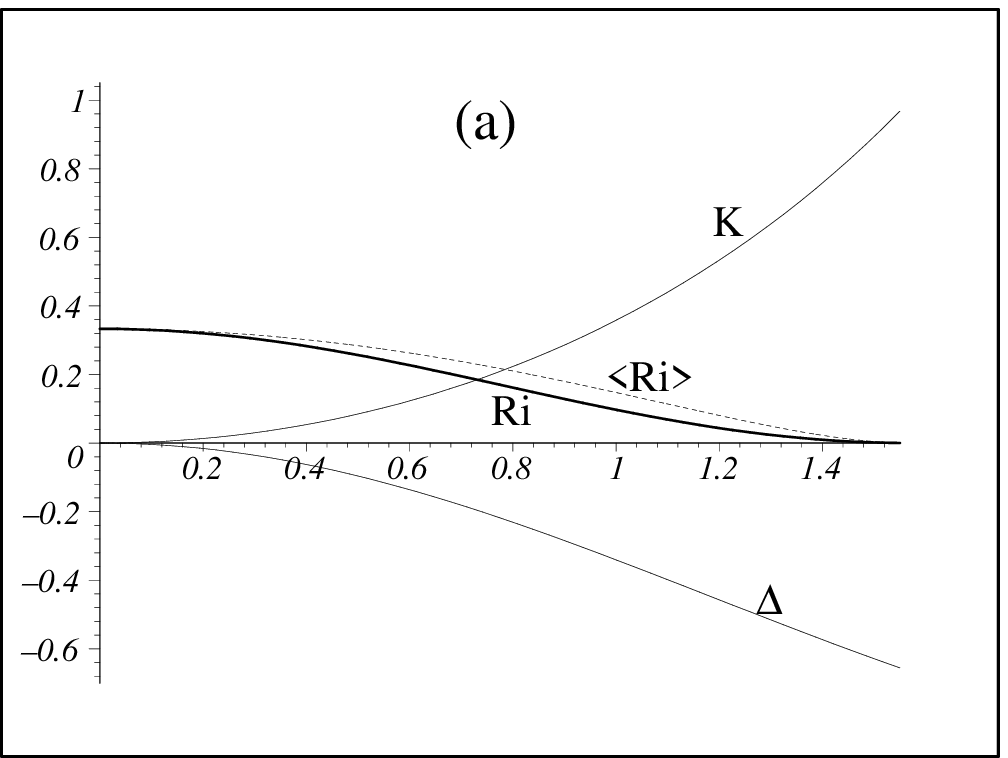 scaled 850}}&{\BoxedEPSF{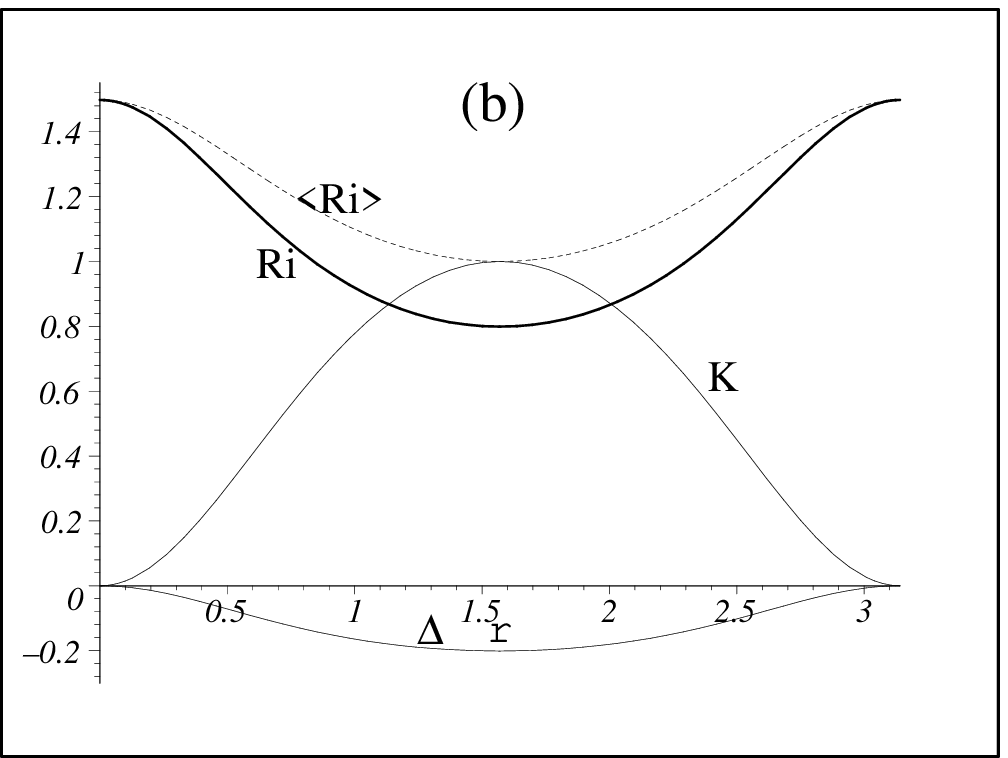
scaled 850}}\cr}\nonumber\ea
\begin{center}
\BoxedEPSF{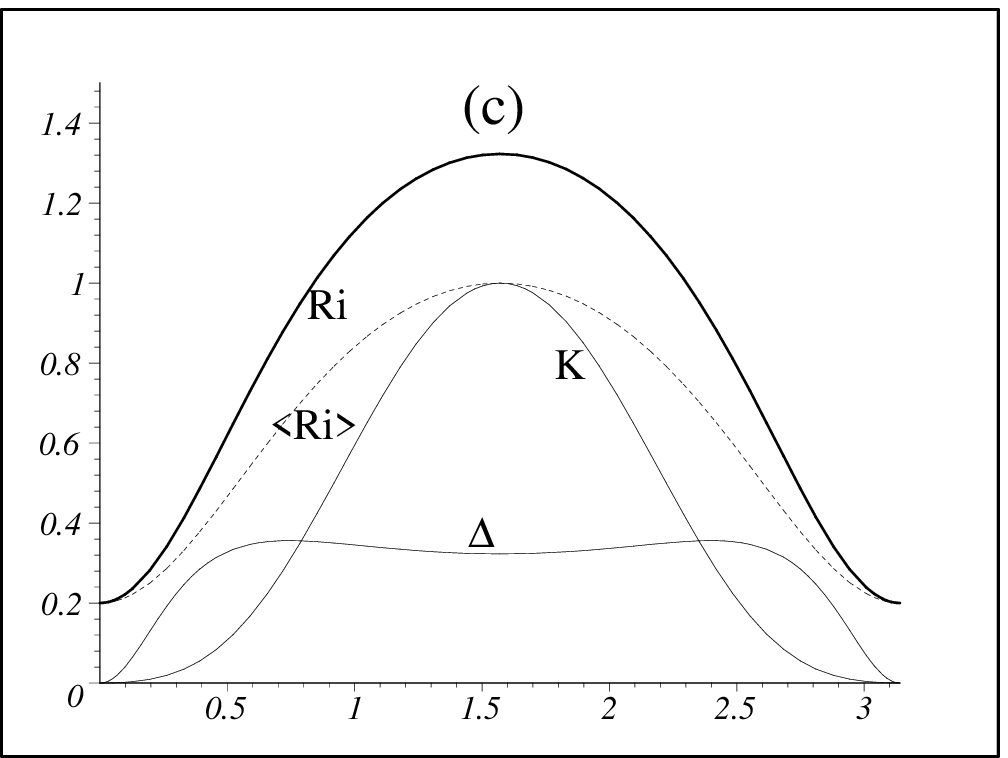 scaled
850}
\end{center}
\caption{$\underline{\hbox{Initial lumps and voids of positive 3-curvature}}$.
The figures display initial value functions $^{(3)}{\cal{R}}_i$ (thick curves), $\langle
^{(3)}{\cal{R}}_i\rangle$ (dotted curves), $K$ and $\Delta_i^{(k)}$ (marked by the symbol $\Delta$)
associated with an initial 3-curvature profile with
$^{(3)}{\cal{R}}_i>0$, hence the dynamics is necessarily elliptic. Lumps or voids of the 3-curvature
correspond to local maxima or minima of $^{(3)}{\cal{R}}_i$ at the SC's. Figure (a) shows the case
of a 3-curvature lump with one SC ($f=\tan\,r$), while (b) and (c) respectively depict lumps and
voids with two SC's ($f=\sin\,r$).  Models with elliptic dynamics must comply with regularity
conditions (48) and (54). For the functions of section VIII these
conditions are equivalent to (127), hence the constant parameters $b_1,\,b_2,\,S_i$
and ${}^{(3)}{\bar{\cal{ R}}}_i$ must satisfy condition (128) for the case in (a) and
condition (129) for the cases in (b) and (c). Notice that $K\to 1$ as $r\to\pi/2$ in
(a) and $K(\pi/2)=1$ in (b) and (c). The contrast function $\Delta_i^{(k)}$ is positive/negative for
voids/lumps, while $K$ vanishes at the SC's and has the same qualitative behavior for lumps and
voids. The initial density profiles that match the configuration in (a) is either one of those
depicted by figures 3a or 3b, while for the cases in (b) and (c)
it would be either one shown in figures  3c or  3d.}
\label{initfunc_pcurv}
\end{figure}

\begin{figure}
\ba\matrix{ {\BoxedEPSF{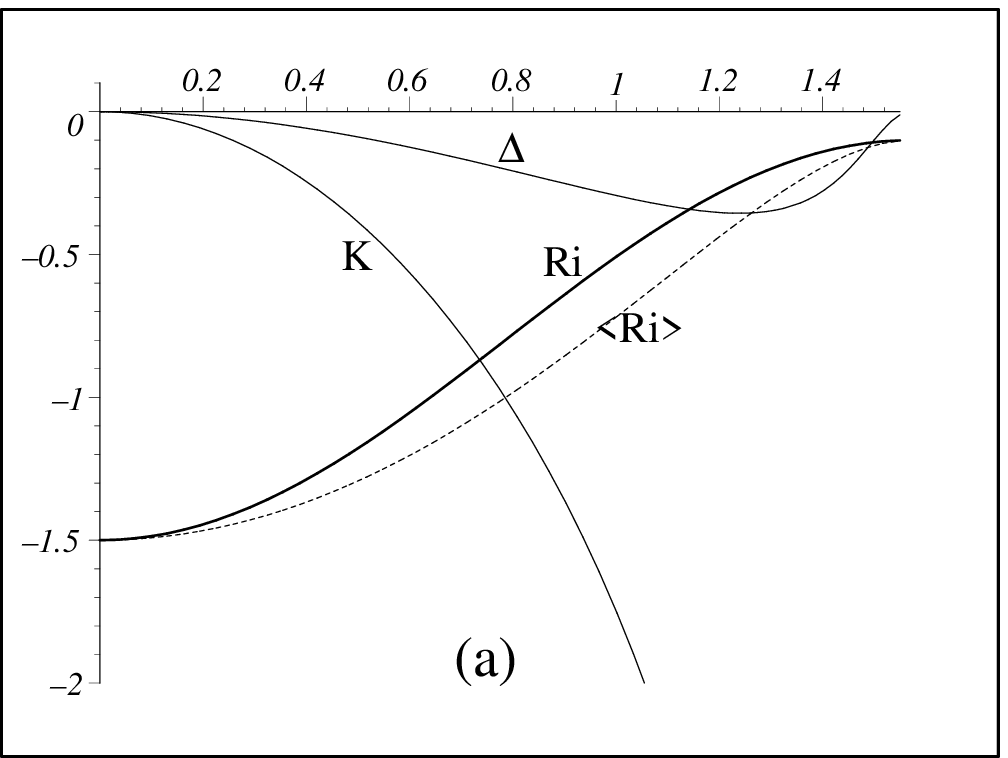 scaled 850}}&{\BoxedEPSF{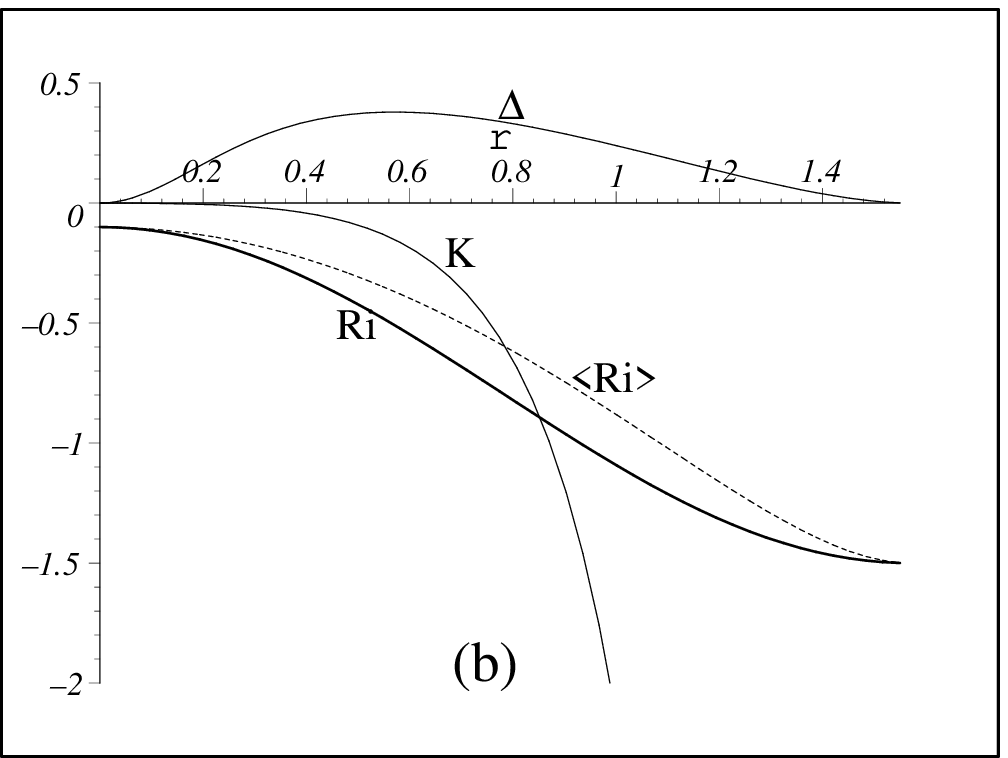
scaled 850}} }\nonumber\ea
\caption{$\underline{\hbox{Initial lumps and voids of negative 3-curvature}}$.  These figures
show $^{(3)}{\cal{R}}_i$ (thick curves), $\langle ^{(3)}{\cal{R}}_i\rangle$ (dotted curves), $K$ and
$\Delta_i^{(k)}$ (marked by the symbol $\Delta$), defined by (24), (36), (40) and (42), for
the case of negative 3-curvature  $^{(3)}{\cal{R}}_i\leq 0$, corresponding to hyperbolic dynamics.
Figure (a) displays a lump and figure (b) a void. Notice that now lumps/voids are characterized by
local minima/maxima of $^{(3)}{\cal{R}}_i$ at the SC (the pictures depict only the case with one
SC). The plots were obtained with equations (114), (123) and (125), with $f=\tan\,r$,\,
${}^{(3)}{\bar{\cal{ R}}}_i=-1$,\, $S_i=1$,\, $b_1=0.1,b_2=1.5$ in (a) and $b_1=1.5,\,b_2=0.1$ in
(b). The corresponding initial density profile for these figures would be either one of those shown
in figures 3a or 3b.}
\label{initfunc_ncurv}
\end{figure}

\begin{figure}
\ba\matrix{ {\BoxedEPSF{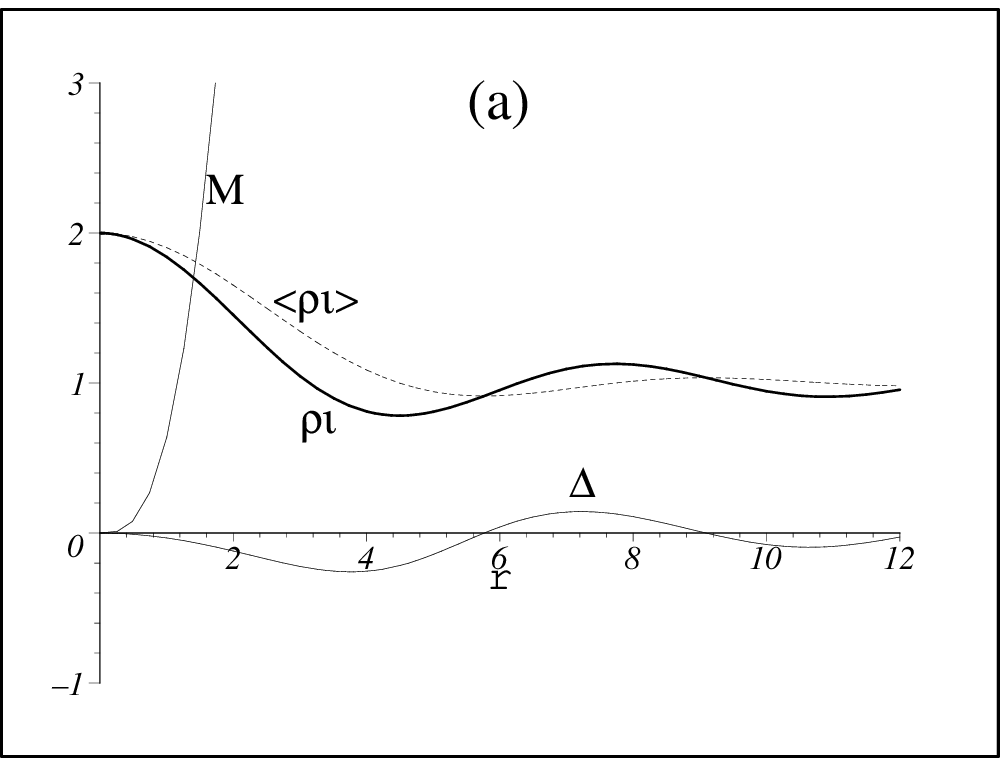 scaled 850}}&{\BoxedEPSF{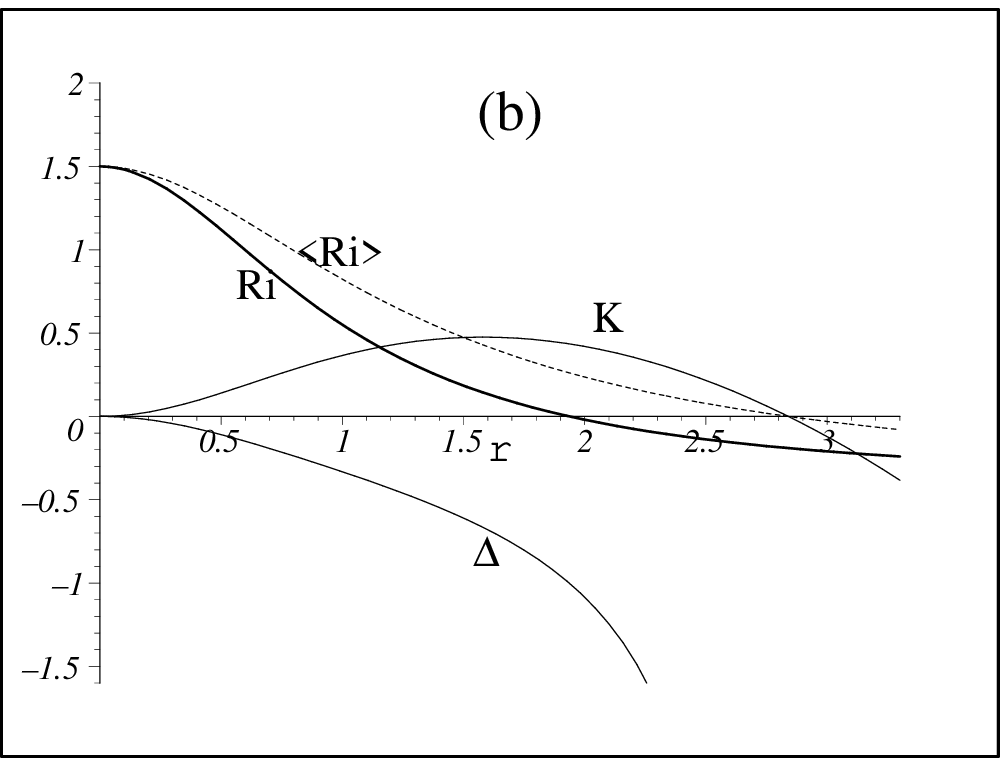
scaled 850}} }\nonumber\ea
\caption{$\underline{\hbox{Other initial conditions}}$. Figure (a) depicts the case where 
$\rho_i$ has many local maxima and minima, so that $\rho_i'$ changes sign many times (ie
``ripples''), then the contrast function $\Delta_i^{(m)}$ changes sign at each turning value. The
function used in this example is $\rho_i=\bar\rho_i(1+\sin\,r/r)$ with $\bar\rho_i=1$ and there is
only one SC. From (48), (54) and the no-shell-crossing conditions derived
in section VII and summarized in Table 1, such configurations must have elliptic
dynamics (with $K=1$ at each turning value) and might be very prone to develop shell crossings.
Figure (b) displays $^{(3)}{\cal{R}}_i$ (thick curve), $\langle ^{(3)}{\cal{R}}_i\rangle$ (dotted
curve), $K$ and $\Delta_i^{(k)}$, defined by (24), (36), (40) and
(42), for a 3-curvature lump in which an initialy positive $^{(3)}{\cal{R}}_i$ becomes
negative. The functions $\langle ^{(3)}{\cal{R}}_i\rangle$ and $K$ also change sign (change from
elliptic to hyperbolic dynamics). Notice that the passage from elliptic to hyperbolic dynamics does
not coincide with the zero of $^{(3)}{\cal{R}}_i$, hence there is a range of $r$ in which
$^{(3)}{\cal{R}}_i<0$ but $\langle ^{(3)}{\cal{R}}_i\rangle$ and
$K$ are still positive. The plots were obtained with equations (114), (123)
and (125) with $f =r$,\, ${}^{(3)}{\bar{\cal{ R}}}_i=1$,\, $S_i=1$,\, $b_1=-0.4$ and
$b_2=1.5$. Since $\langle ^{(3)}{\cal{R}}_i\rangle$ and $K$ vanish at the value of $r$ marking the
transition from elliptic to hyperbolic dynamics, we have (because of (43))
$\Delta_i^{(k)}\to -\infty$ for this value of
$r$.}
\label{initfunc_other}
\end{figure}

The following auxiliary relation, analogous
to (\ref{aux_rhoi}), follows from integration by parts of (\ref{def_Ri_av2}) 
\be \Ri \ - \ \Riav \ = \
\frac{1}{Y_i^3}\,\int_{r_c}^r{\Ri\,'\,Y_i^3\,dr},\label{aux_Ri}
\ee
In all intervals of $r$ containing a SC, $\Ri'\ne 0$ implies
$\Dki\ne 0$, except at the SC itself where both quantities vanish (from (\ref{central1})
and (\ref{central1})). If $\Ri=0$ for a domain of $r$ containing a
SC, then $\Riav = K = 0$ and this domain is a parabolic region (or a parabolic
solution if it is the full domain of $r$). However, we might have
$\Ri=0$ and/or $\Ri\,'=0$ in an open subset of the domain of $r$ containing no
SC's without $K$ or $\Dki$ vanishing. We look at various cases below. The discussion below is
illustrated by figures \ref{initfunc_pcurv}, \ref{initfunc_ncurv} and \ref{initfunc_other}b.

\bi
\item $\Ri > 0$ for a domain of $r$ containing a SC.

The relation between $\Ri$, $\Riav$, $\K$ and $\Dki$ is analogous to that
between $\rho_i$, $\rhoav$, $M$ and $\Dmi$ (see figures \ref{initfunc_pcurv}a,
\ref{initfunc_pcurv}b and \ref{initfunc_pcurv}c). Since  $\Riav$ and $K$ are everywhere positive,
the dynamics of the dust layers is elliptic.  

The results for the cases of curvature ``lumps'' or ``voids''(sign of
$\Ri'$) and/or when there is one or two centers are analogous to those discussed
earlier for $\rho_i$:
\ba \hbox{curvature lumps:}\qquad \Ri \ \leq \ \Riav, \qquad -\frac{2}{3} \ \leq \
\Dki
 \leq \ 0, \cr\cr
 \hbox{curvature voids:}\qquad \Ri \ \geq \ \Riav, \qquad \Dki
 \geq \ 0,\qquad\label{Dki_range}
\ea
where the minimal value $-2/3$ follows from $K>0$ and from the regularity of the
left hand side in (\ref{grad_k}).  Notice that condition (\ref{cond_K2}) is not
trivialy satisfied. We examine below the cases with one and two SC's.

\bi
\item{} One SC. (See figure \ref{initfunc_pcurv}a).

From (\ref{defK}) and (\ref{central2}), the functions $Y_i$ and $K$ vanish at
the center and increase as $r\to \rmax$, but (\ref{cond_K2})
requires that $K\leq 1$ for all the domain of $r$. Condition (\ref{cond_K_3}) can
be avoided by selecting $Y_i=S_i\,r$ so that $Y_i'$ never vanishes. Necessary and
sufficient condition for complying with (\ref{cond_K2}) and (\ref{ell2}) are
\be \Ri \ \to 0\qquad \hbox{and}\qquad K \ = \
\frac{1}{2\,Y_i}\int_{r_{c_1}}^{r}{\Ri\,Y_i^2\,Y_i'\,dr}
\ \to \ 1 \quad\hbox{as}\quad r \ \to \
\rmax,\label{cond_K_1sc}\ee
so that $\ell_i\to\infty$ as $r\to \rmax$. The model examined in
\cite{bon4} is a particular case of this type of configuration.  
 
\item{} Two SC's. (See figures \ref{initfunc_pcurv}b and \ref{initfunc_pcurv}c).

In this case all the functions $K,\,K',\,\Ri-\Riav$ and $\Dki$
must vanish at both centers. Also, $\Ri',\,\Riav',\,K'$ must all vanish for a
value $r^*$ in the open domain $r_{c_1} < r < r_{c_2}$ where $Y_i'$ vanishes. In
order to satisfy  (\ref{cond_K2}) and (\ref{cond_K_3}), the functions $\Ri$ and
$Y_i$ to be selected so that the following conditions hold  
\ba \int_{r_{c_1}}^{r_{c_2}}{\Ri\,Y_i^2\,Y_i'\,dr} \ = \
2\,K(r_{c_2})\,Y_i(r_{c_2}) \ = 0,\cr\cr\cr
\frac{1}{2\,Y_i(r^*)}\int_{r_{c_1}}^{r^*}{\Ri\,Y_i^2\,Y_i'\,dr} \ =  \ K(r^*) \ =
1,\label{cond_K_2sc}
\ea
The graphs of $\Ri,\,\Riav,\,K$ and $\Dki$ are shown in figures \ref{initfunc_pcurv} and
qualitatively analogous to those of $\rhoi,\,\rhoav,\,M$ and $\Dmi$ in figures
\ref{initfunc_dens}. The function $K$ must vanish at the SC's (like $M$) and  comply with
(\ref{cond_K_2sc}).    

\ei 

\item $\Ri < 0$ for a domain of $r$ containing a SC. (See figures \ref{initfunc_ncurv}a and
\ref{initfunc_ncurv}b).

In this case we have $\Riav$ and $\K$ negative for all the domain of $r$,
condition (\ref{cond_K2}) holds trivialy and dust layers follow hyperbolic
dynamics. However, the characterization of ``lump'' or ``void'' in
terms of the sign of $\Ri'$ is the inverse to that when $\Ri$ is nonegative: lump and void
are now respectively characterized by  $\Ri'>0$ and $\Ri<0$. Besides this point, the range of
values of $\Dki$ for lumps and is the same as for the non-negative case in
(\ref{Dki_range}).   

\item $\Ri$ changes sign but $K$ and $\Riav$ do not. (See figure \ref{initfunc_other}b).

Since $\Riav$ and $K$ are obtained from an integral involving $\Ri$ evaluated
from a SC to a given $r$, it is quite possible to have situations in which
$\Ri$ vanishes or pases from positive to negative (or viceverza) but $\Riav$ and
$K$ remains positive (or negative) at least for an interval of the domain of $r$
not containing a SC, so that the model remains elliptic or hyperbolic in this domain. The
situation in which $\Riav$ or $K$ vanish for a given
$r=r^*$ in an open domain of $r$ that excludes a SC is qualitatively different
and more complicated, since the dynamic evolution of dust layers (governed by
(\ref{freq1}) or (\ref{freq2})) passes from elliptic to parabolic or hyperbolic
(or viceverza). This case will be examined in a separate paper\cite{sussgar}.    
      
\ei

\section{The function $\Gamma$, shell crossing singularities and bang
times.}\label{Gamma_etc}

The function $\Gamma$ defined in (\ref{Gamma}) is essential for calculating 
relevant quantities characterizing LTB solutions, such as $\rho$, $\Theta$, 
$\sigma$, $\R$ and $\psi_2$ ((\ref{Theta1}), (\ref{sigma1}),
(\ref{def_R}), (\ref{def_E}) and (\ref{rho2})). The exact functional form of
$\Gamma$ is also required in order to find the restrictions on initial conditions
that fulfil the important regularity condition: 
\be \Gamma \ \geq \ 0 \quad \hbox{for}\quad \rhoi \ \geq \ 0\quad
\hbox{and}\quad K \ \ne \ 1,\label{shx2}\ee
with $\Gamma=0$ only if $\rhoi=0$ and/or $1-K=0$. This condition garantees that:
(a) the metric coefficient $g_{rr}$ is regular, (b) $\rho\geq  0$ for a given
choice of $\rhoi\geq 0$ and (c) provides the necessary and/or sufficient
restrictions on initial value functions in order to avoid shell crossing
singularities (these conditions are summarized in Table 1).  We show in this section that the
conditions for avoiding these singularities based on (\ref{shx2}) are equivalent to those found
by Hellaby and Lake \cite{he_la2} (see also \cite{humph_phd} and \cite{HM1}).

The function $\Gamma$ can be obtained in terms of $y$ and the initial
value functions $\mu,\,\kappa,\,\Dmi,\,\Dki$ by eliminating $y'/y$ from an
implicit derivation with respect to $r$ of equations (\ref{parab_can2}),
(\ref{ell_can21}), (\ref{ell_can22}) and (\ref{hyp_can2}). For hyperbolic and
elliptic dynamics, $\Gamma$ in terms of $\eta$ follows from implicit derivation of
(\ref{ell_par_t2}) and (\ref{hyp_par_t2}), using (\ref{ell_par_Y2}), 
(\ref{hyp_par_Y2}), (\ref{ell_def_eta}) and (\ref{hyp_def_eta}) in order to
eliminate $\eta'$ in terms of $y'/y$. We provide below the most compact forms of this function
for parabolic solutions (in terms of $y$) and for hyperbolic and elliptic solutions (in terms of
$\eta$)  
\be \Gamma \ = \ 1 \ + \ \Dmi \ - \
\frac{\Dmi}{y^{3/2}},\qquad\hbox{parabolic,}\label{Gamma_par}\ee
\be  \Gamma \ = \
1 \ + \ 3\,(\,\Dmi - \Dki\,)\left[\,1 \ - \ \frac{P}{P_i}\,\right] \ - \ 3\,P
\,\left[\,Q-Q_i\,\right]\,\left(\Dmi-\frac{3}{2}
\Dki\right),\qquad\hbox{hyperbolic and elliptic,}\label{Gamma_gen}\ee
with $P(\eta)$ and $Q(\eta)$ given by
\bi
\item{}Hyperbolic dynamics
\ba P(\eta) \ = \ \frac{\sinh\,\eta}{(\cosh\,\eta - 1\,)^2} \ = \
\frac{\sqrt{\,2\,+\,x\,y}}{(\,x\,y\,)^{3/2}},\qquad\qquad\qquad\qquad\cr\cr\cr Q(\eta) \
=
\
\sinh\,\eta - \eta \ = \ \sqrt{x\,y\,(2+x\,y\,)} \ - \ \hbox{arccosh}\,(1+x\,y),
\label{funcs_PQ_h}\ea
\ba P_i \ = \ P(\eta_i) \ = \ \frac{\sqrt{2+x}}{x^{3/2}},\qquad\qquad Q_i \ = \ Q(\eta_i) \ =
\ \sqrt{x\,(2+x)} \ - \ \hbox{arccosh}\,(1+x),\cr\cr\cr \eta \ = \
\hbox{arccosh}\,(\,1+x\,y\,) \ \geq \ 0, \qquad \eta_i \ = \
\hbox{arccosh}\,(\,1+x\,) \ \geq \ 0,\qquad x \ \equiv \
\frac{|\kappa|}{\mu},\label{etax_h}\ea
\item{} Elliptic dynamics
\ba P(\eta) \ = \ \frac{\sin\,\eta}{(\,1-\cos\,\eta)^2} \ = \
\frac{\sqrt{x\,y\,(2\,-\,x\,y\,)}}{(\,x\,y\,)^2} \ = \ \pm
\frac{\sqrt{\,2\,-\,x\,y}}{(\,x\,y\,)^{3/2}},\qquad\qquad 
\cr\cr\cr Q(\eta) \ = \  \eta \ - \ \sin\,\eta \ = \
\arcsin\,(\,1-x\,y\,) \ - \ \sqrt{x\,y\,(2-x\,y\,)},
\label{funcs_PQ_e}\ea
\ba P_i \ = \ P(\eta_i) \ = \
\frac{\sqrt{2-x}}{x^{3/2}},\qquad\qquad Q_i \ = \ Q(\eta_i) \ =
 \ \arccos\,(1-x) \ - \ \sqrt{x\,(2-x)},\cr\cr\cr \eta \ = \
\arccos\,(\,1-x\,y\,), \qquad 0 \ \leq \ \eta \ \leq \ 2\pi 
\qquad\qquad\qquad\qquad\qquad\qquad\qquad\cr\cr \eta_i \ = \
\arccos\,(\,1-x\,), \qquad 0 \ < \ \eta_i \ < \ \pi, \qquad \qquad x \ \equiv \
\frac{\kappa}{\mu}, \qquad\qquad\qquad 
\label{etax_e}\ea
\ei
where all functions of $\eta$ and $\eta_i$ have been constructed with (\ref{ell_def_eta}) and
(\ref{hyp_def_eta}) and the gradients of the form $\mu'/\mu$ and $\kappa'/\kappa$ were
eliminated in terms of $\Dmi$ and $\Dki$ by means of equations (\ref{grad_m}), (\ref{grad_k})
and (\ref{grad_m32k}).  Notice that the \,$\pm$\, sign in (\ref{funcs_PQ_e}) follows from the
fact that \,$\sin\eta$\, becomes negative in the collapsing phase ($\pi \leq\eta <2\pi$),
whereas \,$\sinh(\eta)$\, is positive for all $\eta\leq 0$. It is possible to use
(\ref{etax_h}) and (\ref{etax_e}) in order to transform $\Gamma$ in (\ref{Gamma_gen}) as a
function of $y$ instead of $\eta$, but the obtained expressions are more complicated than
(\ref{Gamma_gen}). Also, the form of $\Gamma=\Gamma(y,r)$ for elliptic solutions consists of
two branches, making it very impractical for most purposes.  

Since the central and shell crossing singularities are (in
general) not simmultaneous with respect to $t$, the existence of regular initial
hypersurfaces $t=t_i$ is not a trivial issue. Such hypersurface must avoid the
coordinate locii of shell crossing and central singularities. The conditions
for avoiding shell crossings will be obtained below, not just for $t=t_i$, but for
all the time evolution, but assuming that the no-shell-crossing conditions have been
met, the hypersurface $\Ti$ must still avoid the central singularity. The conditions
for this can be given straight away as 
\ba t_i \ - \ \tb \ > \ 0,\quad \hbox{for expanding layers},\cr\cr
t_i \ - \ \tb \ < \ 0,\quad \hbox{for collapsing layers}.\label{no_tb}\ea      
These regularity considerations provide an insight on the relation of the
``bang time'' $\tb$ with the initial value functions $\rhoi$, $\Ri$ and their
corresponding average and contrast functions. From the forms of $\tb$ that
will be obtained in this section and displayed in figures \ref{bangtimes}, the fact that
$\tb'\ne 0$ leads to an $r$ dependent ``age difference'' for the comoving observers, a
difference whose maximal value is very sensitive to $\Dmi$ and $\Dki$, and so to the
ratios of maximal to minimal $\rhoi$ and $\Ri$ (but not to local values $\rhoi$ or
$\Ri$). This feature is illustrated in figures \ref{bangtimes}. 

The fulfilment of conditions (\ref{shx2}), as well as a comparison with
no-shell-crossing conditions of Hellaby and Lake \cite{he_la2}, is examined below for the
cases with parabolic,  hyperbolic and elliptic dynamics.     
 
\subsection{Parabolic solutions.}\label{Gamma_etc_p} 

From (\ref{Gamma_par}) and (\ref{shx2}), shell crossing singularities 
are avoided if
\ba (1+\Dmi)\,y^{3/2} \ - \ \Dmi \ \geq \ 0 \qquad \Rightarrow\qquad y^{3/2} \
\geq \ \frac{\Dmi}{1+\Dmi},\ea
Therefore, necessary and sufficient conditions for avoiding shell crossings
are
\be -1\ \leq \ \Dmi \ \leq \ 0,\qquad \rhoi \ \geq \ 0,\label{shx_st_p}\ee
conditions implying that we have initial density lumps.    

Notice that condition (\ref{shx_st_p}) will not hold, in general, for
density voids ($\Dmi \geq 0$). However, since $\Dmi/(1+\Dmi)\leq 1$ and given a
choice of a regular $t=t_i$, condition (\ref{shx_st_p}) is satisfied for initial
lumps or voids and an evolution free of shell crossings is always possible for all
$y\geq 1$ (or $t\geq t_i$ in expanding configurations). This fact allows for a
well posed intial value study of LTB parabolic solutions with density voids,
a feature that has been exploited in papers (\cite{bon_ch},\cite{cel}) dealing
with LTB solutions in which shell crossings do occur, but their coordinate locus
can be confined to earlier times $\tb < t <t_i$ (see figure \ref{reg_LTBvoids}a).   

The no-shell-crossing conditions of Hellaby and Lake (\cite{he_la2},
\cite{humph_phd}, \cite{HM1}) for the parabolic case and
for expanding layers are given by 
\be \pm\, Y' \ \geq \ 0 \quad \Rightarrow \quad \{\,\pm\,M'\ \geq 0, \quad
\pm\,c\,\tb' \ \leq \ 0
\},\label{shx_hl_p}\ee
where the $\pm$ sign accounts for the possibility that $Y'$ might change sign
regularly (only if $M'$ vanishes). However, the equality of the signs of $Y'$ and
$M'$ in (\ref{shx_hl_p}) follows from (\ref{signsYiM}) and and $m'=\rho Y^2Y'=\rhoi
Y_i^2Y_i'$, and is already contained in (\ref{shx2}).  Regarding the sign of $\tb$ we
have, from setting $y=0$ in (\ref{parab_can2}), the following expression for $\tb$  
\be c\,\tb \ = \ c\,t_i \ \mp \ \frac{2}{3\,\sqrt{2\,\mu}},\label{tb_p}\ee
where the $\mp$ sign accounts for expanding ($-$) or collapsing ($+$) layers.
In order to compare with (\ref{shx_hl_p}), we consider $\tb'$ the case of
expanding layers 
\be c\,\tb' \ = \ \frac{1}{3\sqrt{2\,\mu}}\,\frac{\mu'}{\mu} \ = \
\frac{1}{\sqrt{2\,\mu}}\,\frac{Y_i'}{Y_i}\,\Dmi,\label{tbr_p} \ee        
This equation clearly shows that $\tb'\leq 0$ for $Y_i'\geq 0$ only occurs if   
$\Dmi\leq 0$. Thus, the no-shell-crossing conditions of Hellaby-Lake 
(\ref{shx_hl_p}) are equivalent to (\ref{shx_st_p}). However, we argue that
phrasing these conditions in terms of $\rhoi,\,Y_i$ and $\Dmi$ is more intutive
than to do it in terms of $Y',\,M'$ and $\tb'$. The function $\Gamma$ for a regular model with
parabolic dynamics is qualitatively analogous to the plot displayed in figure \ref{reg_LTB}a.

\begin{figure}
\begin{center}
\BoxedEPSF{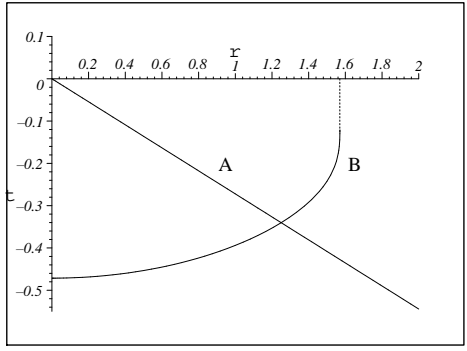 scaled 600}
\end{center}
\caption{$\underline{\hbox{Ill-suited initial value functions}}$. The figure displays the bang
time $t_{_{bb}}$ for ``bad'' choices of the initial value function $\rho_i$. The initial hypersurface
${\cal{T}}_i$ is marked by $t=0$, we have chosen $Y_i\propto r$ and $^{(3)}{\cal{R}}_i=0$ (parabolic
dynamics), so that $t_{_{bb}}$ is given by (84). The curve ``A'' correponds to\,
$\rho_i\propto 1/r$,\, a form that is singular at the SC ($r=0$), hence ${\cal{T}}_i$ is also
singular at $r=0$ (it is intersected by $t_{_{bb}}$ at $r=0$). The curve ``B'' corresponds to
$\rho_i\propto 1/\cos\,r$, diverging as $r\to\pi/2$. In this case ${\cal{T}}_i$ is singular at
$r=\pi/2$ (it is intersected by $t_{_{bb}}$). For this choice of $\rho_i$ we have a void with
infinite contrast and so shell crossings necessarily emerge. Notice (from figure 7)
that ${\cal{T}}_i$ is perfectly regular for all initial value functions obtained from the
expressions of section VIII. We need to select truly pathological initial value functions in
order to have a singular
${\cal{T}}_i$. }
\label{initfunc_badf}
\end{figure}

 Given a choice of $t=t_i$, it is evident from (\ref{shx_st_p}) that
the intial hypersurface avoids shell crossings, either for lumps or voids  (as
long as $\Dmi$ is finite). Regarding the central singularity, the form of
$\tb$ in (\ref{tb_p}) implies that conditions (\ref{no_tb}) are satisfied
as long as $\mu$ does not diverge, a condition that is, in turn, satisfied
as long as $\rhoi$ and $\rhoi'$ are finite. As shown by figure \ref{initfunc_badf}, selecting
unphysical initial conditions associated with ``bad'' choices of
$\rhoi$ (diverging at the SC or for $r>r_c$ in a void) leads to a singular $\Ti$ for which
(\ref{no_tb}) is violated. As shown by figures \ref{bangtimes}a and \ref{bangtimes}b (the
parabolic case is analogous to the hyperbolic one), for reasonable initial value functions it is
always possible to garantee that a regular hypersurface $t=t_i$ exists. In general, a ``good''
choice of $M$ and $\tb$, complying with (\ref{shx_hl_p}), will yield a ``good'' form of $\rhoi$
and $Y_i$ complying with (\ref{shx_st_p}) and (\ref{no_tb}), however, we feel that selecting a
`good' form for $\rhoi$ is easier and more intuitive that choosing adequate forms for $\tb$ and
$M$. 

\subsection{Hyperbolic solutions.}\label{Gamma_etc_h}

In order to examine the fulfilment of (\ref{shx2}), we assume that $\rhoi\geq 0$
and $\Ri\leq 0$ and rewrite (\ref{Gamma_gen}) as 
\be \Gamma \ = \ 1 \ + \ 3\,\left[\,A\,
\Dmi \ - \ B\,\Dki\,\right] \ - \ \frac{3\,P}{P_i}
\,\left[\,A_i\,\Dmi \ - \ B_i\,\Dki\,\right],\label{Gamma_hyp1}\ee
\be A(\eta) \ = \ 1 \ - \ P(\eta)\,Q(\eta),\qquad B(\eta) \ = \ 1 \ - \
\frac{3}{2}\,P(\eta)\,Q(\eta), \qquad A_i \ = \ A(\eta_i),\quad
B_i \ = \ B(\eta_i),\ee  
where $P(\eta)$ and $Q(\eta)$ are given by (\ref{funcs_PQ_h}).
The fulfilment of (\ref{shx2}) obviously depends on the behavior of $A,\,B$ and
$P$ in the full range $\eta\geq 0$. We have $P\geq 0$ with $P\to\infty$ as
$\eta\to 0$, while in the same limit $A\to 2/3$ and $B\to 0$. Therefore,the
following are necessary conditions for (\ref{shx2})
\be A_i\,\Dmi \ - \ B_i\,\Dki \ \leq \ 0,\label{shx_st_h1}\ee
\be 1 \ + \ 3\,\left[\,A\,\Dmi \ - \ B\,\Dki\,\right]\ \geq \
0,\label{shx_st_h2}\ee   
However, $A$ and $B$ satisfy $0< A\leq 2/3$ and $-1/2< B\leq 0$ for all
$\eta\geq 0$, with $A\to 0$ and $B\to 1/2$ as $\eta\to\infty$. The functions
$A_i$ and $B_i$ are bounded by the same values. Further restrictions
follow by inserting these limiting values into (\ref{shx_st_h1}) and
(\ref{shx_st_h2}). This leads to the following necessary and sufficient
conditions for an evolution free of shell crossings
\be \rhoi \ \geq \ 0,\qquad \Ri \ \leq \ 0,\qquad  -1 \ \leq \ \Dmi \ \leq \ 0,
\qquad  -\frac{2}{3} \ \leq \ \Dki \ \leq \ 0,\label{shx_st_h}\ee
Since (\ref{shx_st_h1}) and (\ref{shx_st_h2}) imply (\ref{shx_st_h}) and the
latter also implies the former, conditions (\ref{shx_st_h}) are, indeed, necessary
and sufficient.  Notice that, just as for parabolic solutions, shell crossings
cannot be avoided (in general, see subsection \ref{Gamma_etc_sbb} below) for positive
contrast functions (initial density or 3-curvature voids).  The functions $\Gamma$ and $\rho$ for
a regular model with hyperbolic dynamics are shown in figures
\ref{reg_LTB}a and \ref{reg_LTB}b.      

The no-shell-crossing conditions of Hellaby and Lake (\cite{he_la2},
\cite{humph_phd}, \cite{HM1}) for the hyperbolic case and
for expanding layers are given by 
\be \pm\, Y' \ \geq \ 0 \quad \Rightarrow \quad \{\,\pm\,M'\ \geq 0, \quad
\pm\,K'\ \leq 0, \quad \pm\,c\,\tb' \ \leq \ 0
\},\label{shx_hl_h}\ee
where, as in the parabolic solutions, the $\pm$ sign accounts for the  
possibility that $Y'$ might change sign regularly and we have used (\ref{E_K}) in
order to translate a positive $E$ (in the original LTB variables) into a negative $K$. As in
the parabolic solutions, the condition that $Y',\,Y_i'$  and $M'$ must have the same sign
follows from (\ref{signsYiM}), $\rhoi\geq 0$ and $m'=\rho
Y^2Y'=\rhoi Y_i^2Y_i'$. Opposite sign for $Y'$ and $K'$ follows from (\ref{grad_k})
and $\Ri\leq 0$ (bearing in mind that $K< 0$ for hyperbolic solutions).  Regarding
the sign condition on $\tb'$, we compute $\tb$ by setting $\eta=0$ and $y=0$ in
equations (\ref{hyp_can2}) and (\ref{hyp_par_t2}), leading to
\be c\,\tb \ = \ c\,t_i \ \mp \
\frac{\mu\,Q_i}{|\kappa|^{3/2}} \  = \ c\,t_i \
\mp \ \frac{1}{\sqrt{|\kappa|}}\,\left[\left(1+\frac{2}{x}\right)^{1/2}
-\frac{\hbox{arccosh}\,(1+x)}{x}\right],\qquad x \ \equiv \
\frac{|\kappa|}{\mu}\label{tb_h}\ee
where the $\pm$ sign accounts for expanding ($-$) and collapsing ($+$) layers.
Considering expanding layers and the expression given in terms of $\eta_i$, we
obtain
\be c\,\tb' \ = \
\frac{3Y_i'}{Y_i}\,\frac{\mu}{|\kappa|^{3/2}\,P_i}\,\left[\,A_i\,\Dmi
\ - \ B_i\,\Dki\,\right],\label{tbr_h}\ee                    
but, because of (\ref{shx_st_h1}), the factor in square brackets must be negative
or zero for all $\eta_i>0$, thus implying that $\tb'$ and $Y_i'$ must have
opposite signs. Looking at the maximal and minimal values of $A_i$ and $B_i$,
together with (\ref{grad_m}) and (\ref{grad_k}) leads then to conditions
(\ref{shx_st_h}). Therefore, Hellaby-Lake no-shell-crossing conditions are
equivalent to (\ref{shx_st_h}).    

Assuming that (\ref{shx_st_h1}) and (\ref{shx_st_h2}) hold, a regular
$\Ti$ must satisfy conditions (\ref{no_tb}) in order to avoid
the central singularity. Since the term in square brackets in (\ref{tb_h}) tends,
respectively, to $0$ and $1$ as $x\to 0$ and $x\to \infty$, sufficient conditions
for (\ref{no_tb}) to hold are that both $\mu$ and $|\kappa|> 0$ be finite
everywhere. These conditions are satisfied if $|\Ri|>0$ and $\rhoi$ are finite. 
As with the parabolic case, conditions (\ref{no_tb}) are violated only with very
unphysical initial conditions that might also violate the no-shell-crossing
conditions (see figures \ref{bangtimes}a and \ref{bangtimes}b), since diverging $\rhoi$ and
$|\Ri|$ can be associated with for voids with large values of $\Dmi$ and $\Dki$. Hence, a regular
hypersurfaces $\Ti$ always exist for physicaly reasonable intial conditions and
a `good' selection of functions $M,\,E=-K,\,\tb$ leads to `good' forms of $\Ri,\,\rhoi,\,
Y_i$. However, we argue again, that finding a `good' choice of the latter is easier and more
intuitive.

Shell crossing singularities will always emerge for initial conditions with
density and/or 3-curvature voids (see figure \ref{bangtimes}b), however for reasonable initial
conditions characterized by finite $\rhoi\geq 0$ and $\Ri\leq 0$, we must have finite $\mu>0$
and $\kappa<0$ in all $\Ti$ (see section \ref{ex_LTB} for examples). Thus, unless we choose
pathological  initial value functions as in figure \ref{initfunc_badf}, quantities like $P_i$
and $Q_i$ defined in (\ref{etax_h}) will be finite and non-zero, leading to $\Gamma\to
1-(3/2)\Dki$ as $\eta\to\infty$ asymptoticaly, therefore a regular evolution of intial voids is
possible in the domain $t\geq t_i$ as long as $\Dki\leq 2/3$ (for expanding layers). An LTB
model with initial voids and hyperbolic dynamics following this type of evolution is depicted in
figure \ref{reg_LTBvoids}a.  

\subsection{Elliptic solutions.}\label{Gamma_etc_e} 

As with the hyperbolic case, we assume that $\rhoi\geq 0$ and $\Ri\geq 0$ and
rewrite (\ref{Gamma_gen}) as 
\be  \Gamma \ = \
1 \ + \ 3(\,\Dmi\,-\,\Dki\,) \ - \ 3\,P\,Q
\,\left(\,\Dmi\,-\,\frac{3}{2} \Dki\,\right) \ - \
\frac{3P}{P_i}\,\left[\,A_i \,\Dmi\,-\,B_i\,
\Dki\,\right],\label{Gamma_ell}\ee
\ba 
A_i \ = \ 1 \ - \ P_i\, Q_i,\qquad B_i \ = \ 1 \ - \
\frac{3}{2}\,P_i\, Q_i, \qquad P_i
\ = \ P(\eta_i), \qquad Q_i \ = \ Q(\eta_i), 
\nonumber\ea
where $P(\eta)$ and $Q(\eta)$ are given by (\ref{funcs_PQ_e}). Necessary and
sufficient conditions for avoiding shell crossings can be obtained from
(\ref{shx2}) and (\ref{Gamma_ell}). Following Hellaby and Lake \cite{he_la2}, we
explore the sign of the functions $P$ and $PQ$ near the big bang ($\eta\approx 0$)
and near the big crunch ($\eta\approx 2\pi$). As $\eta\to 0$ we have
$P\to +\infty$ and $PQ\to2/3$, while  $PQ\to 2\pi P\to -\infty$ as $\eta\to 2\pi$,
therefore necessary conditions for $\Gamma\geq 0$ are
\be A_i\,\Dmi \ - \ B_i\,\Dki \ \leq \ 0,\label{shx_st_e1}\ee
\be 2\pi\,P_i\,\left(\,\Dmi\,-\,\frac{3}{2} \Dki\,\right)+A_i\,\Dmi-B_i\,\Dki \
= \ (2\pi -Q_i)\,P_i\,\left(\,\Dmi\,-\,\frac{3}{2}
\Dki\,\right)+\Dmi-\Dki \ \geq \ 0,\label{shx_st_e2}\ee
Since the range of $\eta_i$ is $0<\eta_i\leq \pi$, we have $1/3\leq A_i\leq 1$,
\,$0\leq B_i\leq 1$ and $0\leq Q_i\leq \pi$, a third necessary (but not
sufficient) condition that follows  from (\ref{shx_st_e1}) and (\ref{shx_st_e2})
is
\be \Dmi\,-\,\frac{3}{2} \Dki \ \geq \ 0, \label{shx_st_e3}\ee
Following the arguments of Hellaby and Lake \cite{he_la2}, we prove the 
sufficiency of (\ref{shx_st_e1}) and (\ref{shx_st_e2}) by assuming that 
\ba 1 \ + \ 3\left(\,\Dmi\,- \,\Dki\,\right) \ = \  3\alpha\,(\eta_i)\,
\left(\,\Dmi \ - \ \frac{3}{2}\Dki\right),\cr\cr \Dmi \ - \ \frac{3}{2}\,\Dki \ =
\  -\frac{\beta(\eta_i)}{P_i}\,
\left(A_i\,\Dmi \ - \ B_i\,\Dki\right),\label{aux_shx_e}\ea
for unspecified functions $\alpha(\eta_i),\,\beta(\eta_i)$,  so that 
\ba \Gamma \ = \ 
-\frac{A_i\,\Dmi\,-\,B_i\,\Dki}{P_i}\left[\,\beta\,\left(\alpha\,-\,
PQ\,\right)\,+\,P\,\right],\nonumber
\ea
and the fulfilment of $\Gamma\geq 0$ requires (\ref{shx_st_e1}) and \, $\beta\geq
-P/(\alpha-PQ)$.  However, from the forms of $P$ and $Q$ in (\ref{Gamma_ell}),
this latter condition can only be satisfied in the whole evolution range
$0<\eta<2\pi$ if the functions $\alpha,\,\beta$ satisfy $\alpha \geq 2/3$ and
$\beta \geq 1/(2\pi)$. Inserting these inequalities into (\ref{aux_shx_e}) we
obtain (\ref{shx_st_e2}) together with $\Dmi\geq -1$. Conditions
(\ref{shx_st_e1}) and (\ref{shx_st_e2}) are then necessary and sufficient, but we
still need to test the compatibility of these conditions with the signs of $\Dmi$
and $\Dki$ (initial lumps or voids). Clearly, (\ref{shx_st_e1})
is compatible with either $\Dmi\leq 0$ and $\Dki\geq 0$ (density lump and
3-curvature void) or with $\Dmi\leq 0$ and $\Dki\leq 0$ (density and 3-curvature
lumps). However, (\ref{shx_st_e2}) is only compatible with the second choice.
Therefore, the full set of necessary and sufficient conditions for avoiding shell
crossing singularities are (\ref{shx_st_e3}), together with
\be  -\frac{2}{3} \ \leq \ \Dki \ \leq \ 0,\qquad  -1 \ \leq \ \Dmi \ \leq \ 0,
\label{shx_st_e4}\ee     
and the two inequalities (\ref{shx_st_e1}) and (\ref{shx_st_e2}) which can be
combined into the single expression
\be 2\,\pi\,P_i\,\left(\Dmi\,-\,\frac{3}{2} \Dki\right) \ \geq \
P_i\,Q_i\,\left(\Dmi\,-\,\frac{3}{2} \Dki\right) \ - \ \left(\Dmi\,-\,\Dki\right)
\ \geq \ 0,\label{shx_st_e5}\ee
A graphical representation of this condition is provided in figures \ref{initfunc_test}a and
\ref{initfunc_test}b.    
     
The no-shell-crossing conditions of Hellaby and Lake (\cite{he_la2},
\cite{humph_phd}, \cite{HM1}) for the elliptic case are given by 
\be \pm\, Y' \ \geq \ 0 \quad \Rightarrow \quad \{\,\pm\,M'\ \geq 0, 
\quad \pm\,c\,\tb' \ \leq \ 0,\quad \pm\,c\,\tb' \ \geq \ \frac{-2\pi
M}{K^{3/2}}\,\left(\frac{M'}{M}-\frac{3}{2}\frac{K'}{K}\right)
\},\label{shx_hl_e}\ee
where the $\pm$ sign accounts for the possibility that $Y'$ might change sign
regularly and (\ref{E_K}) has been used. As in the parabolic and hyperbolic
solutions, the condition that $Y',\,Y_i'$  and $M'$ must have the same sign
follows from $\rhoi\geq 0$, (\ref{grad_m}), (\ref{signsYiM}) and $m'=\rho\,
Y^2Y'=\rhoi\, Y_i^2Y_i'$. The same sign for $Y'$ and $K'$ follows from
$K\geq 0$, (\ref{grad_k}) and (\ref{shx_st_h1}).  Regarding the sign condition on
$\tb'$, we compute $\tb$ by setting $\eta=0$ and $y=0$ in equations
(\ref{ell_can21}) and (\ref{ell_par_t2}), leading to
\ba c\,\tb \ = \ c\,t_i \ - \ \frac{\mu\,Q_i}{\kappa^{3/2}} \ = \ c\,t_i \ - \
\frac{1}{\sqrt{\kappa}}\,\left[\frac{\arccos\left(\,1\,-\,
x\,\right)}{x} \ - \ \left(\frac{2}{x}-1\right)^{1/2}
\right],\cr\cr\cr 0\ \leq \kappa \ \leq
2\,\mu,\qquad x \ \equiv \ \frac{\kappa}{\mu},\qquad 0 \ \leq \ x \ \leq
\ 2\label{tb_e}
\ea
We differentiate the term depending on $\eta_i$ with respect to
$r$ 
\be c\,\tb' \ = \
\frac{3Y_i'}{Y_i}\,\frac{\mu}{\kappa^{3/2}\,P_i}\,\left[\,A_i\,\Dmi
\ - \ B_i\,\Dki\,\right], \label{tbr_e}
\ee                    
where $A_i,\,B_i$ and $P_i$ are the same functions given in
(\ref{Gamma_ell}). Comparing $\tb'$ above with (\ref{shx_st_e1}), it is evident
that the Hellaby-Lake conditions (\ref{shx_hl_e}) involving $\tb'$ are the same as
(\ref{shx_st_e1}) and (\ref{shx_st_e2}), and so can be reduced to
(\ref{shx_st_e3}) - (\ref{shx_st_e5}) if one translates these conditions in
terms of $\Dmi$, $\Dki$ and $\eta_i$.     

Assuming that (\ref{shx_st_e1}) - (\ref{shx_st_e5}) hold, a regular
hypersurface $\Ti$ must satisfy both conditions (\ref{no_tb}) in order to avoid
the central singularity (big bang and big crunch). The term in square brackets
depending on $x$ in (\ref{tb_e}) is bounded between $0$ (as $x\to 0$) and
$\pi/2$ (as $x\to 2$), also, we have $c(t_i-\tb)\to 2/(3\sqrt{2\mu})$ as $\kappa\to
0$. On the other hand, the big crunch time is
$c\,t_{_{bc}}=c\,\tb+2\pi\mu/\kappa^{3/2}$. Therefore, conditions (\ref{no_tb}) can
only be violated if both $\kappa$ and $\mu$ diverge but complying with $0\leq
\kappa/\mu\leq 2$, a situation that can be avoided if both
$\mu$ and $\kappa$ or $\rhoi$ and $\Ri$ are finite.   As shown by figures
\ref{bangtimes}c, \ref{bangtimes}d and \ref{bangtimes}e, conditions (\ref{no_tb}) are satisfied
for reasonable initial value functions, even in the cases displayed in \ref{bangtimes}d and
\ref{bangtimes}e in which shell crossing occur.  In general,  a `good' selection of functions
$M,\,K=-E,\,\tb$ leads to `good' forms of $\Ri,\,\rhoi,\, Y_i$ and viceverza. 

As with parabolic and hyperbolic solutions, initial conditions with density
and/or 3-curvature voids lead, in general, to shell crossings but a regular
evolution with these type of initial conditions is possible in the domain $t\geq
t_i$. In order to illustrate this point, we consider $\Gamma$ given by 
(\ref{Gamma_gen}) and assume that $\rhoi\geq 0$ and $\Ri \geq 0$ are everywhere
finite (in order to comply with (\ref{no_tb})). Bearing in mind that:\,
$3P(Q_i-Q)\geq -2$ \, and \, $1-P/P_i\geq 0$ \, for \, $\eta\geq \eta_i$ \, (or
$t\geq t_i$), a sufficient condition for
$\Gamma\geq 0$ is given by
\be \Dki \ \leq \ \Dmi \ \leq \ \frac{1}{2} \ + \
\frac{3}{2}\Dki,\label{reg_voids_e}\ee
This condition is illustrated graphicaly by figure \ref{initfunc_test}c, leading to the model
displayed in figure \ref{reg_LTBvoids}b, an elliptic LTB model with an initial
density void that has a regular evolution for $t\geq t_i$. 

\subsection{Simmultaneous big bang.}\label{Gamma_etc_sbb}

An important special class of initial conditions is that characterized by a
simmultaneous big bang: ie $y=0$ taking place in a singular hypersurface marked
by $t=$ constant.  The conditions for a symmultaneous big bang  readily follow by
setting $\tb'=0$ in (\ref{tbr_p}), (\ref{tbr_h}) and (\ref{tbr_e}).  Clearly, a
symmultaneous big bang is incompatible with parabolic solutions, since $\tb'=0$ in
(\ref{tbr_p}) implies $\Dmi=0$, but this case is the FLRW limit (see section
\ref{limits}). From (\ref{tbr_h}) and (\ref{tbr_e}), a  symmultaneous big bang
for hyperbolic and elliptic solutions implies the vanishing of the term in square
brackets in the right hand sides of (\ref{Gamma_hyp1}) and (\ref{Gamma_ell}), that is 
\be A_i\,\Dmi \ = \ B_i\,\Dki,\label{shx_stb1}\ee
so that $\Gamma$ becomes
\be \Gamma \ = \ 1 \ + 3\,\left(\,A\,\Dmi\,-\,B\,\Dki\right)\label{Gamma_stb}\ee
The conditions for avoiding shell crossings depend on the forms of $A$ and $B$
given by (\ref{Gamma_hyp1}) and (\ref{Gamma_ell}). We assume that $\rhoi\geq 0$
and $\Ri\geq 0$ (elliptic) and $\Ri\leq 0$ (hyperbolic) are all finite and examine
the hyperbolic and elliptic cases below:

\bi
\item{} For hyperbolic solutions, both $A$ and $B$
are monotonously decreasing with
$0\leq A\leq 1/3$ and $-1/2\leq B\leq 0$, while $A\to 1/3,\,B\to 0$ as $\eta\to 0$ and
$A\to 0,\,B\to -1/2$ as $\eta\to \infty$. For the asymptotic limits, $\eta\approx
0$ and $\eta\to \infty$, we have respectively  $\Gamma\approx 1+\Dmi$ and
$\Gamma\approx 1+(3/2)\Dki$. Since $A_i$ and $B_i$ have also opposite signs,
these asymptotic limits, together with (\ref{shx_stb1}), lead to the following
necessary conditions for avoiding shell crossings
\be \Dmi \ \geq \ -1,\qquad \Dki \ \geq \ -\frac{2}{3},\qquad \Dmi\Dki \ \leq \
0,\label{shx_stb2}\ee   
The functions $\mu,\,\kappa,\,\Dmi,\,\Dki$ are related by  (\ref{tb_h}) and
(\ref{shx_stb1}), which can be expressed as
\be \Dki \ = \
\frac{x^{3/2}-\sqrt{2+x}\left[\,\sqrt{x}\,\sqrt{2+x}-\hbox{arccosh}\,(1+x)\right]}
{x^{3/2}-\frac{3}{2}\sqrt{2+x}\left[\,\sqrt{x}\,\sqrt{2+x}-\hbox{arccosh}\,(1+x)
\right]}\,\Dmi,\label{D_smbb_h}\ee
\be \frac{1}{\sqrt{|\kappa|}}\,\left[\left(1+\frac{2}{x}\right)^{1/2}
-\frac{\hbox{arccosh}\,(1+x)}{x}\right] \ = \ c\,(t_i\ - \
\tb^{(0)}),\label{simbb_h}\ee
where $\tb^{(0)}<t_i$ is the constant value of $\tb$ and $x=|\kappa|/\mu$.     
\item{}For elliptic solutions both $A,\,B$ are non-negative, with $A\to 1/3 $ and
$B\to 0$ as $\eta\to 0$, while $B\to (3/2)A\to\infty$ as $\eta\to 2\pi$. By looking
at the asymptotic behavior as $\eta\to 0$ and $\eta\to 2\pi$ and considering
(\ref{shx_stb1}), we obtain the necessary conditions
\be \Dmi \ \geq \ -1,\qquad \Dmi \ - \ \frac{3}{2}\Dki \ \geq \ 0, \qquad 
\Dmi\Dki \ \geq \ 0,\label{shx_stb3}\ee
As with the hyperbolic case, for $\tb=\tb^{(0)}<t_i$ a constant, and
$x=\kappa/\mu$, the functions $\mu,\,\kappa,\,\Dmi,\,\Dki$ follow from  (\ref{tb_e})
and (\ref{shx_stb1})
\be \Dki \ = \
\frac{x^{3/2}-\sqrt{2-x}\left[\,\arccos\,(1-x)-\sqrt{x}\,\sqrt{2-x}\right]}
{x^{3/2}-\frac{3}{2}\sqrt{2-x}\left[\,\arccos\,(1-x)-\sqrt{x}\,\sqrt{2-x}
\right]}\,\Dmi,\label{D_smbb_e}\ee 
\be \frac{1}{\sqrt{\kappa}}\,\left[\frac{\arccos\left(\,1\,-\,
x\,\right)}{x} \ - \ \left(\frac{2}{x}-1\right)^{1/2}
\right] \ = \ c\,(t_i\ - \ \tb^{(0)}),\label{simbb_e}\ee
\ei    
Since both sets of conditions (\ref{shx_stb2}) and (\ref{shx_stb3}) can be
obtained by applying (\ref{shx_stb1}) to the no-shell-crossing conditions of
the general cases with $\tb'\ne 0$ (ie (\ref{shx_st_h}), (\ref{shx_st_e4}) and
(\ref{shx_st_e5})), conditions (\ref{shx_stb2}) and (\ref{shx_stb3}) are necessary
and sufficient.   

\begin{figure}
%
%\ba\matrix{{\BoxedEPSF{susgar_12a.eps scaled 850}}&{\BoxedEPSF{susgar_12b.eps
%scaled 850}}\cr }\nonumber\ea
%
\ba\matrix{{\BoxedEPSF{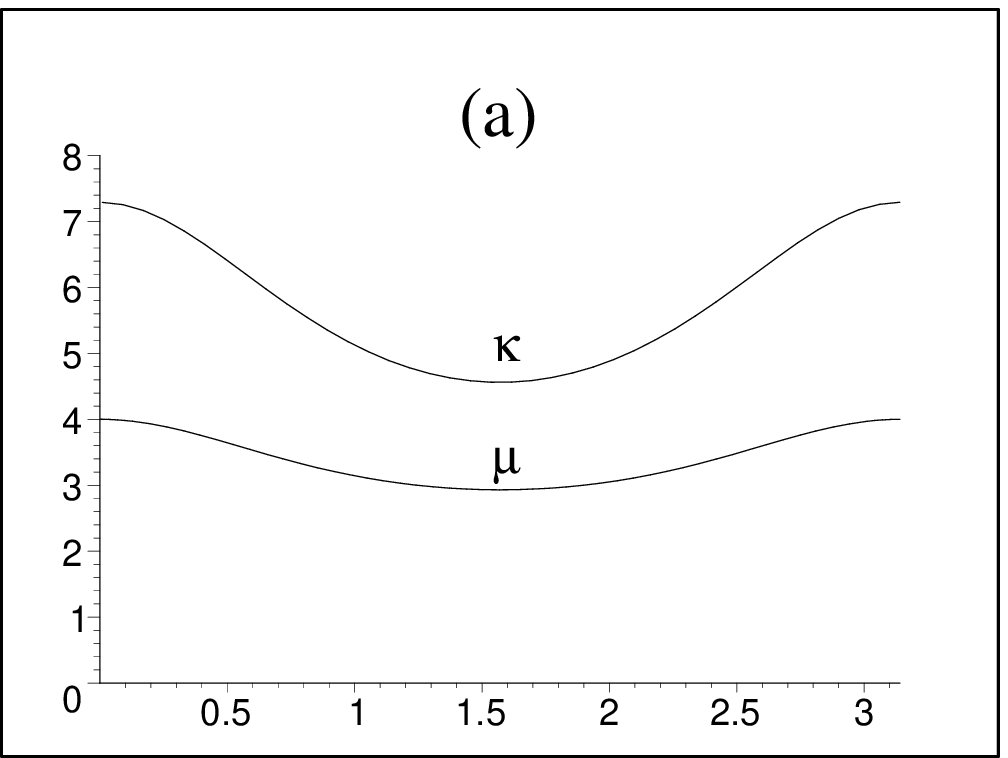 scaled
700}}&{\BoxedEPSF{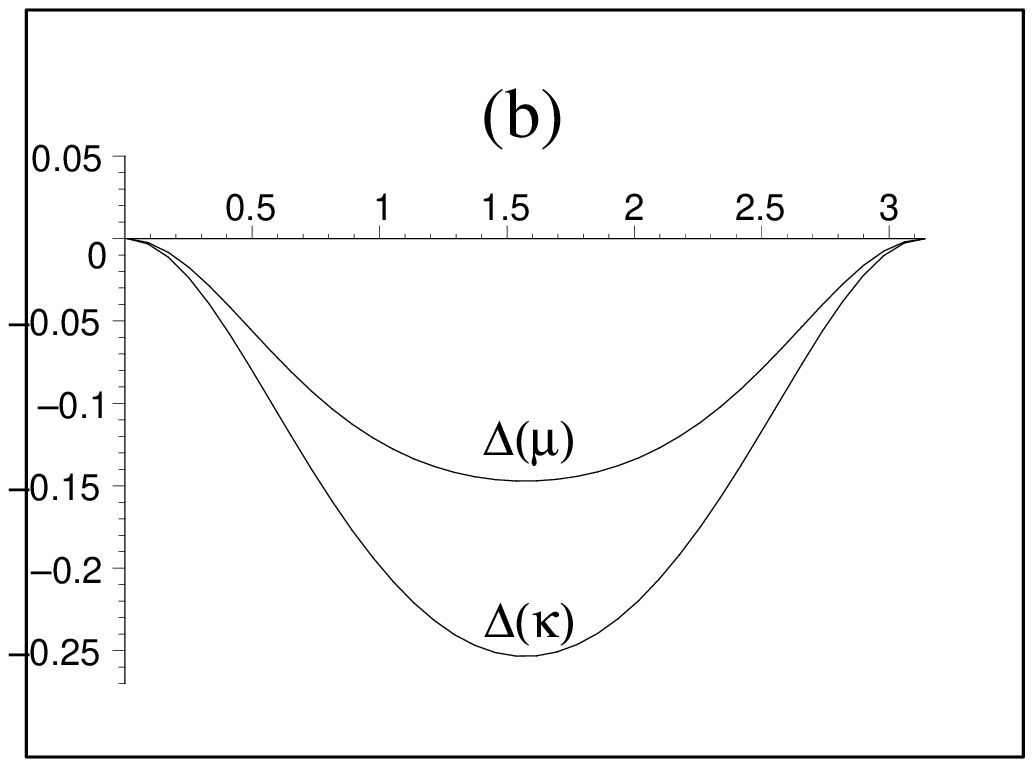 scaled 700}}\cr\cr {\BoxedEPSF{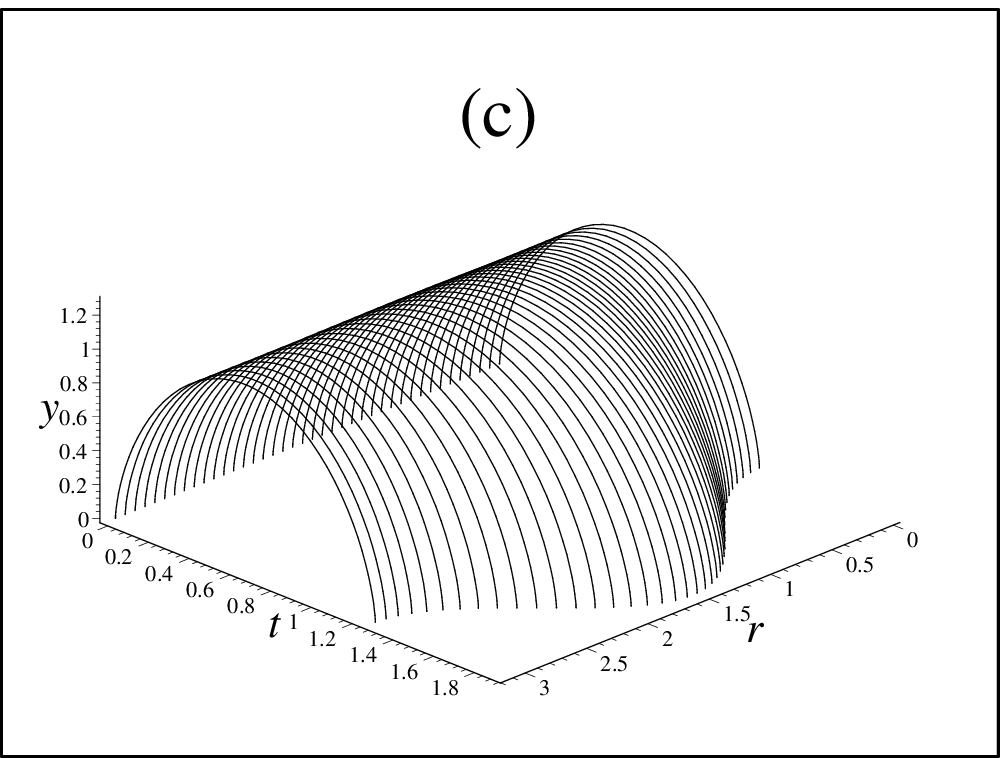 scaled
700}}&{\BoxedEPSF{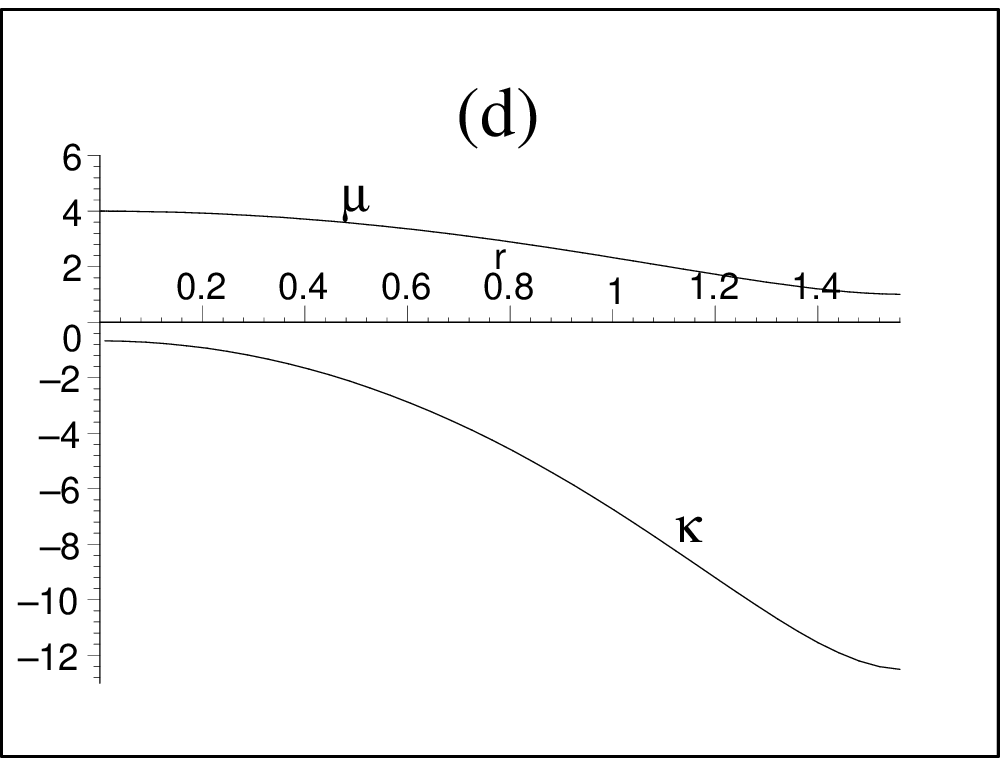 scaled 700}}\cr\cr {\BoxedEPSF{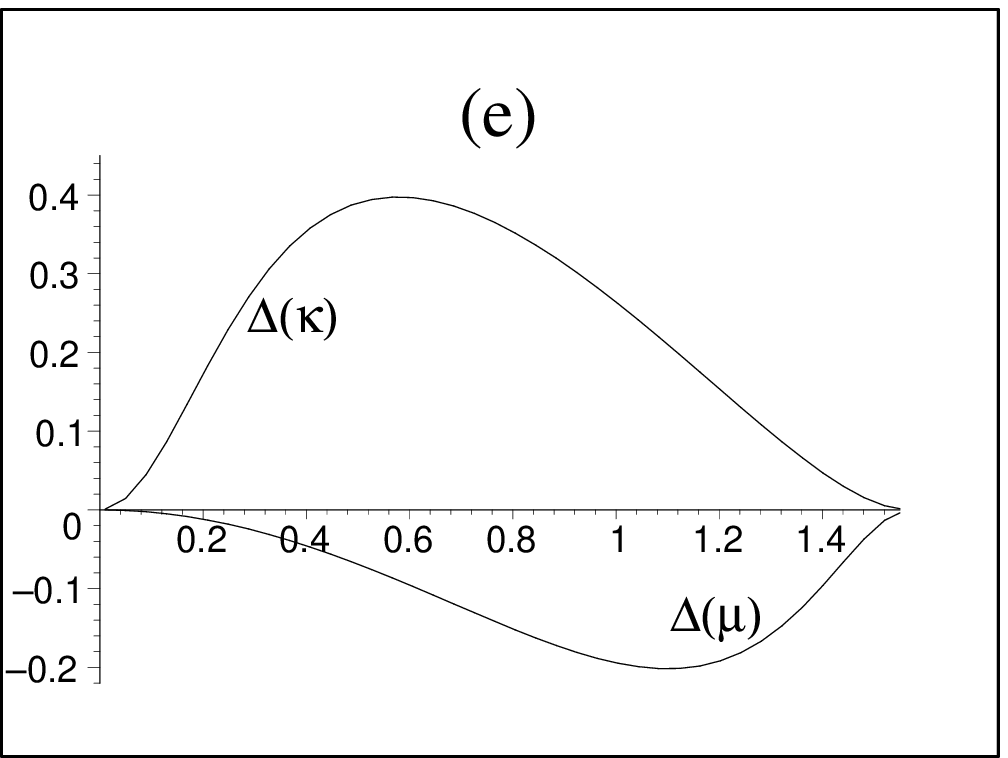
scaled 700}}&{\BoxedEPSF{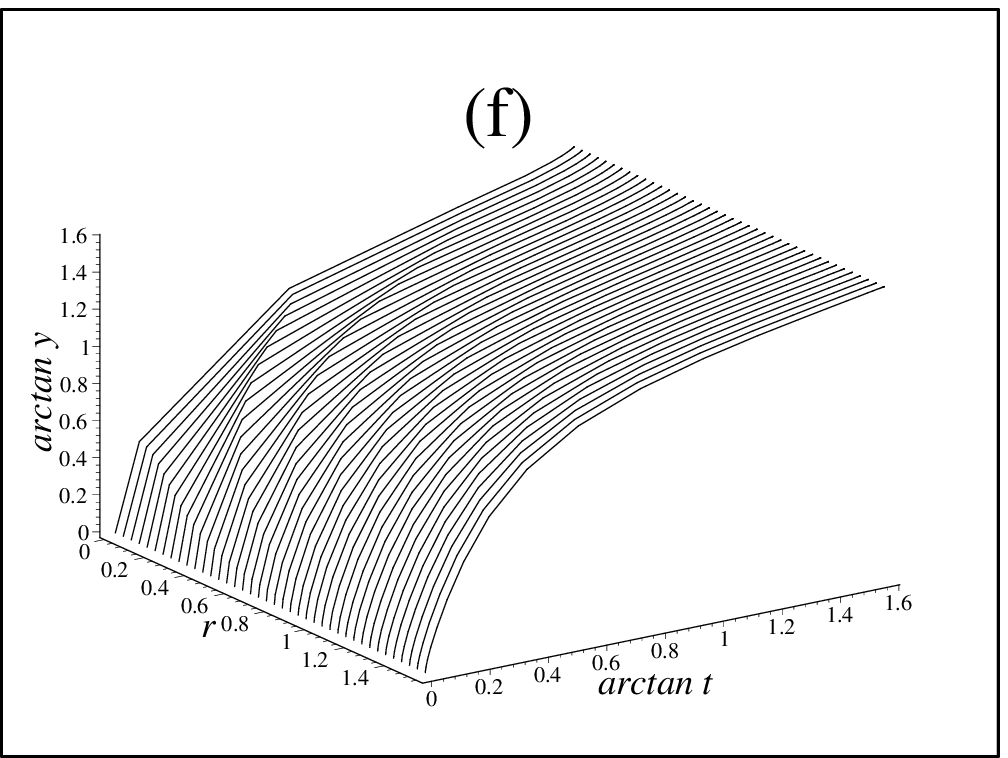 scaled 700}}\cr }\nonumber\ea
\caption{$\underline{\hbox{Models with a simmultaneous big bang}}$. We consider models with
elliptic (figures (a), (b) and (c)) and hyperbolic (figures (d), (e) and (f)) dynamics. We
select $\mu$ and $\Delta_i^{(m)}$ from equations (122) and (124) and find
$\kappa$ and $\Delta_i^{(k)}$ by solving numericaly equations (108), (109) for
the hyperbolic case and (111), (112) for the elliptic case. The functions $\mu$
and $\Delta_i^{(m)}$ are characterized by $a_1=1,\, a_2=4$ and $f=\sin\,r,\,\tan\,r$, respectively,
for the elliptic and hyperbolic case  (we can assume that $4\pi G\bar\rhoi/c^4=1$). For the
numerical solutions we have taken the constant bang time in (109) and
(112) to be $t_{_{bb}}^{(0)}=0.23$ (hyperbolic) and $t_{_{bb}}^{(0)}=0.4$ (elliptic), for an
initial hypersurface marked by $t=t_i=1$. The values of $t_{_{bb}}^{(0)}$ are not arbitrary and
depend on the choice of $t_i$,  $\mu$ and $\Delta_i^{(m)}$. Figures (a) and (b) display the obtained
$\kappa$ and $\Delta_i^{(k)}$ plotted next to the selected $\mu$ and $\Delta_i^{(m)}$ (both contrast
functions have the same sign, as required by (110)). Figure (c) displays the parametric
3-d plot $y(t,r)$ that corresponds to the full set of intial value functions. Notice that a
simmultaneous big bang does not imply (in general) a simmultaneous ``big crunch''. Figures (d) and
(e) display the corresponding initial value functions for the hyperbolic model (notice that the
contrast functions have opposite signs, as required by (107)). The  parametric 3-d plot
$y(t,r)$ for the hyperbolic case is shown in (e).}
\label{sbb_mods}
\end{figure}

A simmultaneous big bang greatly restricts the forms of $\mu$ and $\kappa$. Notice
that it is not possible to obtain a closed form $\kappa=\kappa(\mu)$
from (\ref{simbb_h}) or (\ref{simbb_e}). Given a choice of
$\rhoi$ leading to $\mu$ and $\Dmi$ from (\ref{defmu}), (\ref{defM}) and
(\ref{defDmi}), the constraints (\ref{simbb_h}) or (\ref{simbb_e}) must be solved
numericaly in order to determine a very specific and unique form of
$\kappa$ for a given $\tb^{(0)}$. Having obtained $\kappa$ numericaly, we can use it
to construct $x=\kappa/\mu$ so that $\Dki$ can be computed from
(\ref{D_smbb_h}) or (\ref{D_smbb_e}). The remaining functions, such as $K$ and $\Ri$
follow from (\ref{D_vs_vbls}). This numerical evaluation was done in ploting 
$\kappa$ and $\Dki$ in figures \ref{sbb_mods}. A simmultaneous big bang leads to a much simpler
expression for $\Gamma$, and correspondingly, to a significantly more lenient no-shell-crossing
conditions (\ref{shx_stb2}) and (\ref{shx_stb3}). Since (\ref{shx_stb2}) and (\ref{shx_stb3}) are
not incompatible with positive $\Dmi$ or $\Dki$, these conditions can accomodate,
besides lumps, some combinations of density and 3-curvature voids.
Unfortunately, this particular class of initial conditions has a complicated
description with the initial value functions that we have used. However,  a
simmultaneous big bang is a restrictive type of initial conditions, like for  example
to assume parabolic evolution ($\Ri=0$), or zero density or 3-curvature contrasts
($\Dmi=0$ or $\Dki=0$). We tend to conclude that the small possibility of a dust
configuration exactly attaining these very special sets of intial conditions makes
them somehow unrealistic. Graphical examples of initial value functions and the function
$y(t,r)$ for LTB models with a simmultaneous big bang and hyperbolic and elliptic dynamics are
displayed in figures \ref{sbb_mods}. The no-shell-crossing conditions obtained and discussed in
this section are summarized in Table 1.

\section{LTB models `a la carte'.}\label{ex_LTB}

The functions $Y_i,\,\rho_i$ and $\Ri$ play the same role as the
traditional variables $M,\,E,\,\tb$, that is, basic ``building blocks'' whose
specification fuly determines any given LTB model. It is then entirely equivalent
to work with the functions $M,\,E,\,\tb$ and the solutions of (\ref{freq1}) (as is
usualy done) or to use the initial value functions $Y_i,\,\rhoi,\,\Ri$ together
with the solutions of (\ref{freq2}). Given a choice of old variables $M,\,E=-K,\,\tb$, we can
find $\rho,\,\Ri,\,Y_i$ from equations (\ref{def_rhoi}), (\ref{def_Ri}), (\ref{tb_p}), (\ref{tb_h}) and
(\ref{tb_e}). For either set of variables it is always possible to use one of
the functions in order to fix the radial coordinate, leaving the other two functions as free
parameters. If the traditional variables are used in this process, then
$\rho$ and $\rhoi$ can be computed from (\ref{rho1}) and (\ref{def_rhoi}) once $Y$ and
$Y'$ are known from the integration of (\ref{freq1}). The alternative approach we
suggest is to fix the radial coordinate with $Y_i$ as in section \ref{Yi_etc}, so that
$\rhoi$ and $\Ri$ become the free parameters to be used for computing
$\mu,\,\kappa$ and the contrast functions, $\Dmi,\,\Dki$ by means of equations (\ref{defM}) to
(\ref{grad_k}). Next we can obtain $\Gamma$ defined
in (\ref{Gamma}) from the solutions of (\ref{freq2}) (see section \ref{Gamma_etc}), leading to
$\rho$ (from (\ref{rho2})) and the remaining geometric and kinematic quantities provided in the
Appendix: \,$\R,\,\sigma,\,\Theta$, etc. The old functions $M$ and
$E=-K$ can be determined from (\ref{defM}) and (\ref{defK}),
while $\tb$ can be obtained by setting $y=0$ in equations (\ref{parab_can2}) to
(\ref{hyp_def_eta}) (see equations (\ref{tb_p}), (\ref{tb_h}) and
(\ref{tb_e}) in section \ref{Gamma_etc}).

In the remaining of this section we provide a simple and practical, step-by-step, guideline on
how to construct a given LTB model in terms of mathematicaly simple ansatzes for the
new variables.\footnote{These ansatzes are (obviously) not the most
general initial conditions and are not based on any realistic observational data, we suggest
them simply for illustrating the utility of the new variables. The curves plotted in all figures
(except \ref{initfunc_other}a and \ref{initfunc_badf}) were obtained from expressions derived in
this section and generated with Maple.}. We also show how the
characteriztion of initial conditions as lumps or voids, as well as regularity conditions and
relevant quantities (like
$y,\,Y,\,\Gamma,\,\rho$, etc ), can all be illustrated graphicaly by plots that can be easily
generated by a standard Computer Algebra System (like Maple or Mathematica).\\ 

\noindent
\underline{$\hbox{Step (1). Prescribe the initial value functions.}$}
Consider the following simple ansatzes 
\be \rhoi \ = \ \bar\rhoi\,
\frac{a_2 \ + \ a_1\,f^2}{1 \ + \ f^2},\label{ex_rhoi}\ee
\be \Ri \ =  \bRi\,\frac{b_2 \ + \ b_1\,f^2}{1 \ + \ f^2},\label{ex_Ri}\ee
\be Y_i \ = \ S_i\,f,\label{ex_Yi}\ee
where 

\bi
\item{}$S_i$ is a constant characteristic length and $\bar \rhoi,\,\bRi$ are suitable constant 
reference values of $\rhoi,\,\Ri$, 
\ei

\bi
 \item{}$f=f(r)$ is a $C^0$ non-negative and adimensional function (see section
\ref{Yi_etc}) that satisfies: 
\bi
\item{} $f(r_c)=0$, where $r=r_c$ marks a SC, $f(r)>0$ for all $r\ne r_c$.
\item{} $f\to\infty$ as $r\to \rmax$, with $\rmax
$ defined by (\ref{ell2}),  
\item{} If there exist a ``turning value'' $r=r^*$ defined by $f'(r^*)=0$, all
radial derivatives of $\rhoi,\,\Ri,\,\mu,\,\kappa,\,K,\,M$ vanish at $r=r^*$.  
\ei
\ei
and the adimensional real constants $a_1,\,a_2,\,b_1,\,b_2$ determine the following generic
properties.   

\bi

\item{} Central and asymptotic values
 
The central values $\rhoi(r_c),\,\Ri(r_c) $ can be related to
$a_2,\,b_2$ by 
\be \rhoi(r_c) \ = \ a_2\,\bar\rhoi,\qquad \Ri(r_c) \ = \
b_2\,\bRi,\label{ex_cvals}\ee
while in the case when there is a maximal value $\rmax$, the
asymptotic values $\rhoi(\rmax),\,\Ri(\rmax)$
relate to $a_1,\,b_1$ by
\be \rhoi(\rmax) \ = \ a_1\,\bar\rhoi\,\qquad
\Ri(\rmax) \ = \ b_1\,\bRi,\label{ex_asvals}\ee
Hence, we can consider the constants $a_1,\,a_2,\,b_1,\,b_2$ as comparative adimensional ratios
of special values of $\rhoi$ and $\Ri$ with respect to the reference values $\bar\rhoi$
and $\bRi$, for example, values associated to a FLRW background (see further below). Numerical
values of these constants used in all figures follow this interpretation, hence the curves
display comparative values of the plotted quantities, say $\rhoi$, $\Ri$, etc,  with respect to
specific reference values (for example, with respect to central values if $a_1=b_1=1$).

\item{} Type of dynamics 
\bi
\item{} Parabolic solutions: \,\, $b_1=b_2=0$, \,\, so that
$\Ri=\Riav=\kappa=K=0$.
\item{} 
If $b_1,\,b_2$ are both positive, the sign of $\Ri$ is given by the sign of $\bRi$.
We have elliptic or hyperbolic dynamics if $\bRi\geq 0$ or $\bRi\leq 0$.  
\item{} Mixed dynamics: If $b_1$ and $b_2$ have different signs, $\Riav$ and $K$ might
pass from positive to negative or viceverza (see figure \ref{initfunc_other}b). 
\ei
\item{} Lumps and voids

Since $a_2/a_1$ and $b_2/b_1$ are the ratios of central to asymptotic values of
$\rhoi$ and $\Ri$, the characterization of (\ref{ex_rhoi}) and (\ref{ex_Ri})
as lumps or voids follows from these constants.  
\ba 0\ \leq \ a_1 \ < \ a_2, \quad\hbox{density lumps,}\qquad\qquad  0\ \leq \ a_2 \
< \ a_1, \quad\hbox{density voids,}\cr\cr 0\ \leq \ b_1 \ < \ b_2,
\quad\hbox{curvature lumps,}\qquad\qquad  0\ \leq \ b_2 \ < \ b_1,
\quad\hbox{curvature voids,}\label{ex_l&v}\ea
If $\rhoi(r_c)>0$ and $\Ri(r_c)\ne 0$, we can identify $\bar\rhoi =\rhoi(r_c)$ and
$\bRi=\Ri(r_c)$ by setting $a_2=b_2=1$. In this case, we have lumps if $a_1<1,\,b_1<1$
and voids if $a_1>1,\,b_1>1$. Of course, $\bar\rhoi$ and $\bRi$ can also be
identified with asymptotic values or with values associated with a FLRW background.  For
example, in a present day cosmological context ($t_i=t_0$, present cosmic time), we can
identify $\bar\rhoi=\bar\rho_0$ and $\bRi={}^{(3)}{\bar{\cal{ R}}}_0$ as constant values
associated with a FLRW background at $t=t_0$. If we set $a_1=b_1=1$, we can identify these
constant parameters in terms of the observational parameters
$\bar\Omega_0,\,\bar H_0$ of the FLRW background by
\ba \frac{8\pi G}{3c^4}\,\bar\rho_0 \ = \ \bar H_0^2\,\bar\Omega_0,\qquad 
\frac{1}{6}\,{}^{(3)}{\bar{\cal{ R}}}_0 \ = \ \bar H_0^2\,(\bar\Omega_0-1).\nonumber\ea
The values of $\rho_0$ and ${}^{(3)}{{\cal{ R}}}_0$ depend on $r$ (from (\ref{ex_rhoi}) and
(\ref{ex_Ri})), but    the other parameters $a_2,\,b_2$ provide the ratio of the central
and background values:
\ba \frac{8\pi G}{3c^4}\,\rho_0(r_c) \ = \ \frac{8\pi G}{3c^4}\,\bar\rho_0\,a_2 \ = \ a_2\,
\bar H_0^2\,\bar\Omega_0,\qquad 
\frac{1}{6}\,{}^{(3)}{{\cal{ R}}}_0(r_c) \ = \ \frac{1}{6}\,{}^{(3)}{\bar{\cal{ R}}}_0\,b_2 \
= \ b_2\,\bar H_0^2\,(\bar\Omega_0-1)\label{ex_Fvals}\ea
From this example, we clearly have a density lump/void if $a_2$ is greater/smaller than $1$
and a 3-curvature lump/void if $b_2$ is greater/smaller than $1$. In all figures
we can assume that the numerical values of $a_1,\,a_2$ and $b_1,\,b_2$ can be interpreted as
comparative adimensional ratios with respect to these FLRW reference values. If desired, units
can be selected so that $\bar\rhoi$ and $\bRi$, as well as terms like $4\pi G/c^4 $ can be
normalized to unity.\\  

\ei

\noindent
\underline{$\hbox{Step (2). Compute auxiliary variables:}$} $\mu,\,\kappa$, the volume averages
and contrast functions. We need to evaluate the integrals (\ref{defM}) and (\ref{defK}) for free
functions $\rhoi,\Ri,\,Y_i$ given by (\ref{ex_rhoi}), (\ref{ex_Ri}) and (\ref{ex_Yi}).
Let $\M$ and $\K$ be defined as   
\be \M \ = \ 3\,\bar\rhoi\,S_i^3\,\int{\frac{(a_2+a_1\,f^2)\,f^2\,f'\,dr}{1+f^2}} \ = \
a_1\,f^3 \ + 3\,(a_2-a_1)\,(f-\arctan\,f)\ee
\be \K \ = \ 3\,\bRi\,S_i^3\,\int{\frac{(b_2+b_1\,f^2)\,f^2\,f'\,dr}{1+f^2}} \ = \
b_1\,f^3 \ + 3\,(b_2-b_1)\,(f-\arctan\,f)\ee   
From (\ref{defmu}), (\ref{defkappa}) and (\ref{defM})-(\ref{defDki}) we obtain
\be M \ = \ \frac{4\pi G\,\bar\rhoi\,S_i^3}{3c^4}\,\M, \qquad \mu \ =
\ \frac{4\pi G \,\bar\rhoi}{3c^4}\,\frac{\M}{f^3} \ = \ \frac{4\pi
G}{3c^4}\,\rhoav,\label{ex_Mmu}\ee
\be K \ = \ \frac{\bRi\,S_i^2}{6\,f}\,\K, \qquad \kappa \ =
\ \frac{\bRi}{6}\,\frac{\K}{f^3} \ = \ \frac{1}{6}\,\Riav, \label{ex_Kkappa}\ee
\be \Dmi \ = \ 
\frac{(a_2-a_1)\,\left[f^3+(1+f^2)\,\left(f-\arctan\,f\right)\right]}{(1+f^2)\,\M},
\label{ex_Dmi}\ee
\be\Dki \ = \
\frac{(b_2-b_1)\,\left[f^3+(1+f^2)\,\left(f-\arctan\,f\right)\right]}{(1+f^2)\,\K},
\label{ex_Dki}\ee \\
Since $\mu$ is simply the average $\rhoav$ in units $\hbox{cm}^{-2}$, while $\kappa$ and $\Riav$
differ by a factor of $1/6$, all plots displaying $\rhoav$ and $\Riav$ are qualitatively
equivalent to plots displaying $\mu$ and $\kappa$. Numerical factors like $\bRi/6$ or $4\pi
G\bar\rhoi/c^4$ can always be absorbed into the numerical values of the constants
$a_1,\,a_2,\,b_1,\,b_2$.\\

\noindent
\underline{$\hbox{Step (3). Evaluate bang times.}$}

\begin{figure}
\ba\matrix{{\BoxedEPSF{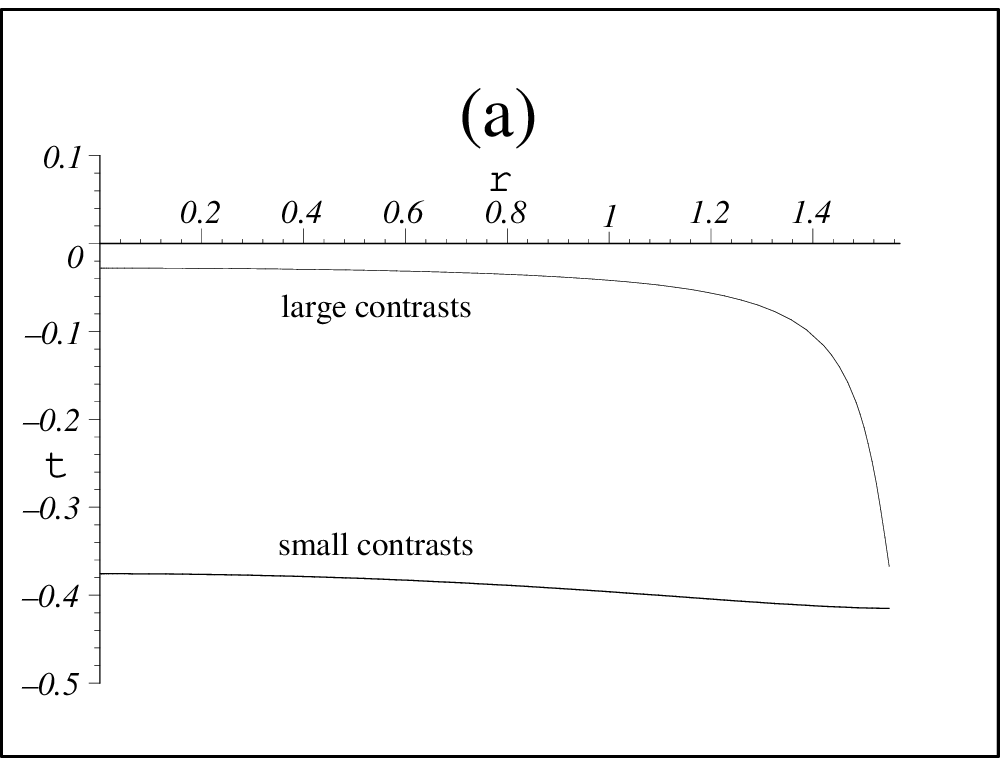 scaled 650}}&{\BoxedEPSF{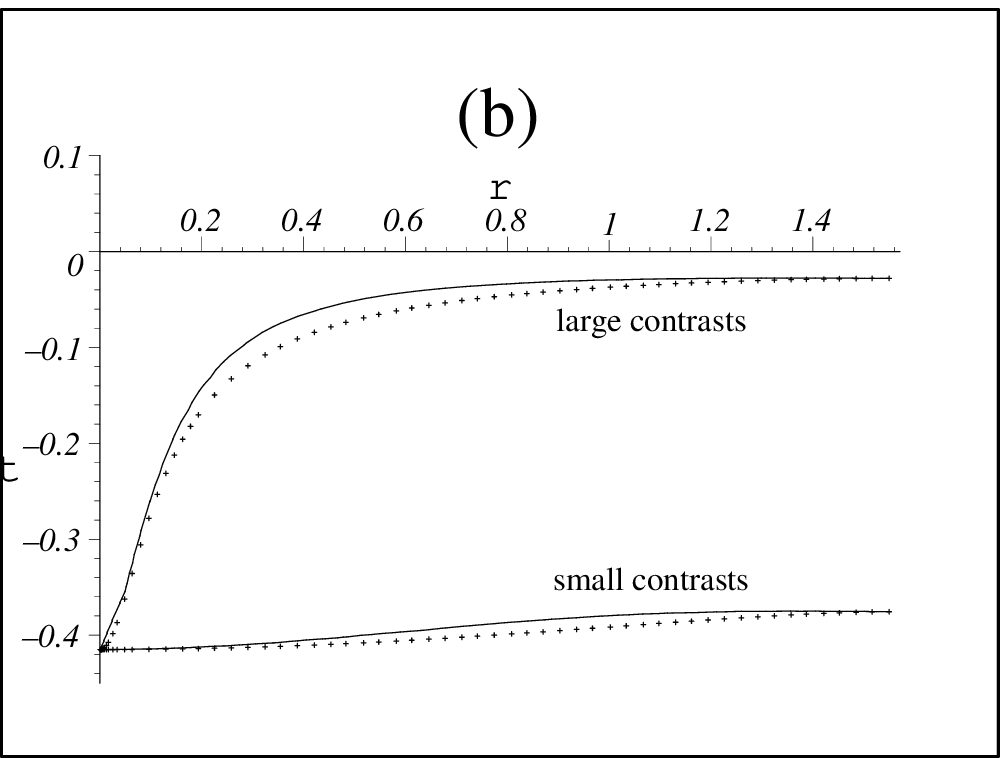
scaled 650}}\cr\cr {\BoxedEPSF{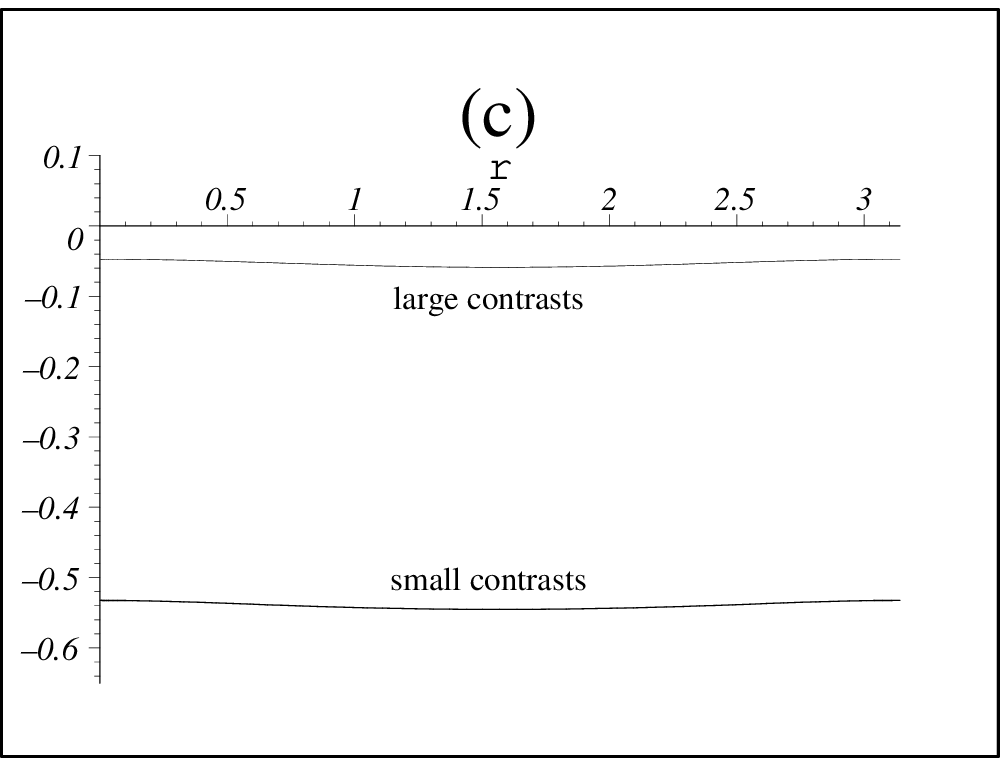 scaled
650}}&{\BoxedEPSF{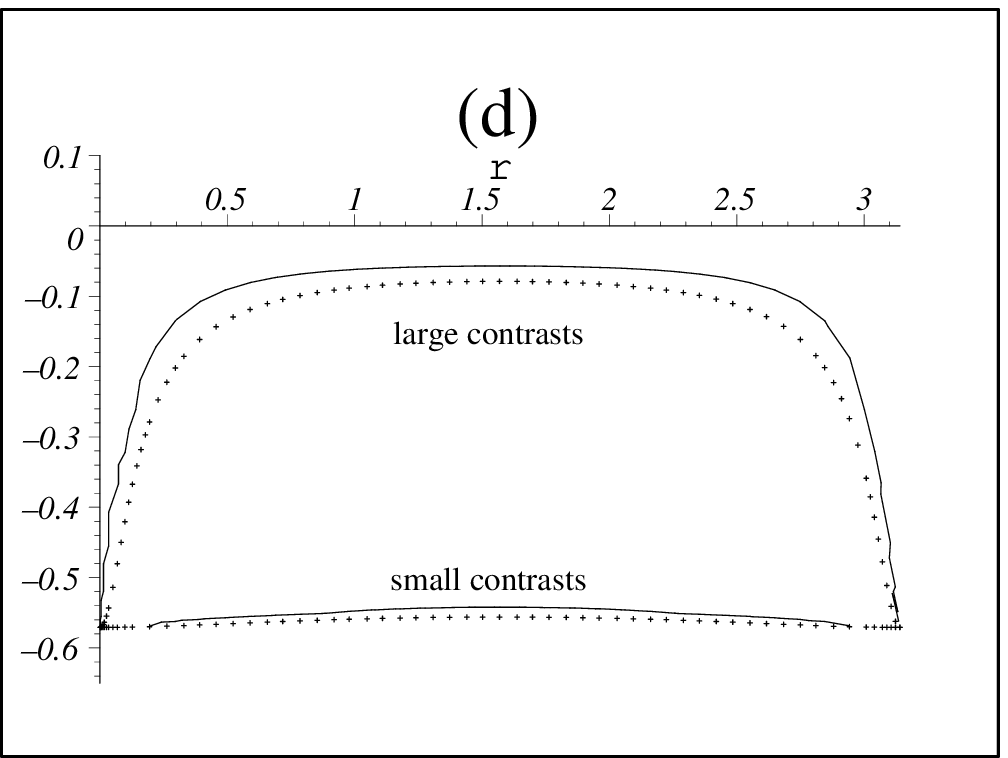 scaled 650}}\cr\cr {\BoxedEPSF{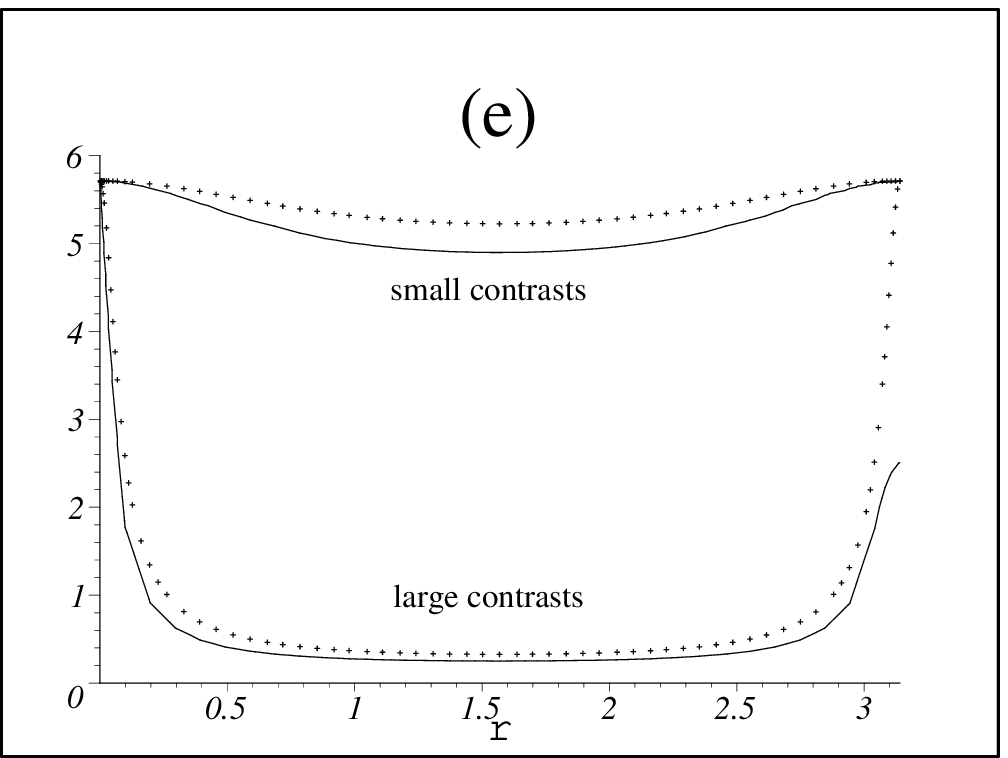
scaled 650}}&{}\cr }\nonumber\ea
\caption{$\underline{\hbox{Bang and shell crossing times.}}$  The figures display plots of
the bang time function $t_{_{bb}}(r)$, equivalent to displaying the locus of the central singularity
($y=0$) in the coordinate patch $(t,r)$. The locus of the shell crossing ($\Gamma=0$) singularity
is also displayed in those cases that exhibit this feature (solid curves in figures (b), (d) and
(e)). Figures (a) and (c) correspond to initial density and 3-curvature lumps with 
hyperbolic dynamics and one SC (figure (a)) and elliptic dynamics and two centers 
(figure (c)). Configurations with initial voids are shown in figures (b) (with
hyperbolic dynamics), (d) and (e) (elliptic dynamics, expanding and collapsing phases).  All the
plots were made with initial value functions obtained from section
VIII, with $f=\tan\,r$ and $f=\sin\,r$ for hyperbolic and elliptic dynamics. Lumps with
``small contrasts'' are characterized by $a_1=b_1=1,\,a_2=1.2,\,b_2=1.3$, while ``large
contrasts'' correspond to $a_1=b_1=1,\,a_2=200,\,b_2=300$. The same values characterize voids,
exchanging $a_1,\,b_1$ for $a_2,\,b_2$. Notice how configurations with small contrasts are
``older'', in the sense of larger time lapse between $t_{_{bb}}$ and $t_i=0$. Also, this ``age''
varies with $r$  due to the dependence of $t_{_{bb}}$ on $r$, this variation is larger for large
contrasts and hyperbolic dynamics (figures (a) and (b)), while for small contrasts and elliptic
dynamics it is almost insignificant. In figures (b), (d) and (e), the curve of $t_{_{bb}}$ is marked
by crosses while those of shell crossings are solid lines. Comoving observers ($r=$ const.) in
all these cases cannot avoid the shell crossings, but in the hyperbolic case (figure (b)) this
singularity is confined to the past of the initial hypersurface $t_i=0$. As shown by figures (d)
and (e), the shell crossing singularity might exhibit two branches for elliptic dynamics, unless
we choose initial conditions satisfying (104) as shown in figures
9c and 11b. The locii of singularities in cases with parabolic
dynamics is qualitatively analogous to those of hyperbolic dynamics shown in (a) and (b).}
\label{bangtimes}
\end{figure}

\bi
\item{} For parabolic dynamics, the bang time $\tb$ can be computed directly from inserting
$\mu$ given by (\ref{ex_Mmu}) into (\ref{tb_p}). For hyperbolic and elliptic dynamics with
$\tb'\ne 0$ the bang time, $\tb$, is furnished by (\ref{tb_h}) and (\ref{tb_e}) in terms of
$\kappa$ and $x$. The variable $x=\kappa/\mu$ (or $x=|\kappa|/\mu$ in hyperbolic dynamics)
follows directly from (\ref{ex_Mmu}) and (\ref{ex_Kkappa}) as
\ba x \ = \ \frac{c^4}{8\pi G}\,\frac{\bRi}{\bar\rhoi} \ \
\frac{\K}{\M},\label{ex_x}\ea
where we must consider $|\bRi|$ for the hyperbolic case. The bang time $\tb$ follows
directly from inserting $\kappa$ from (\ref{ex_Kkappa}) and $x$ from (\ref{ex_x}) into
(\ref{tb_h}) and (\ref{tb_e}). This has been done in the plots displayed in figures
\ref{bangtimes}. 

\item{} For the particular case with a simmultaneous big bang, the constraints
(\ref{simbb_h}) and (\ref{simbb_e}) cannot be satisfied by $\mu$ and
$\kappa$ both having the forms (\ref{ex_Mmu}) and (\ref{ex_Kkappa}). However, we can choose
either one of these functions and find the other one from (\ref{D_smbb_h}) to (\ref{simbb_e}).
We can choose $\rhoi,\,\mu,\,\Dmi$ from (\ref{ex_rhoi}), (\ref{ex_Mmu})
and (\ref{ex_Dmi}) and then find $\kappa=\mu\,x$ and $\Dki$ by numericaly solving two of
the equations (\ref{D_smbb_h}) to (\ref{simbb_e}). This is the way
the plots of figure \ref{sbb_mods} were done.

\ei

\noindent
\underline{$\hbox{Step (4). Verify fulfilment of regularity conditions}$}

\bi

\begin{figure}
\ba\matrix{{\BoxedEPSF{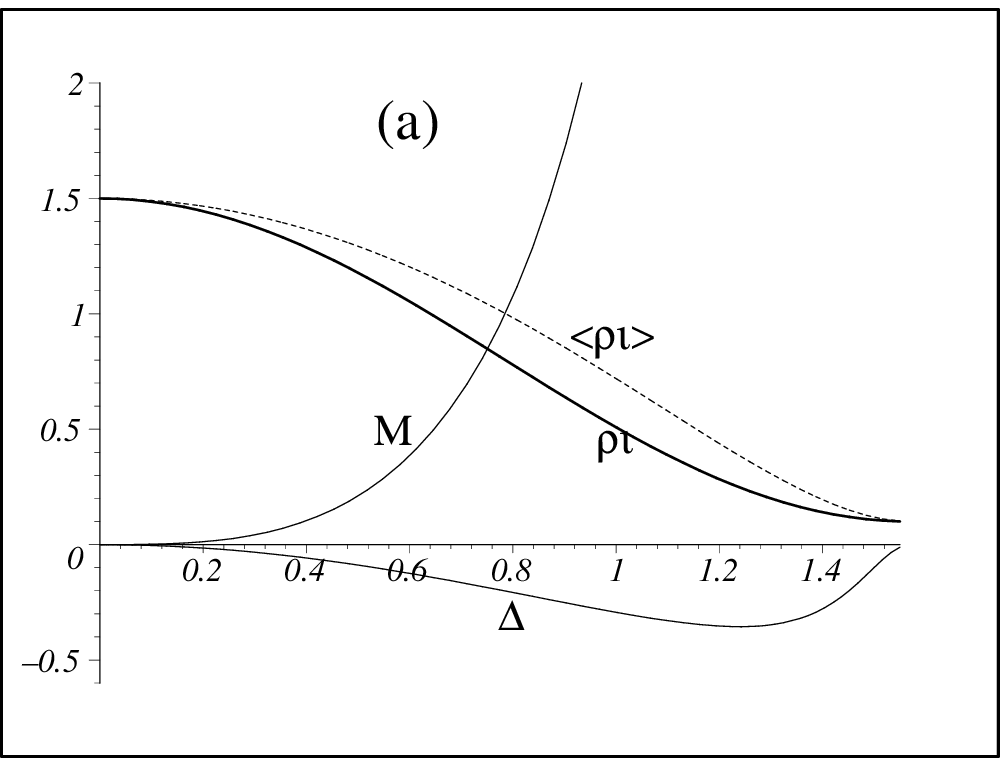 scaled 850}}&{\BoxedEPSF{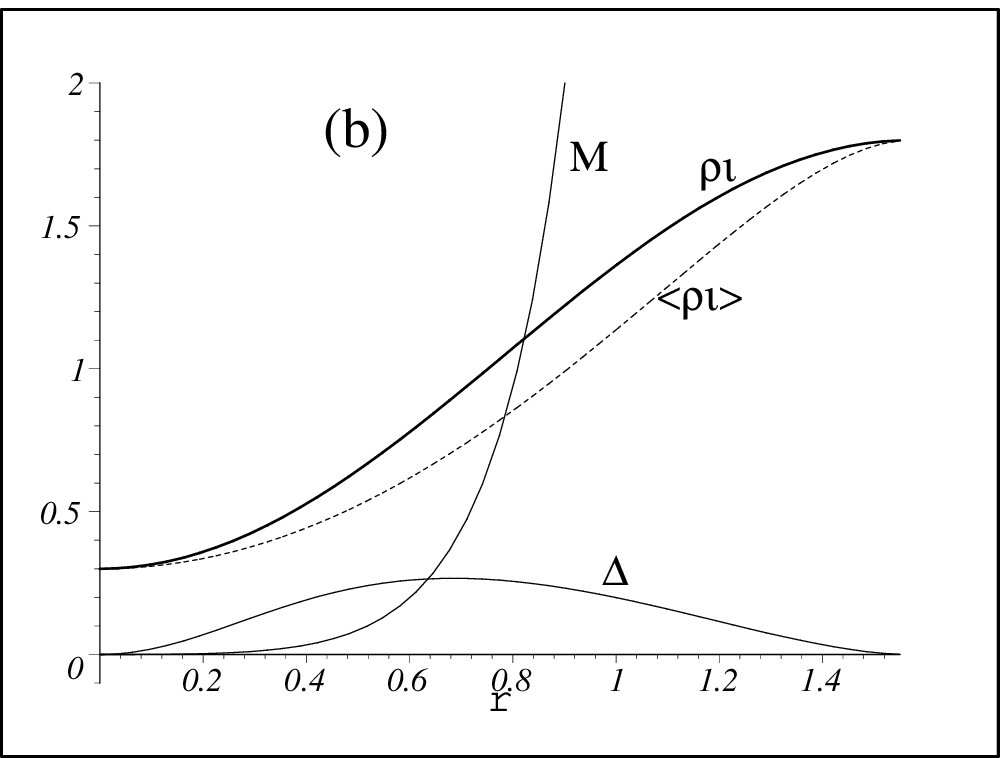
scaled 850}}\cr\cr {\BoxedEPSF{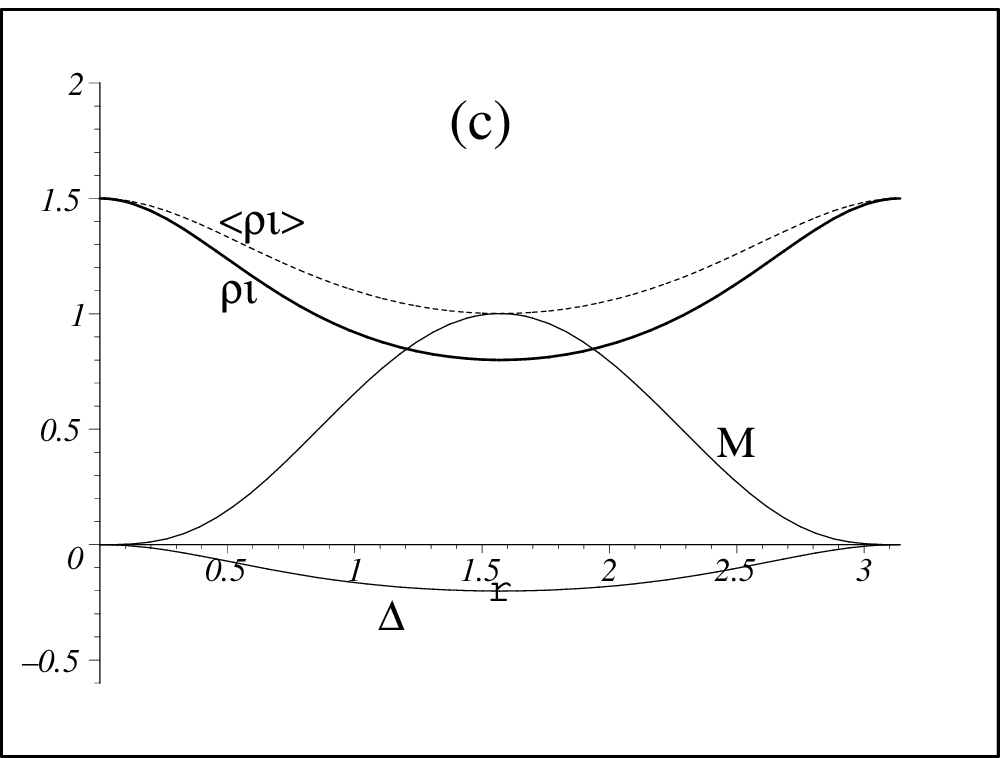 scaled
850}}&{\BoxedEPSF{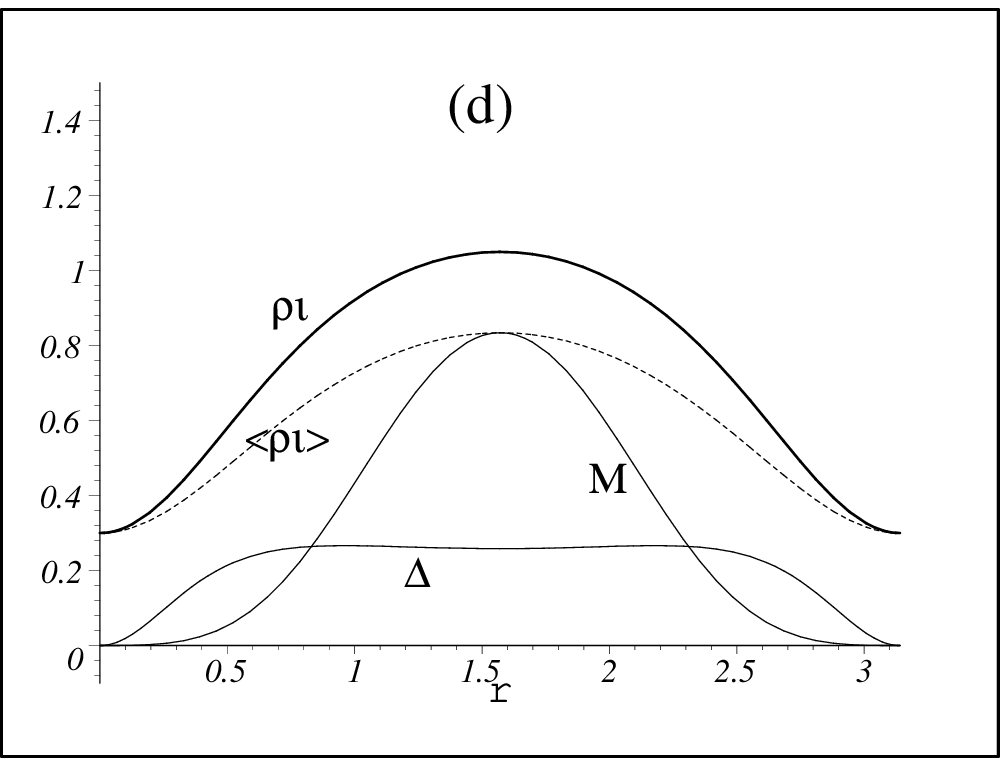 scaled 850}}\cr }\nonumber\ea
\caption{$\underline{\hbox{Initial density lumps and voids}}$. Initial value functions $\rho_i$
(thick curve), $\langle \rho_i \rangle$ (dotted curve), $M$ and $\Delta_i^{(m)}$ (marked by the
symbol $\Delta$) are depicted for a hypersurface ${\cal{T}}_i$ with one SC (figures (a) and (b)) and
two SC's (figures (c) and (d)). Figures (a) and (c) depict initial density lumps in which $\rho_i$ is
a local maximum at the SC's, while figures (b) and (d) correspond to voids in which $\rho_i$ is a
local minimum at the SC's. These functions were obtained from section VIII with parameters:
$[a_1=0.1,\,a_2=1.5]$ for the lumps ((a) and (c)) and $[a_1=1.8,\,a_2=0.3]$ for the
voids ((b) and (d)), with $f=\tan\,r$ in (a) and (b) and $\sin\,r $ in 
(c) and (d). Notice that $\langle \rho_i \rangle\geq \rho_i$ for lumps ((a) and (c)) and  
$\langle \rho_i \rangle\leq\rho_i$ for voids ((a) and (c)), hence $\Delta_i^{(m)}$ is negative for
lumps and positive for voids, clearly illustrating how the sign of this contrast function
characterizes the nature of the inhomogeneity in density along an initial hypersurface
${\cal{T}}_i$. In all cases 
$M$ vanishes at the centers and $M'=0$ at the turning value $r=r^*=\pi/2$ (in (c) and
(d)), showing the same qualitative behavior for lumps or voids. Since the type of dynamics
(parabolic, elliptic or hyperbolic) depends on $^{(3)}{\cal{R}}_i$ and not on $\rho_i$, the initial
density profiles in (a) and (b) can correspond to any dynamics (see figures 4a,
5a and 5b). Regular LTB models (without surface
layers) having the initial density profiles shown (c) and (d) must have the  3-curvature profiles
with two SC's  shown in figures 4b and 4c.}
\label{initfunc_dens}
\end{figure}

\begin{figure}
\ba\matrix{{\BoxedEPSF{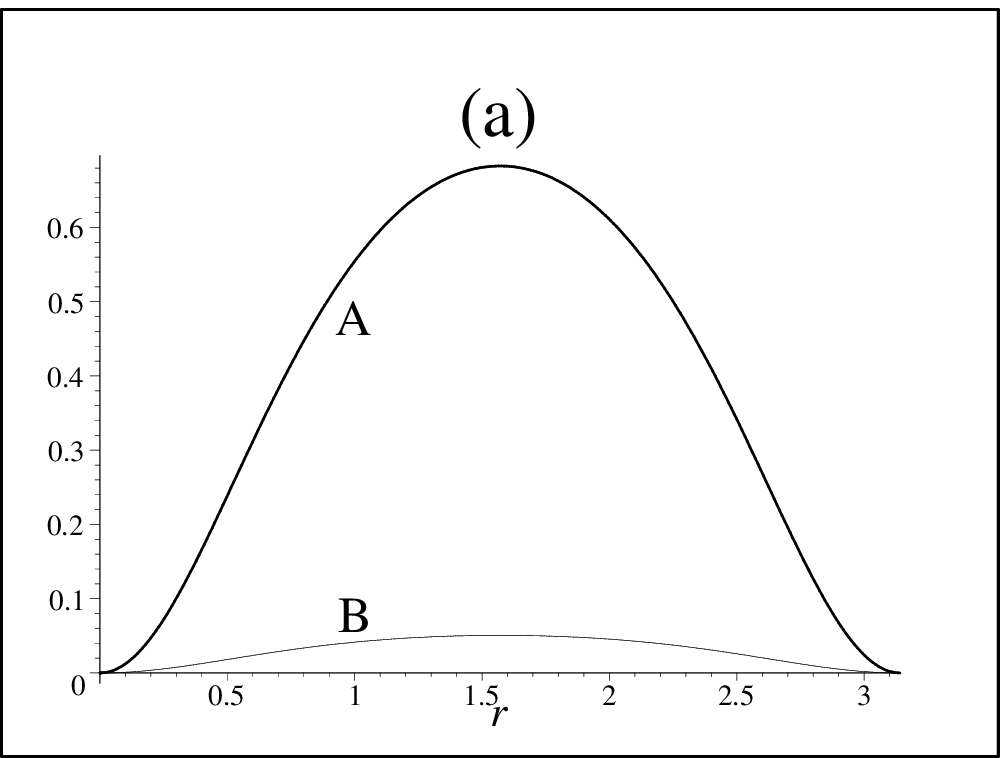 scaled 575}}&{\BoxedEPSF{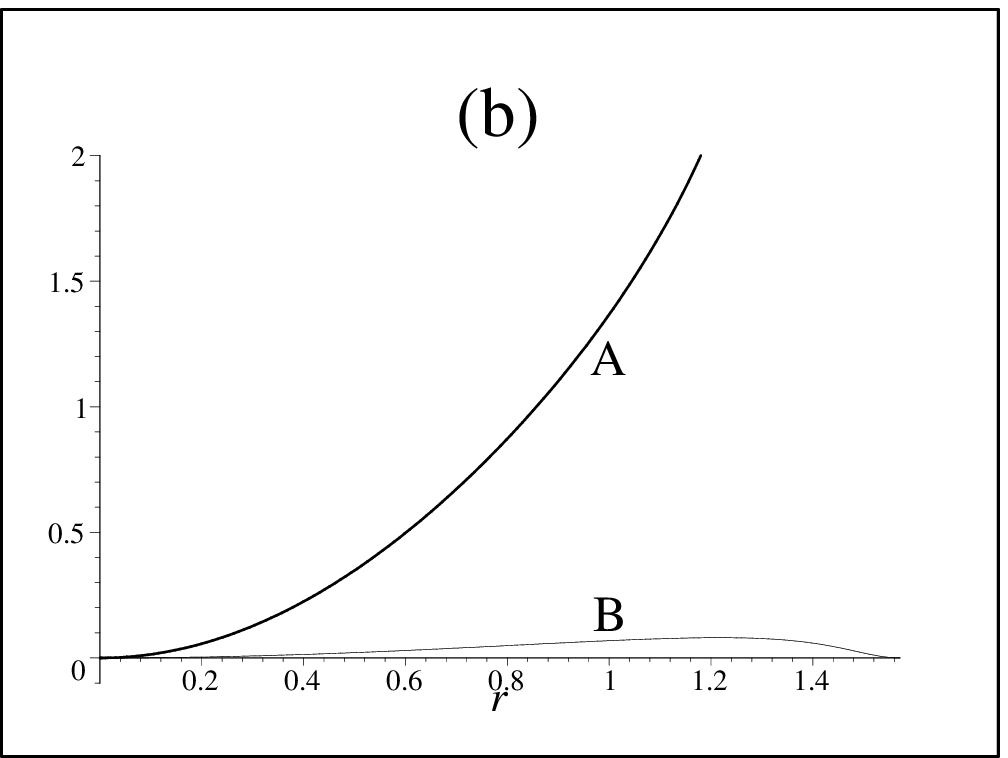
scaled 575}}&{\BoxedEPSF{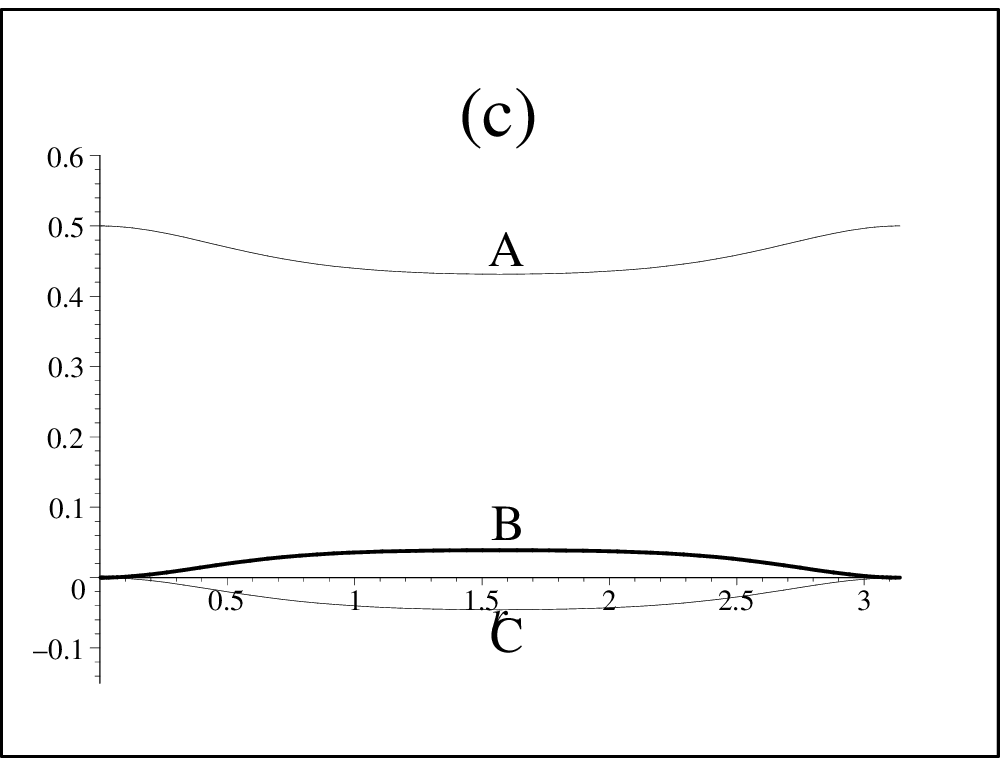
scaled 575}}\cr}\nonumber\ea
\caption{$\underline{\hbox{Testing initial conditions for regularity}}$.  A graphical
representation is provided in (a) and (b) for the no-shell-crossing condition of
elliptic dynamics (100). Figure (c) ilustrates the condition (104)
allowing for the possibility of initial voids with elliptic dynamics evolving free from shell
crossings for $t\geq t_i$. This type of graphical illustration is not necessary for parabolic
and hyperbolic dynamics, since the no-shell-crossing conditions in these cases (equations
(82) and (90), see also Table 1) only depend on the signs of $\rho_i$,
$^{(3)}{\cal{R}}_i$ and the contrast functions and can be verified simply by plotting these
functions (as in figures 3, 4 and 5). However, for elliptic dynamics the verification of regularity
conditions like (100)  and (104) is not trivial. In figures (a) and (b) we
have density and 3-curvature lumps characterized by the expressions of section VIII. The
corresponding parameters are:  $[a_1=b_1=1,\,a_2=15,\,b_2=20,\,f=\sin\,r]$ and two SC's in figure
(a) and $[a_1=1,\,b_1=0,\,a_2=15,\,b_2=20,\,f=\tan\,r]$ and one SC in figure (b). The
curves marked by ``A'' and ``B'' respectively represent the terms\, $2\pi
P_i[\Delta_i^{(m)}-(3/2)\Delta_i^{(k)}]$\, and\,
$P_iQ_i[\Delta_i^{(m)}-(3/2)\Delta_i^{(k)}]-[\Delta_i^{(m)}-\Delta_i^{(k)}]$ appearing in
(100). For this regularity condition to hold we need to verify that, for all $r$, the
curve ``A'' must be larger than that marked by ``B'' and both curves have to be larger than zero. As
shown in (a) and (b) this is what happens for the selected initial value functions (these initial
conditions are then used in the elliptic models depicted in figures 10c,
10d, 10e and 10f).  In figure (c), the curves marked by ``A'', ``B'' and ``C'' respectively
denote the terms $1/2+(3/2)\Delta_i^{(k)}$,\,\,$\Delta_i^{(m)}$ and $\Delta_i^{(k)}$ appearing in
(104), calculated for initial value functions characterized by $f=\sin\,r$,\,
$a_1=1.3,\,a_2=1$ (density void) and $b_1=1,\,b_2=1.4$ (3-curvature lump). For (\ref{reg_voids_e})
to hold, we need (for all $r$) the curve ``C'' to be larger than ``B'' and ``B'' to be larger than
``C'', as it happens in the figure. These intial value functions are used in the elliptic model
depicted in figure 11b.  In general, one might choose the parameters
$a_1,\,a_2,\,f$ characterizing $\rho_i,\,\langle \rho_i \rangle$ and $\Delta_i^{(m)}$ from section
VIII, then we can always find, by trial and error, the appropriate parameters $b_1$ and
$b_2$ that comply with (100) and (104). }
\label{initfunc_test}
\end{figure}

\begin{figure}
%
%\ba\matrix{{\BoxedEPSF{susgar_10a.eps scaled 850}}&{\BoxedEPSF{susgar_10b.eps
%scaled 850}}\cr }\nonumber\ea

\ba\matrix{{\BoxedEPSF{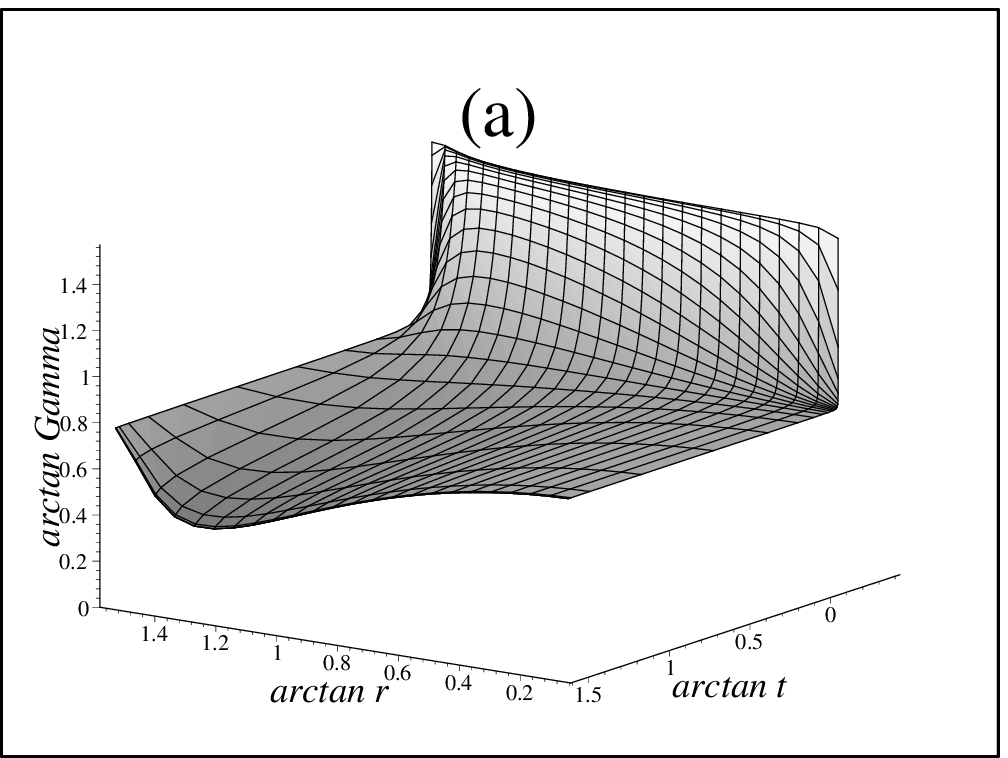 scaled
850}}&{\BoxedEPSF{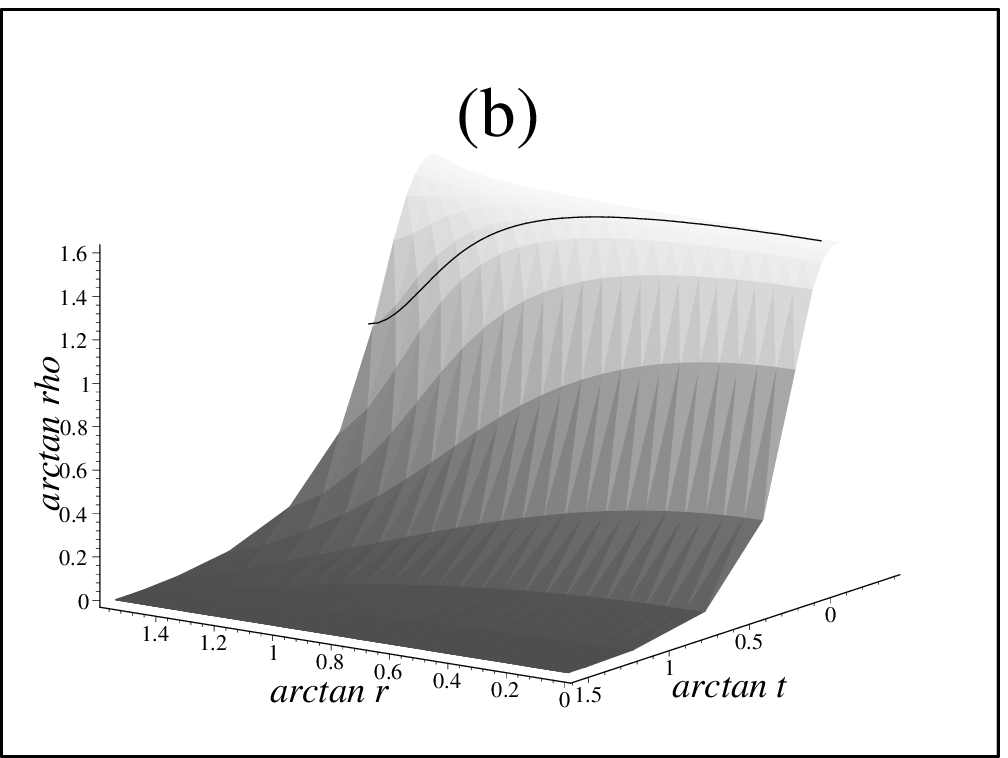 scaled 850}}\cr\cr {\BoxedEPSF{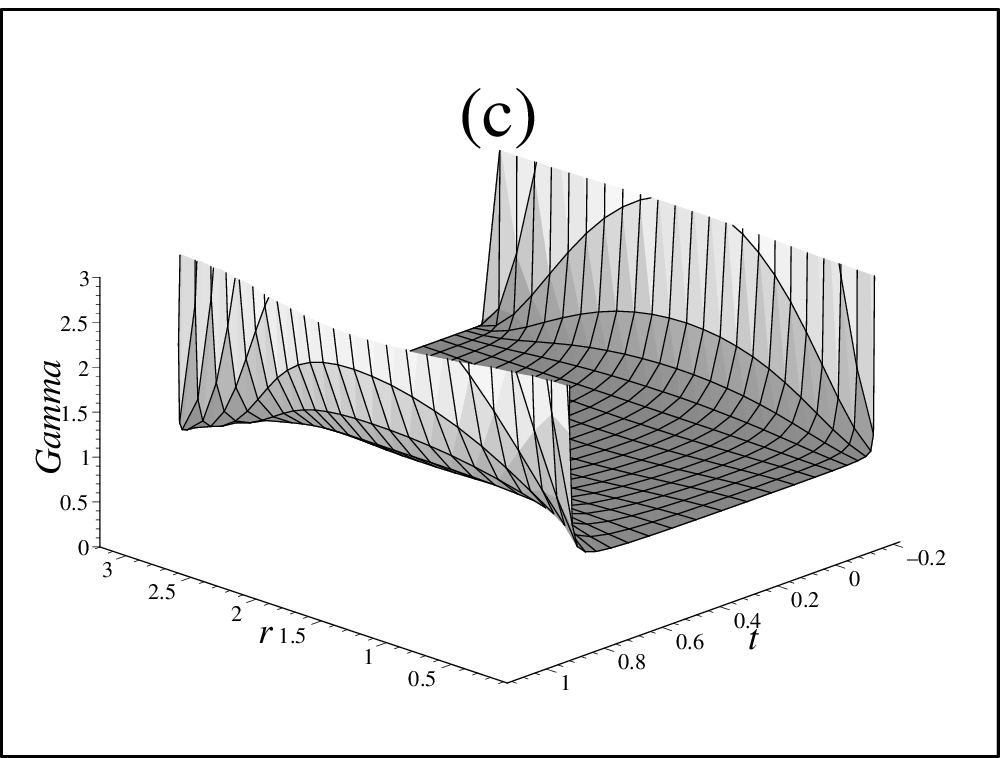 scaled
850}}&{\BoxedEPSF{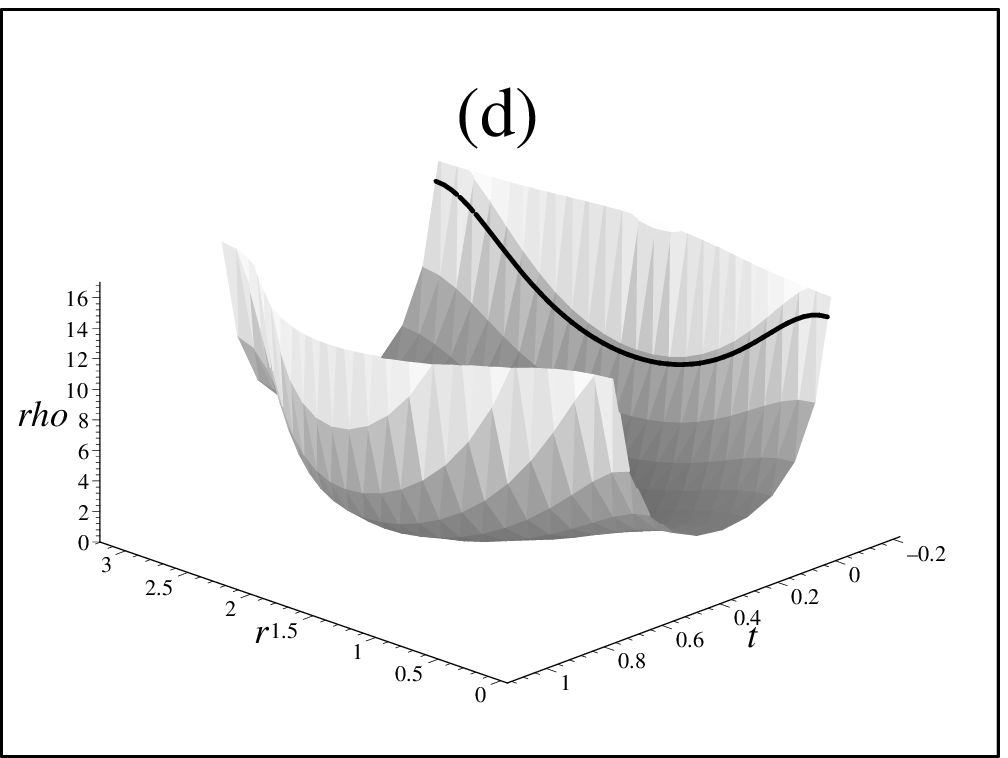 scaled 850}}\cr\cr {\BoxedEPSF{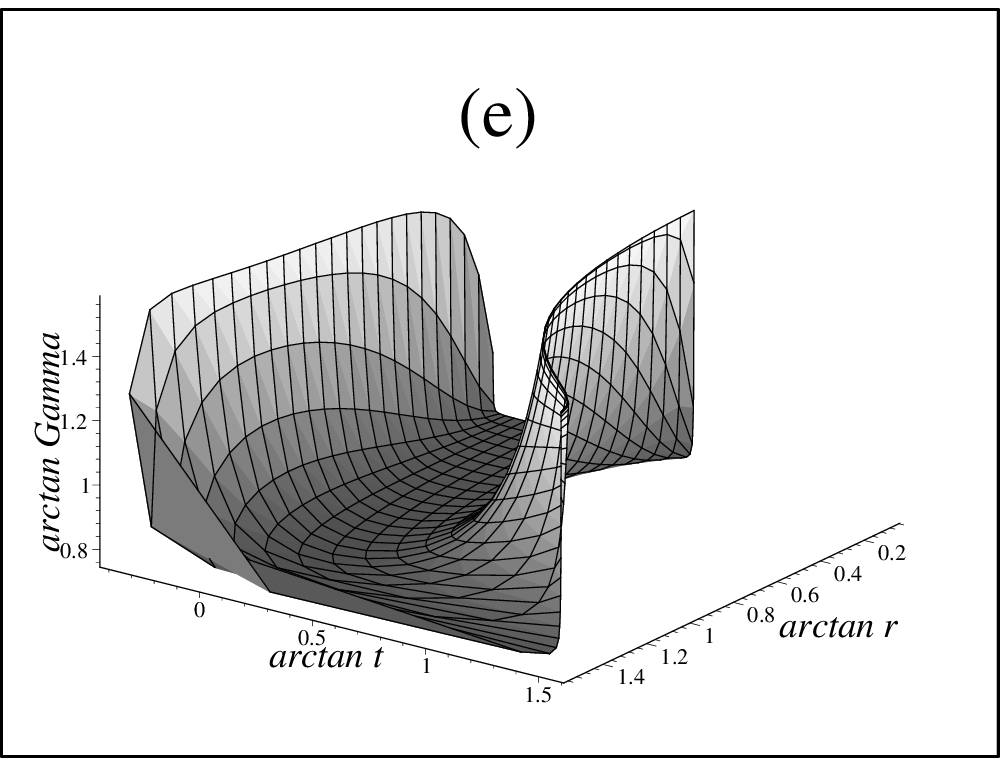
scaled 850}}&{\BoxedEPSF{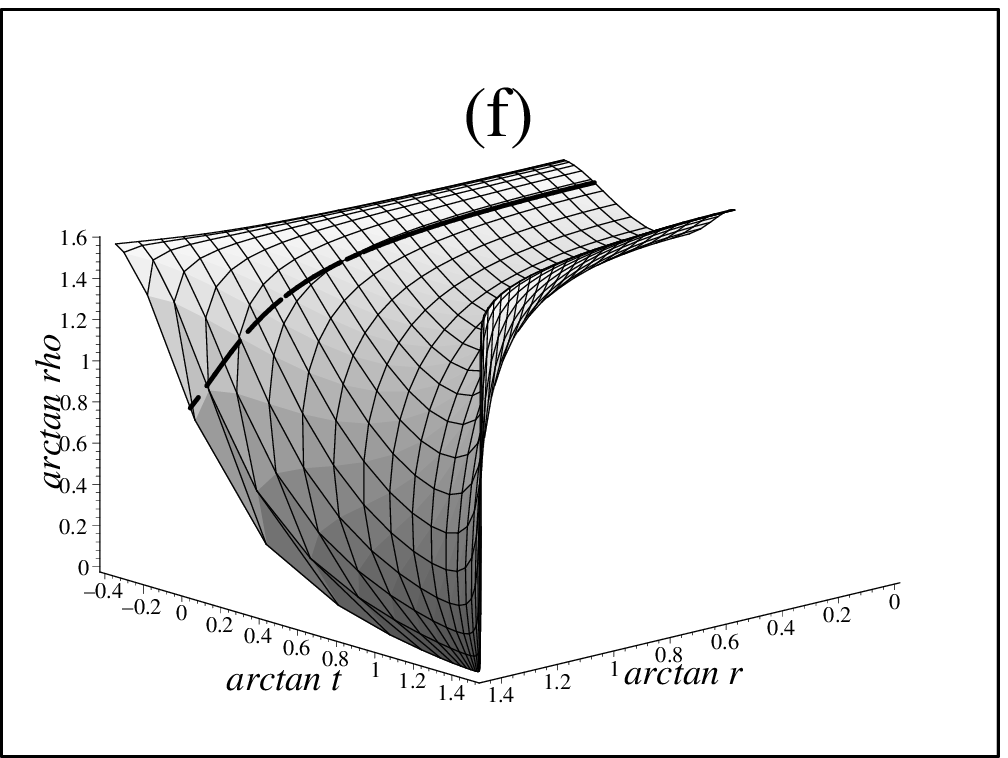 scaled 850}}\cr }\nonumber\ea
\caption{$\underline{\hbox{LTB models free from shell crossings}}$. The function $\Gamma$
(figures (a), (c) and (e)) and the rest mass density $\rho$ (figures (b), (d) and (f)) are
depicted as functions of $(t,r)$ by parametric 3-d plots like in (130). The initial
value functions have been selected so that shell-crossings are absent for all the evolution of
the models (see Table 1). Figures (a) and (b) correspond to the hyperbolic model characterized by
$a_1=b_1=1,\,a_2=15,\,b_2=20, \,f=\tan\,r$ (initial density and 3-curvature lumps). Notice in (b)
the initial density profile (thick curve in the plotted surface) associated with the initial
hypersurface $t=t_i=0$. Regular models with parabolic dynamics have the same qualitative
features as hyperbolic models. The intial value functions associated with the elliptic models in
(c), (d), (e) and (f) are those tested in figures 9a and 9b.
The models in (c) and (d) have two SC's (hence $f=\sin\,r$), while those of (e) and (f) have one
SC ($f=\tan\,r$ as those in figures 2e and 2f). Notice in (d) and (f) the
initial density profile $\rho_i$ as the thick curve marked by restrcting the 3-d plot to the
initial hypersurface $t=t_i=0$. }
\label{reg_LTB}
\end{figure}

\item{}The no-shell-crossing conditions (\ref{shx_st_p}), (\ref{shx_st_h})
and (\ref{shx_st_e4}), involving only $\mu$, $\Dmi$ and $\Dki$, can be
tested and illustrated directly with (\ref{ex_Mmu})-(\ref{ex_Dki}) (see
Table 1 and figures \ref{initfunc_dens}, \ref{initfunc_pcurv} and \ref{initfunc_ncurv}). However,
there is no simple prescription on how to choose $a_1,a_2,b_1,b_2$ in order to comply with the
no-shell-crossing condition (\ref{shx_st_e5}) of elliptic solutions. In this case, it is
convenient to express the functions $P_i,\,Q_i$ in terms of $x=\kappa/ \mu$ by means of
(\ref{etax_h}), (\ref{funcs_PQ_e}) and then in terms of $a_1,\,a_2,\,b_1,\,b_2$
and $f$ by (\ref{ex_x}).  As shown by figures \ref{initfunc_test}a, \ref{initfunc_test}b and
\ref{reg_LTB}c-e, it is not difficult to select the appropriate values of these parameters in
order to comply with this regularity condition.
  
\item{}Regularity conditions, such as (\ref{cond_K2}) and (\ref{cond_K_3}), take the
form

 \be{\K \over f} \ = \ b_1f^2 \ + \ 3\,(b_2-b_1)\,\left( {1-{{\arctan \,f}  \over f}}
\right) \ \leq \ {6 \over {\bRi\,S_i^2}},
 \label{ex_cond_K23}.
\ee
where the equality holds for every turning value such that $f'=0$. Condition (\ref{ex_cond_K23})
places strong restrictions on the values of $\bRi$ and $\Ri$ in models with elliptic dynamics. 
\ei

\begin{figure}

%\ba\matrix{{\BoxedEPSF{susgar_2a.eps scaled 750}}&{\BoxedEPSF{susgar_2b.eps
%scaled 750}}\cr }\nonumber\ea
%
\ba\matrix{{\BoxedEPSF{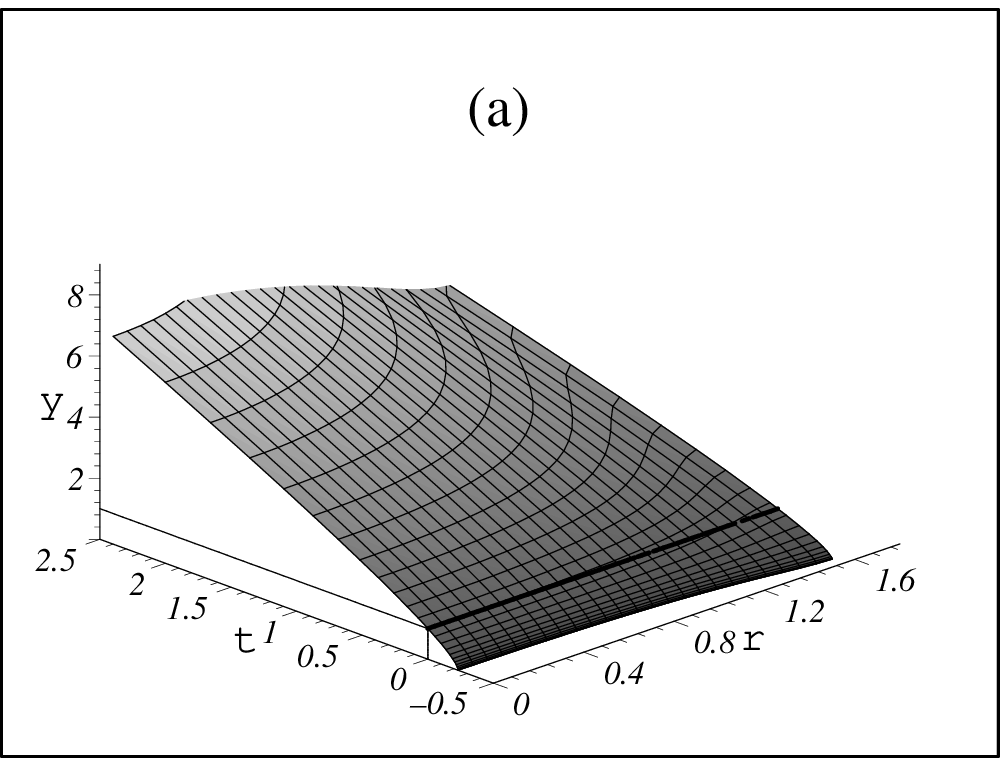 scaled 700}}&{\BoxedEPSF{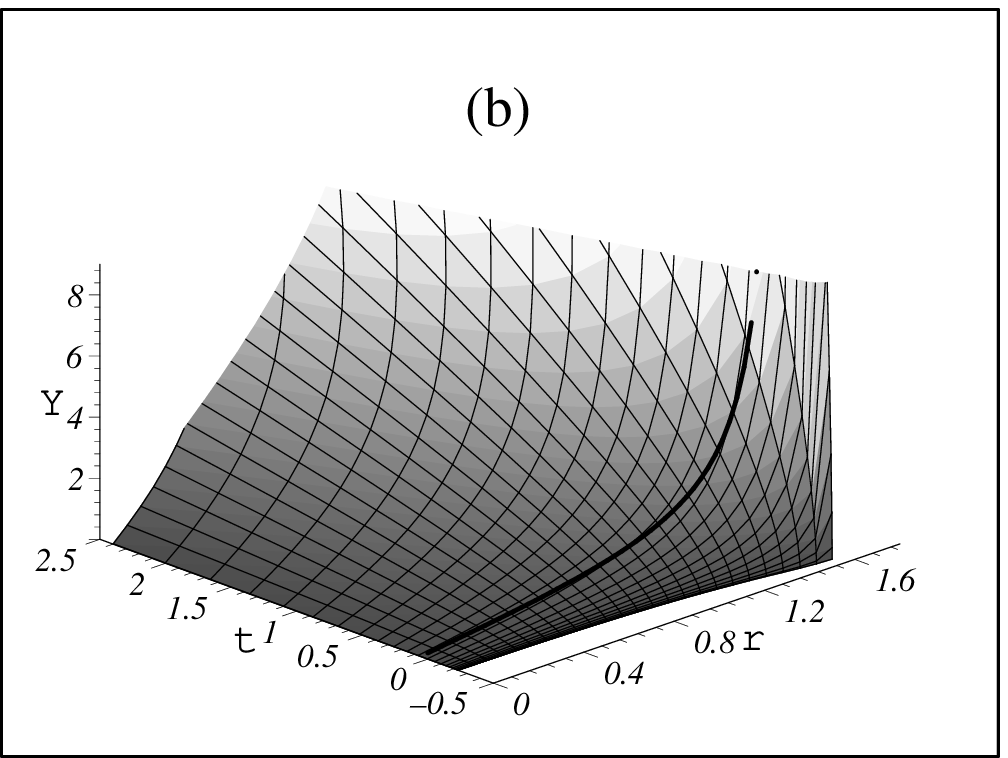
scaled 700}}\cr\cr {\BoxedEPSF{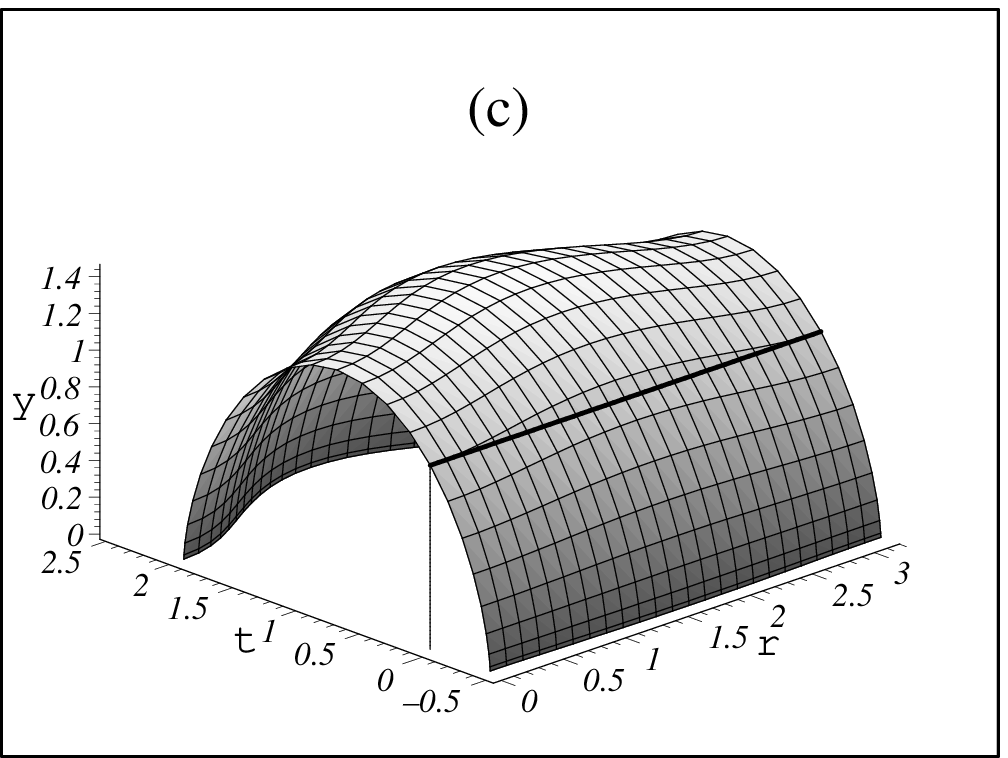 scaled
700}}&{\BoxedEPSF{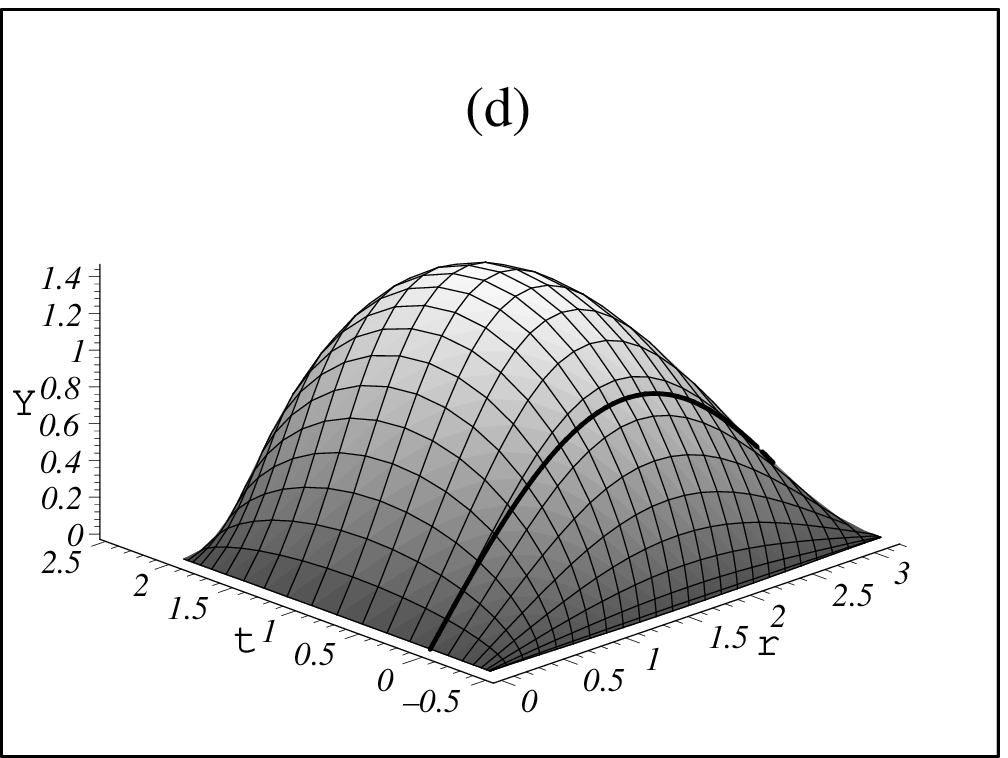 scaled 700}}\cr\cr {\BoxedEPSF{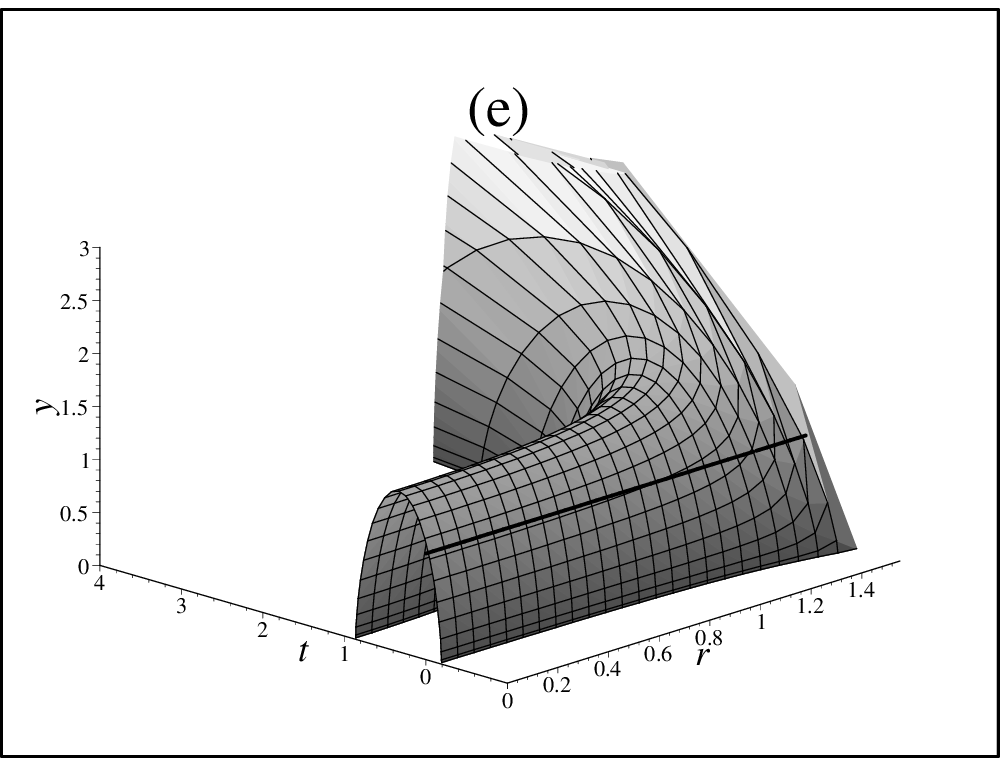
scaled 700}}&{\BoxedEPSF{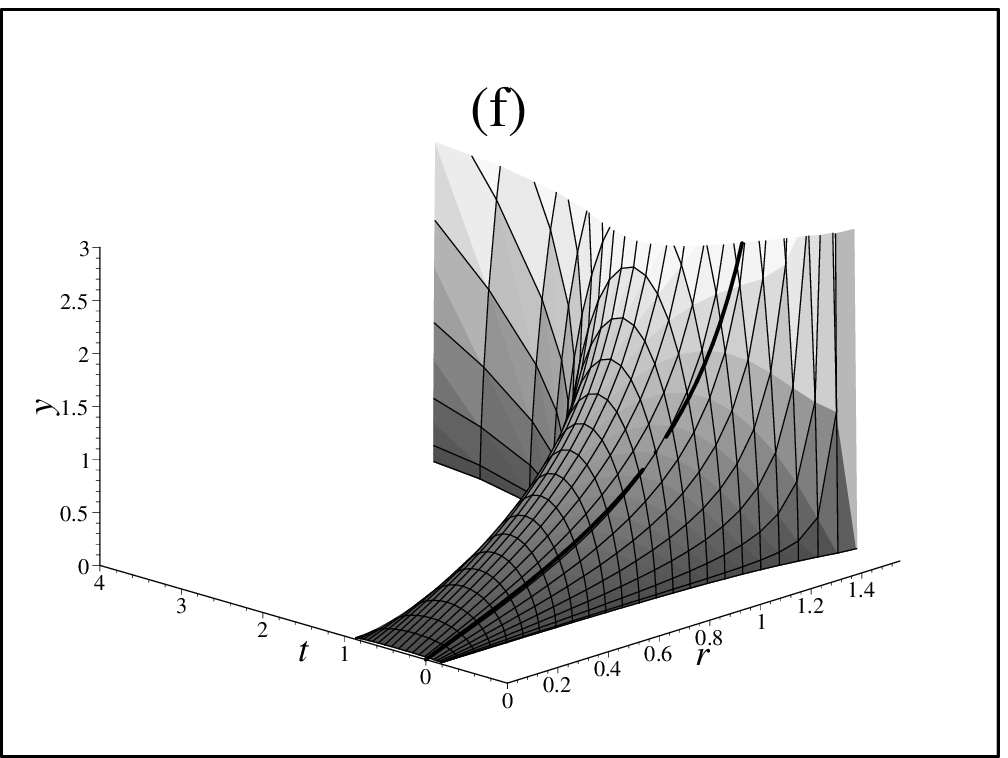 scaled 700}}\cr }\nonumber\ea
\caption{$\underline{\hbox{Comoving layers in terms of $Y(t,r)$ and $y(t,r)$}}$.
These figures examine the difference between $y$ and $Y$ by looking at the functions $y(t,r)$
and $Y(t,r)$ obtained as parametric 3-d plots for three LTB models: one with hyperbolic
dynamics and one SC (figures (a) and (b)), one with elliptic dynamics and two SC's (figures (c)
and (d)) and one with elliptic dynamics and one SC (figures (e) and (f)). Figures (a), (c)
and (e) depict $y(t,r)$, while $Y(t,r)$ is shown in figures (b), (d) and (f). Both $y$ and $Y$
have been plotted as suggested in (130), with initial value functions given in
section VIII having the following parameters: figures (a) and (b)
$[a_1=1,\,a_2=3,\,b_1=1,\,b_2=4,\,f=\tan\,r]$, figures (c) and (d)
$[a_1=1,\,a_2=2,\,b_1=1,\,b_2=3,\,f=\sin\,r]$, figures (e) and (f)
$[a_1=1,\,a_2=15,\,b_1=0,\,b_2=20,\,f=\tan\,r]$. The difference between $y$ and
$Y$ emerges by comparing (a) vs (b),\, (c) vs (d) and (e) vs (f). In figures (b) and (f)
we have $Y=0$ for the central layer marked by $r=0$, while figures (a) and (e) clearly
illustrate the fact that $y>0$ for this layer. The same happens in figures (c) and (d), but now
also along the second SC at $r=\pi$.  In figure (f) the function $Y$ diverges along
all hypersurfaces of constant $t$ as $r$ tends to the maximal value $r=\pi/2$ (since 
$f=\tan\,r\to\infty$). For the same model, as shown by figure (e), there are hypersurfaces of
constant $t$, marked by $t<1$, for which $y$ remains everywhere bounded as $r\to\pi/2$. All
these  features reinforce the interpretation of $y$ as a local ``scale factor'' for a given
layer, in contrast with the curvature radius $Y$. The grids in the figure correspond to
hypersurfaces of constant $r$ (comoving layers) and of constant $\eta$. Notice how the latter do
not coincide (in general) with hypersurfaces of constant $t$. The initial hypersurface, marked
by $t=t_i=0$, is shown as a thick line in figures (a), (c) and (e). Notice that $y=1$ for all
layers at $t=0$. The thick curves shown in figures (b), (d) and (f) are the ``profile'' of
$Y_i=Y(t_i,r)$ that corresponds to $f$ in each case. }
\label{y_vs_Y}
\end{figure}

\begin{figure}
%
%\ba\matrix{{\BoxedEPSF{susgar_11a.eps scaled 950}}&{\BoxedEPSF{susgar_11b.eps
%scaled 950}}\cr }\nonumber\ea
%
\begin{center}
\BoxedEPSF{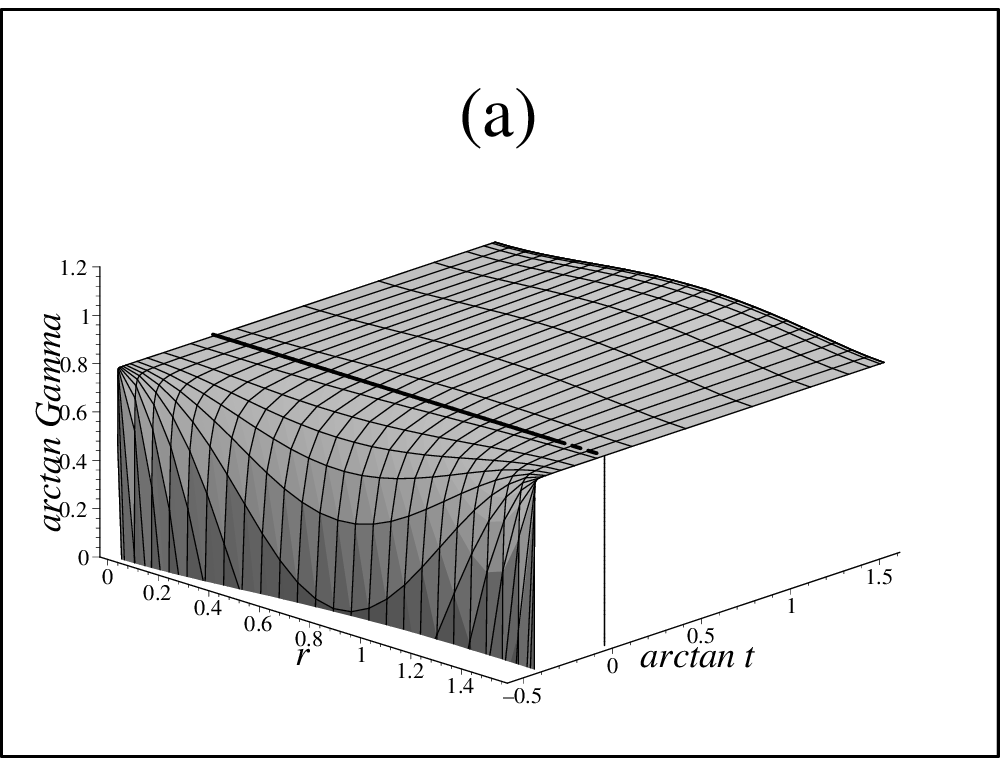 scaled 950}
\end{center}
\begin{center}
\BoxedEPSF{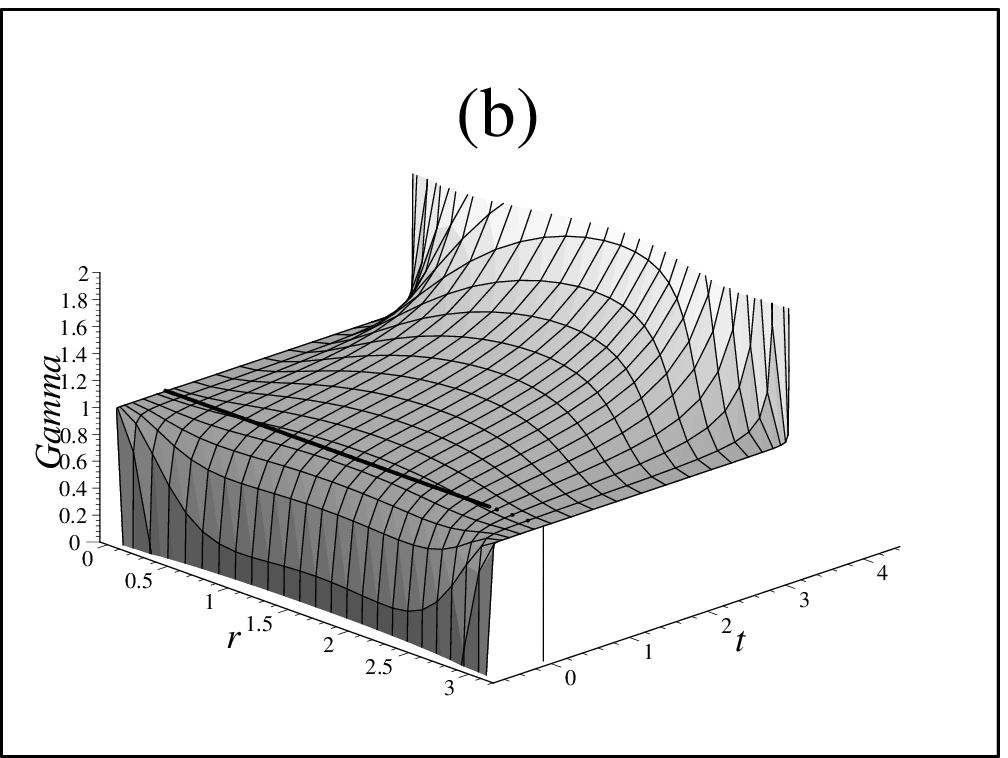 scaled 950}
\end{center}
\caption{$\underline{\hbox{Models with initial voids that are regular for $t\geq t_i$}}$.
These figures depict the parametric 3-d plot of $\Gamma(t,r)$ for models with initial density and
3-curvature voids having hyperbolic (a) and elliptic (b) dynamics. The model in (a) is
characterized by $a_1=b_1=1,\, a_2=0.45,\,b_2=0.5 $. Notice that $\Gamma$ is negative
for earlier times but $\Gamma=1$ along the initial hypersurface $t=t_i=0$ and $\Gamma>0$ for all
$t\geq t_i=0$. As long as initial value functions are finite, this type of evolution is possible
for all models with parabolic and hyperbolic dynamics and initial value functions characterized
as voids. The elliptic model in (b) corresponds to the initial value functions tested in figure
9c. As with the hyperbolic model in (a), the locuus of the shell crossing
singularity, $\Gamma=0$, is confined to times earlier than the initial hypersurface $t=t_i=0$
(marked as a thick line in the 3-d surface), likewise we have $\Gamma>0$ for all $t\geq
t_i=0$. As shown in figures 7d and 7e, the locuus of $\Gamma=0$ in
elliptic models with initial voids will exhibit, in general, two branches (as the locuus of
$y=0$). Condition (104), leading to a single branched form for $\Gamma=0$, is a
strong restriction on the initial conditions of elliptic models. We used in (b) the elliptic
model with two SC's but a very similar plot emerges by considering the case with one SC. }
\label{reg_LTBvoids}
\end{figure}

\noindent
\underline{$\hbox{Step (5). Choose a specific model.}$}
A wide variety of models with specific initial conditions can be constructed with
(\ref{ex_rhoi}), (\ref{ex_Ri}) and (\ref{ex_Yi}) by selecting suitable parameters
$\bar\rhoi,\,\bRi,\,S_i$ and constants $a_1,\,a_2,\,b_1,\,b_2$ associated with
comparative ratios with respect to reference values of $\rhoi$ and $\Ri$ (central or FLRW
values, as in (\ref{ex_cvals}), (\ref{ex_asvals}) or (\ref{ex_Fvals})), or to lumps/voids (from
(\ref{ex_l&v})), or to a given type of dynamics, while, following the discussion of section
VI-C, simple forms of
$f$ can be selected for zero, one or two SC's. Ommiting the cases with mixed dynamics we have the
following possibilities of fuly regular LTB models: 

\bi
\item{} Parabolic solutions with one SC:\, $b_1=b_2=\bRi=0$,\, $f=\tan\,r$. Central and maximal
values of $r$ are $r_c=0$ and $\rmax=\pi/2$. Initial density profiles are as shown in figures
\ref{initfunc_dens}a or \ref{initfunc_dens}b. Bang and shell crossing times are as shown in
figures \ref{bangtimes}a and \ref{bangtimes}b. See figures \ref{y_vs_Y}a, \ref{y_vs_Y}b,
\ref{reg_LTB}a, \ref{reg_LTB}b and \ref{reg_LTBvoids}a.  
\item{} Hyperbolic solutions with one SC:\, $\bRi < 0$,\, $f=\tan\,r$. Central and maximal
values of $r$ are as in the parabolic case above. Initial density profiles are as shown in
figures \ref{initfunc_dens}a or \ref{initfunc_dens}b. Initial 3-curvature profiles are as shown
in figures \ref{initfunc_ncurv}a or \ref{initfunc_ncurv}b. Bang and shell crossing times are as
shown in figures \ref{bangtimes}a and \ref{bangtimes}b. See figures \ref{y_vs_Y}a, \ref{y_vs_Y}b,
\ref{reg_LTB}a, \ref{reg_LTB}b and \ref{reg_LTBvoids}a. 
\item{} Elliptic solutions with one SC's: $\bRi > 0$, \, $f= \tan\,r$. Central and maximal
values of $r$ are the same as the parabolic and hyperbolic cases above. Initial density profiles are as shown in
figures \ref{initfunc_dens}a or \ref{initfunc_dens}b. There are no turning
values $r^*$, but (\ref{cond_K_1sc}), and (\ref{ex_cond_K23}) must hold. Necessary and
sufficient conditions for this are
\be b_1 \ = \ 0,\qquad \hbox{so that:} \qquad K \ \to \ 1 \quad
\hbox{as}\quad r\to \rmax=\pi/2 \qquad
\Rightarrow \qquad \bRi \ = \frac{2}{b_2\,S_i^2},\label{ex_cond_K22}\ee
therefore, the initial 3-curvature profiles must be as shown in figure \ref{initfunc_pcurv}a,
clearly indicating that only curvature lumps are possible since
$\Ri=2/[S_i^2(1+\tan^2(r))]\geq 0$ and so, $\Ri'\leq 0$ for all $r$. The bang and shell crossing
times are analogous to those displayed in figures \ref{bangtimes}a and \ref{bangtimes}b. A model
with these characteristics but with $\tb'=0$ was examined in \cite{bon4}. See figures
\ref{y_vs_Y}e, \ref{y_vs_Y}f, \ref{initfunc_test}b, \ref{reg_LTB}e and
\ref{reg_LTB}f.
\item{} Elliptic solutions with two SC's:  $\bRi>0$, \, $f=\sin\,r$.  Central values of $r$ are 
$r_{c_1}=0$ and $r_{c_2}=\pi$. There are no maximal values $\rmax$ and there is a turning value
$r^*=\pi/2$. Initial density profiles must be as shown in figures \ref{initfunc_dens}c or
\ref{initfunc_dens}d. Initial 3-curvature profiles must be as shown in figures
\ref{initfunc_pcurv}b or \ref{initfunc_pcurv}c.  Condition (\ref{ex_cond_K23}) must be satisfied
with the equality holding for the turning value
$r^*=\pi/2$ so that
$f_*=\sin(\pi/2)=1$. Hence the parameters
$b_1,\,b_2,\,S_i,\,\bRi$ are linked by the constraint
\be K(\pi/2) \ = \ 1\qquad\Rightarrow\qquad b_1 \ + \ 3\,(\,b_2\,-\,b_1\,) \,
\left(1\,-\,\frac{\pi}{4}\right)
\ =
\
\frac{6}{\bRi\,S_i^2},\label{ex_cond_K222}\ee
See figures \ref{y_vs_Y}c, \ref{y_vs_Y}d,
 \ref{initfunc_test}a, \ref{initfunc_test}c, \ref{reg_LTB}c, \ref{reg_LTB}d and
\ref{reg_LTBvoids}b.                           
\ei
Since condition (\ref{ex_cond_K23}) cannot hold if $K\leq 0$ (parabolic and hyperbolic
dynamics), surface layers necessarily emerge at turning points $f'(r^*)=0$ for parabolic and
hyperbolic solutions with two SC's. However, these configurations can still be constructed with
the expressions derived in this section. \\

\noindent
\underline{$\hbox{Step (6). Evaluate and plot time dependent quantities.}$}

Once all previous steps have been accomplished, we have all $r$ dependent quantities in
(\ref{parab_can2})-(\ref{hyp_def_eta}) fuly determined by (\ref{ex_rhoi})-(\ref{ex_Yi}) and
(\ref{ex_Mmu})-(\ref{ex_Dki}) and (\ref{ex_x}), so that all relevant regularity conditions, such
as (\ref{shx_st_p}), (\ref{shx_st_h}), (\ref{shx_st_e4}), (\ref{shx_st_e5}) and
(\ref{ex_cond_K23}) can be tested (see Table 1). If this test is passed, we have fuly determined
initial conditions for a regular evolution. We are ready to derive expressions for
$y$ (or $Y$ from (\ref{defy})) for all $t$ beyond the initial hypersurface $\Ti$.  For parabolic
dynamics, we have $y=y(t,r)$ from (\ref{parab_can2}). For hyperbolic and elliptic dynamics, we
have either an implicit solution $t=t(r,y)$ (as in (\ref{ell_can21}), (\ref{ell_can22}) and
(\ref{hyp_can2})) or parametric solutions $[t=(r,\eta),y=y(r,\eta)]$ (as in (\ref{ell_par_Y2}),
(\ref{ell_par_t2}), (\ref{hyp_par_Y2}) and (\ref{hyp_par_t2})). The function $\Gamma$ can be
obtained as $\Gamma(t,r)$ in the parabolic case by inserting (\ref{parab_can2}) into
(\ref{Gamma_par}), then $\rho=\rho(t,r)$ follows from (\ref{rho2}). For hyperbolic or elliptic
dynamics we have $\Gamma=\Gamma(r,\eta)$ from (\ref{Gamma_gen}) and $\rho=\rho(r,\eta)$ also
follows from (\ref{rho2}). All other time-dependent quantities, such as $\R,\,\sigma,\,\Theta$
and other invariant scalars given by equations (\ref{Theta1}), (\ref{sigma1}), (\ref{def_R}),
(\ref{def_E}) and (\ref{def_scalars}), can be transformed as functions of $(t,r)$, for
parabolic dynamics, and as functions of $(r,y)$ or $(r,\eta)$, for hyperbolic and elliptic
cases, by eliminating gradients like $Y'$ and $\dot Y'$ in terms of $\Gamma$ from (\ref{Gamma})
and then insterting the previously known $\Gamma(t,r)$ or $\Gamma(r,\eta)$. For hyperbolic and
elliptic dynamics, we cannot obtain closed functional forms for the time dependent quantities
in terms of $(t,r)$. However, we can obtain 3d plots of $y$ (and thus of all these expressions)
in terms of $(t,r)$, since they are all real functions defined over a domain contained in the
plane $(t,r)$, hence their image can be represented as parametric 3d plots of the form
\ba y \ = \ [r, \,t(r,\eta),\,y(t,\eta)],\qquad  \Gamma \ = \ [r,
\,t(r,\eta),\,\Gamma(t,\eta)],\qquad \rho \ = \ [r,
\,t(r,\eta),\,\rho(t,\eta)],\qquad \hbox{etc.}\nonumber\ea
where $\Gamma(r,\eta)$ is given by (\ref{Gamma_gen}), while $\rho(r,\eta)$ follows from
(\ref{rho2}) once $y(r,\eta)$ and $\Gamma(r,\eta)$ are known (we are assuming, of course, that
all $r$-dependent functions, such as $\rhoi$ have already been determined by steps 1 to
5).  These 3d plots can be easily generated with any general use Computer Algebra
System, like Maple or Mathematica. The corresponding commands for plotting $y(t,r)$ are:
\ba {\tt{plot3d}\,([\,r, \,\,t(r,\hbox{eta}),\,\, y\,(r,\hbox{eta})\,], \ \ r\ = \
0\,.\,.\,\hbox{Pi}, \ \ \hbox{eta}\ = \ 0\,.\,.\,\hbox{2*Pi}\,);} \qquad
\qquad\hbox{Maple},\cr\cr\cr {\tt{ParametricPlot3D}\,[\,\{\,r,\ t[\,r,\,\hbox{eta}\,], \
y\,[\,r,\,\hbox{eta}\,]\,\},
\ \ \{\,r,\,0,\,\pi\}, \ \ \{\,\hbox{eta},\,0,\,2\pi\,\}\,]} \qquad \hbox{Mathematica},
\label{plot_coms}\ea
where the parametric functions $t(r,\eta),\,y(r,\eta)$ must be previously defined in each
package and we considered in the example above the elliptic case with $f=\sin\,r$. The plotting
commands for other quantities follow the same syntaxis. The reader familiar with either Maple or
Mathematica (or other package) can, of course, embellish the plots by adding a wide
variety of plot options concerning axes, fonts, labels, colors, etc. All curves and surfaces
displayed in the figures were drawn with Maple, with those surfaces in figures \ref{layers_e}, 
\ref{y_vs_Y}, \ref{reg_LTB} and \ref{reg_LTBvoids} plotted using commands like that shown in
(\ref{plot_coms}). For the case of a simmultaneous big bang, we can still obtain simmilar 3d
plots, but $t(r,\eta)$ and $y(r,\eta)$ need to be constructed with numerical values of
either one of $\mu$ or $\kappa$.           

\acknowledgements
This work has been supported by the National University of Mexico (UNAM),
under grant DGAPA-IN-122498.

%\[\]

\section*{Table of no-shell-crossing conditions.}

\begin{table}
\begin{center}
\begin{tabular}{|c| c| c|}
\hline 
\hline
\multicolumn{3}{|c|}{In all cases $\rhoi$ and $\Ri$ must be finite. Also: $\rhoi\geq 0$
and:
 $\Ri\geq 0$ (elliptic), $\Ri\leq 0$ (hyperbolic)  }
\\  
\hline
\hline
\multicolumn{3}{|c|}{$t'_{bb}\ne 0$}
\\  
\hline
\hline
{} &{} &{}
\\
{Parabolic} &{$-1 \ \leq \ \Dmi \ \leq \ 0$} &{eq. (\ref{shx_st_p})\qquad} 
\\
{} &{} &{} 
\\
{Hyperbolic} &{$-1 \ \leq \ \Dmi \ \leq \ 0,\qquad -\frac{2}{3} \ \leq \ \Dmi \ \leq \ 0,$}
&{eq. (\ref{shx_st_h})\qquad}
\\
{} &{} &{} 
\\
{Elliptic} &{$-1 \ \leq \ \Dmi \ \leq \ 0,\qquad -\frac{2}{3} \ \leq \ \Dmi \ \leq \ 0,$} &{eq.
(\ref{shx_st_e4})\qquad}
\\
{} &{} &{} 
\\
{} &{$2\pi\,P_i\,\left(\Dmi\,-\,\frac{3}{2}\Dki\right) \ \geq \
P_i\,Q_i\,\left(\Dmi\,-\,\frac{3}{2}\Dki\right) \ - \ \left(\Dmi\,-\,\Dki\right) \ \geq \
0,\qquad$} &{eq. (\ref{shx_st_e5})\qquad} 
\\
{} &{} &{}
\\
\hline
\hline
\multicolumn{3}{|c|}{$t'_{bb}= 0$}
\\
\hline
\hline
{} &{} &{}
\\
{Hyperbolic} &{$-1 \ \leq \ \Dmi,\qquad -\frac{2}{3} \ \leq \ \Dki,\qquad \Dmi\,\Dki \ \leq \
0,$} &{eq. (\ref{shx_stb2})\qquad}
\\
{} &{} &{} 
\\
{Elliptic} &{$-1 \ \leq \ \Dmi,\qquad \Dmi\,-\,\frac{3}{2}\Dki \geq \ 0,\qquad \Dmi\,\Dki \
\geq \ 0,$} &{eq. (\ref{shx_stb3})\qquad}
\\
{} &{} &{}
\\   
\hline 
\hline
\multicolumn{3}{|c|}{No-shell-crossings for $t\geq t_i$}
\\ 
\hline
\hline
{} &{} &{}
\\
{Hyperbolic} &{Always} &{}
\\
{and parabolic} &{possible} &{} 
\\
{} &{} &{}
\\
{Elliptic} &{$\Dki \ \leq \ \Dmi \ \leq \frac{1}{2}\,+\,\frac{3}{2}\Dki $} &{eq.
(\ref{reg_voids_e})\qquad}
\\ 
{} &{} &{}
\\ 
\hline 
\hline
\end{tabular}
\end{center}
\caption{$\underline{\hbox{Summary of no-shell-crossing conditions discussed in section
\ref{Gamma_etc}.}}$}
\end{table}

\section*{Figures.}

\vskip 1cm

\vskip 1cm
\appendix
\section*{Invariants and generic features of LTB solutions.}\label{app}

\bi
\item{$\underline{\hbox{Kinematic invariants}}$}.
For (\ref{ltbmetric}) and (\ref{Tdust}), we have $\dot u_a=u_{a;b}\,u^b=0$, so
that the 4-velocity is a geodesic field, while the expansion scalar and shear
tensor are
\be
u^a\,_{;a}\equiv\Theta= {\dot Y'\over{Y'}}+{{2\dot Y}\over
Y},\label{def_Theta}\qquad\qquad \label{Theta1} 
\ee
\be
\sigma^a\,_b= {\bf{\hbox{diag}}}\left[0,-2\sigma,\sigma,\sigma
\right],\,\qquad \sigma\equiv{1\over 3}\left({\dot Y\over Y}-{\dot Y'\over
Y'}\right),\qquad \sigma_{ab}\,\sigma^{ab}=6\sigma^2,\qquad
\frac{\Theta}{3} + \sigma = \frac{\dot Y}{Y},\label{sigma1}
\ee
\item{$\underline{\hbox{Curvature scalars.}}$}.
The Ricci scalar, $\R$, of the 3-dimensional hypersurfaces of constant $t$ (rest
frames of comoving observers) can be expressed in terms of $\rho$, $\Theta$ and
$\sigma$ by the equation
\be\frac{c^2}{6}\,\R \ = \ \frac{8\pi G}{3c^2}\,\rho \ - \frac{1}{9}\,\Theta^2 \
+ \ \sigma^2 \ = \ -\frac{c^2\,(E\,Y)'}{3\,Y^2Y'},\label{def_R} 
\ee
The 4-dimensional Ricci scalar, Ricci tensor squared and Riemann tensor squared
(the Kretschmann scalar) are
\ba {\cal{R}} \ = \ \frac{8\pi G}{c^4}\,\rho,\qquad\
{\cal{R}}_{ab}\,{\cal{R}}^{ab} \ = \ {\cal{R}}^2\qquad\qquad\qquad\qquad\cr\cr
 {\cal{R}}_{abcd}\,{\cal{R}}^{abcd} \ = \
\frac{4}{3}\,\left(\frac{6\,M}{Y^3}\,-\,{\cal{R}}\right)^2 \ - \
\frac{5}{3}\,{\cal{R}}_{ab}\,{\cal{R}}^{ab}
\ =
\
\left(\frac{8\pi
G}{c^4}\right)^2\,\left[\frac{4}{3}\,\left(\frac{3\,m}{Y^3}\,-\,\rho\right)^2\,+\,
\frac{5}{3}\,\rho^2\right],\label{def_scalars}\ea

while the electric part of the conformal Weyl tensor, $E^a\,_b\equiv C_{abcd}u^au^b$,
is given in terms of the only nonzero invariant Weyl curvature scalar, $\psi_2$, by 
\be E^a\,_b \ = \
\hbox{\bf{diag}}\,[0,-2\psi_2,\psi_2,\psi_2],
\qquad\qquad \psi_2 \ = \ \frac{4\pi G}{3c^4}\,\left[ \ \frac{3\,m}{Y^3 }\ - \
\rho \ \right] \ = \ \frac{M}{Y^3} \ - \ \frac{4\pi G}{3c^4}\,\rho,\label{def_E}
\ee

\item{$\underline{\hbox{Regularity and differentiability.}}$}.

Spherical symmetry is characterized by the group SO(3) acting transitively along
2-spheres (orbits) with proper area $4\pi\,Y^2(t,r)$. The domain of regularity of
LTB solutions can be given then as the set of all spacetime points marked by
$(t,r)$ for which (a) $Y\geq 0$ and (b) all curvature scalars (like those in
(\ref{def_R}), (\ref{def_scalars}) and (\ref{def_E})) are bounded. 

Minimal differentiability requirements for the basic variables can be summarized
as follows \cite{humph_phd}, \cite{HMM1}: $Y$ and all its time derivatives are
continuous, while quantities involving radial derivatives like
$Y',\,\dot Y',\,Y'',$ etc can be piecewise continuous, thus allowing jump
discontinuities for a countable number of values $r=r_1$ so that the right and
left limits as $r\to r_1$ are different but this difference is always finite.
These conditions imply (via (\ref{freq1})) that the basic variables
$Y,\,M,\,E,\,\tb$ must be at least continuous (ie
$C^0$), while $\rho$ and other quantities ($\Theta,\,\sigma,\,\R,\,\psi_2$)
containing terms like $Y'$ or $\dot Y'$ could be piecewise continuous.

From (\ref{rho1}), the weak energy condition leads to
\be \rho \ \geq \ 0 \qquad \Leftrightarrow \qquad \hbox{sign}(Y') \ = \
\hbox{sign}(M'),\label{wec}\ee
while regularity of the metric component $g_{rr}$
requires 
\be 1 \ + \ E \ \geq \ 0,\label{cond_K_2}\ee
a condition that is trivialy satisfied for
parabolic and hyperbolic ($E\geq 0 $), but not for elliptic
solutions or regions ($E< 0$). The equality in (\ref{cond_K_2}) implies that $Y'$
and $E$ must have a common zero of the same order (see below).  

\item{$\underline{\hbox{Zeroes of} \ Y \ \hbox{and} \ Y'}$}

The functions $Y$ and $Y'$ can vanish, either for a value $r=$ const., or for a
coordinate locuus $(t,r)$ with $r\ne$ const. The two cases lead to very different
situations:

\bi
\item{} Zeroes of $Y$ for $r=r_c$ ($\underline{\hbox{Symmetry centers, SC}}$).\\
These regular worldlines marked by $r=r_c$ are the worldlines of fixed points of
SO(3). LTB solutions (like all sphericaly symmetric spacetimes) admit zero, one or two
SC's. They are characterized by $Y(t,r_c)=\dot Y(t,r_c)=0$. All of
$M(r_c),\,M'(r_c),\,E(r_c),\,E'(r_c),\tb'(r_c)$ vanish at a SC, but the order of the
zeroes of these functions (power of leading terms) follows from the regularity of
each term in (\ref{freq1}). Near $r=r_c$ we have \cite{humph_phd}, \cite{HM1},
\cite{HMM1}  
\ba Y\approx Y'(t,r_c)(r-r_c),\qquad M\approx M'''(r_c)(r-r_c)^3,\qquad
E\approx E''(r_c)(r-r_c)^2, \qquad \tb\approx
\tb''(r_c)(r-r_c)^2,\label{central_1}  
\ea 

\item{} Zeroes of $Y$ for non-comoving locii ($\underline{\hbox{Central
singularities}}$).\\ 
All curvature scalars, such as (\ref{def_R}), (\ref{def_scalars}) and (\ref{def_E}),
diverge as the proper area of the orbits of SO(3) vanish, hence the name ``central''
(associated with singular centers). By analogy with FLRW spacetimes and depending on
whether dust layers expand or collapse, these singularities are called a ``big bang''
or a ``big crunch''. These singularities cannot be avoided for any choice of free
parameters, thus they are an essential feature of LTB models.

\item{} Zero of $Y'$ for $r=r^*$ ($\underline{\hbox{Turning values of $Y'$ and
surface layers}}$).\\ 
The extrinsic curvature forms of the 3-dimensional submanifolds $r=$ const. must be
continuous (at least $C^0$) as $r$ increases. If if  $Y'(t,r^*)=0$ for a comoving
$r^*\ne r_c$, these extrinsic curvature forms are continuous as $r\to r^*$ as long
as the following regularity conditions hold \cite{bon3}, \cite{humph_phd}, \cite{HM1},
\cite{HMM1}
\be Y'(t,r^*) \ = \ 0 \qquad \Leftrightarrow \qquad M'(r^*) \ = \ 0, \qquad \ 1 \
+ \ E(r^*) \ = \ 0,\label{cond_K_1}\ee
implying that the common zero, $r=r^*$, of $Y',\,M',\,1+E$ is of the same order in
$r-r^*$. If these conditions fail to hold, the extrinsic curvature of hypersurfaces
$r=$ const. is discontinuous as $r\to r^*$ and we have a surface layer at
$r=r^*$ (see \cite{ZG}, \cite{bon3}, \cite{humph_phd}, \cite{HM1}, \cite{HMM1}). 

\item{} Zeroes of $Y'$ for non-comoving locii ($\underline{\hbox{Shell crossing
singularities}}$).\\ 
From (\ref{rho1}), energy density $\rho$ diverges (and thus all curvature scalars like
(\ref{def_R}), (\ref{def_scalars}) and (\ref{def_E}) diverge). Since $g_{rr}$ vanishes
for (in general) $Y>0$, proper distances between comoving layers vanish, hence the
name ``shell crossing''. Unlike central singularities, shell crossings  are unphysical
and are not esential. Conditions for having LTB models free from these singularities
can be given in terms of $M,\,E,\,\tb$ and their gradients (see section VII, Table 1 and
references \cite{he_la2}, \cite{humph_phd}, \cite{HM1}, \cite{HMM1}). 
\ei 
 
\item{$\underline{\hbox{``Open'' or ``closed'' models}}$}. Hypersurfaces $t=$
const, ${\cal{T}}$, are everywhere orthogonal to the 4-velocity vector field $u^a$,
hence they can be characterized invariantly as the rest frames of comoving
observers associated with the metric 
\be h_{ab} \ = \ u_a\,u_b \ + \ g_{ab}\ee
All ${\cal{T}}$ are Cauchy hypersurfaces \cite{burnet} in LTB solutions. They are
foliated by 2-spheres marked by constant values of $r$. However, the radial coordinate
can be arbitrarily rescaled, and so an invariant measure of distances along its
direction is given by the proper length integral $\ell = \int{\sqrt{g_{rr}}dr}$,
evaluated for spacelike curves with fixed $(\theta,\phi)$. For a regular and maximaly
extended hypersurface, the domain of
$r$ is limited, either by SC's (since $Y\geq 0$) and/or by maximal values
$r=\rmax$ defined by 
\be \ell \ = \ \int{\frac{Y'dr}{\sqrt{1\,+\,E}}} \ \to \ \infty
\qquad\hbox{as}\qquad r \
\to \ \rmax,\label{ell1}\ee
The terms ``open'' and ``closed'' respectively corresponds to an infinite or
finite proper volume of these hypersurfaces, evaluated from (\ref{ltbmetric}) as
\be V \ =\  4\pi\int{\frac{Y^2Y'dr}{\sqrt{1\,+\,E}}},\label{def_V}\ee
where the integral must be evaluated along all the domain of
$r$. The class of homeomorphism of hypersurfaces $t=$ const.
is related the number of SC's (see section VI-C for more details).  From
familiarity with FLRW spacetimes, one would tend to expect a ``natural''
association between closed topologies with elliptic (recollapsing) dynamics and
open topologies with parabolic or hyperbolic dynamics. However, this association
need not be mandatory (see \cite{he_la2}, \cite{hellaby1} and sections VI and VIII):
open topologies might have elliptic dynamics and closed topologies might present
parabolic or hyperbolic evolution, though in this last case a surface layer
appears at $r=r^*$ such that $Y'(t,r^*)=0$ (see \cite{ZG}, \cite{bon3},
\cite{humph_phd} and
\cite{HMM1}). 

	\ei

\end{document}